\RequirePackage[l2tabu, orthodox]{nag}
\documentclass[12pt,a4paper]{article}

\usepackage{fixltx2e}

\usepackage[font=small]{caption}
\usepackage[text={450pt,650pt},headheight={14.5pt},centering]{geometry}

\usepackage{lmodern}

\usepackage[nosort]{cite}

\usepackage{graphicx}
\usepackage{booktabs}
\usepackage{subfig}
\usepackage{tikz,pgfplots}

\usetikzlibrary{plotmarks,calc,decorations,decorations.pathmorphing,patterns,intersections}

\tikzstyle dynkin node=[very thick,shape=circle,draw,inner sep=0pt,minimum size=5mm]
\tikzstyle dynkin line=[very thick]
\tikzstyle inverse line=[gray,line width=1.46pt,line cap=round, dash pattern=on 0pt off 2\pgflinewidth]
\tikzstyle red phase=[thick,red,decoration={snake,amplitude=0.1mm,segment length=1.6mm},decorate]
\tikzstyle blue phase=[thick,blue,decoration={snake,amplitude=0.1mm,segment length=0.9mm},decorate]
\tikzstyle green phase=[thick,green,decoration={snake,amplitude=0.1mm,segment length=0.9mm},decorate]
\tikzstyle brown phase=[thick,brown,decoration={snake,amplitude=0.1mm,segment length=0.9mm},decorate]

\tikzstyle arrow=[thick,rounded corners=18pt,-latex]
\tikzstyle box=[draw,rounded corners,outer sep=4pt]
\tikzstyle B node=[outer sep=0pt]
\tikzstyle Q node=[inner sep=1pt,outer sep=0pt]

\pgfdeclarelayer{background}
\pgfdeclarelayer{foreground}
\pgfsetlayers{background,main,foreground}

\usepackage{dsfont}
\usepackage{collref}

\usepackage{amsmath,amssymb}
\usepackage{mathrsfs}
\usepackage{mathtools}
\usepackage{bbm}
\usepackage{slashed}
\usepackage{braket}

\DeclareMathAlphabet{\mathsfit}{\encodingdefault}{\sfdefault}{m}{sl}

\usepackage{hyperref}

\usepackage{xspace}

\numberwithin{equation}{section}

\makeatletter
 \let\old@startsection=\@startsection
 \let\oldl@section=\l@section
 \renewcommand{\@startsection}[6]{\old@startsection{#1}{#2}{#3}{#4}{#5}{#6\mathversion{bold}}}
 \renewcommand{\l@section}[2]{\oldl@section{\mathversion{bold}#1}{#2}}
\makeatother

\renewcommand{\leq}{\leqslant}

\DeclareMathOperator{\tr}{tr}

\def\XXint#1#2#3{{\setbox0=\hbox{$#1{#2#3}{\int}$}
    \vcenter{\hbox{$#2#3$}}\kern-.5\wd0}}

\newcommand{\AdS}{\text{AdS}}
\newcommand{\CFT}{\mathrm{CFT}}
\newcommand{\Sphere}{S}
\newcommand{\Torus}{T}

\newcommand{\matId}{\mathds{1}}

\newcommand{\Smat}{\mathcal{S}}

\newcommand{\comm}[2]{[#1,#2]}
\newcommand{\acomm}[2]{\{#1,#2\}}

\newcommand{\alg}[1]{\mathfrak{#1}}
\newcommand{\grp}[1]{\mathrm{#1}}

\newcommand{\algSL}{\alg{sl}}

\newcommand{\algSU}{\alg{su}}

\newcommand{\grpU}{\grp{U}}

\newcommand{\algPSU}{\alg{psu}}
\newcommand{\grpPSU}{\grp{PSU}}

\newcommand{\algGL}{\alg{gl}}

\newcommand{\gen}[1]{\mathbf{#1}}
\newcommand{\Ygen}[1]{#1}

\newcommand{\smallL}{\scriptscriptstyle\textit{L}}
\newcommand{\smallR}{\scriptscriptstyle\textit{R}}

\newcommand{\Integers}{\mathbbm{Z}}

\newcommand{\Complex}{\mathbbm{C}}

\newcommand{\order}{\mathcal{O}}

\newcommand{\superN}{\mathcal{N}}

\renewcommand{\Re}{\operatorname{Re}}
\renewcommand{\Im}{\operatorname{Im}}

\newcommand{\ie}{\textit{i.e.}\xspace}
\newcommand{\eg}{\textit{e.g.}\xspace}
\newcommand{\cf}{\textit{cf.}\xspace}

\definecolor{dgreen}{rgb}{0,0.7,0.2}
\newcommand{\HL}{\mbox{\scriptsize HL}}
\newcommand{\BES}{\mbox{\scriptsize BES}}
\newcommand{\AFS}{\mbox{\scriptsize AFS}}

\newcommand{\curvearrowurl}{\curvearrowleft}
\newcommand{\curvearrowulr}{\curvearrowright}
\newcommand{\curvearrowdlr}{\rotatebox[origin=c]{180}{$\curvearrowleft$}}

\newcommand{\inturl}{\;{\int \negthickspace \negthickspace \negthickspace\negthickspace \negthinspace \curvearrowurl}\mbox{ }\,} 
\newcommand{\intulr}{\;{\int  \negthickspace \negthickspace \negthickspace \negthinspace  \curvearrowulr}\mbox{ }\,} 
\newcommand{\intdlr}{\;{\int \negthickspace \negthickspace \negthickspace \negthinspace \curvearrowdlr}\mbox{ }\,}

\begin{document}
\thispagestyle{empty}
\begin{flushright}\footnotesize\ttfamily
{DMUS-MP-16/13\\
Imperial-TP-RB-2016-04\\
NORDITA-2016-79}
\end{flushright}
\vspace{4em}

\begin{center}
\textbf{\Large\mathversion{bold} On the Dressing Factors, Bethe Equations and Yangian Symmetry of Strings on $\AdS_3\times \Sphere^3\times \Torus^4$}

\vspace{2em}

\textrm{\large Riccardo Borsato${}^1$, Olof Ohlsson Sax${}^2$, Alessandro Sfondrini${}^3$,\\ Bogdan Stefa\'nski, jr.${}^4$ and Alessandro Torrielli${}^5$} 

\vspace{2em}

\vspace{1em}
\begingroup\itshape\small
1. The Blackett Laboratory, Imperial College, London SW7 2AZ, United Kingdom

2. Nordita, Stockholm University and KTH Royal Institute of Technology, Roslagstullsbacken 23, SE-106 91 Stockholm, Sweden

3. Institut f\"ur Theoretische Physik, ETH Z\"urich, Wolfgang-Pauli-Str.~27, CH-8093 Z\"urich, Switzerland

4. Centre for Mathematical Science, City University London, Northampton Square, EC1V 0HB London, UK

5. Department of Mathematics, University of Surrey, Guildford, GU2 7XH, UK\par\endgroup

\vspace{1em}

\texttt{r.borsato@imperial.ac.uk, olof.ohlsson.sax@nordita.org, sfondria@itp.phys.ethz.ch, Bogdan.Stefanski.1@city.ac.uk, a.torrielli@surrey.ac.uk}


\end{center}

\vspace{6em}

\begin{abstract}\noindent
Integrability is believed to underlie the $\AdS_3/\CFT_2$ 
correspondence with sixteen supercharges. 
We elucidate the role of massless modes
within this integrable framework. Firstly, we find the dressing factors
that enter the massless and mixed-mass worldsheet S matrix. Secondly, 
we derive a set of all-loop Bethe Equations for 
the closed strings, determine their symmetries and 
weak-coupling limit. Thirdly, we investigate the underlying Yangian 
symmetry in the massless sector and show that it fits into 
the general framework of Yangian integrability. In addition, we
compare our S matrix in the near-relativistic limit with recent perturbative worldsheet calculations of Sundin and Wulff.
\end{abstract}

\newpage

\tableofcontents

\section{Introduction}
\label{sec:introduction}

Over the last few years integrable methods have been extensively 
employed in the context of the spectral problem of the $\AdS_3/\CFT_2$ 
correspondence, see for  example~\cite{David:2008yk,Babichenko:2009dk,OhlssonSax:2011ms,
Sundin:2012gc,Cagnazzo:2012se,Borsato:2012ud,Borsato:2013qpa}. 
Initial progress did not include world-sheet massless 
modes~\cite{Babichenko:2009dk,OhlssonSax:2011ms}, for a review see~\cite{Sfondrini:2014via}.\footnote{For some early 
results on massless modes in the weak- and strong-coupling limits 
see~\cite{Sax:2012jv} and~\cite{Lloyd:2013wza,Abbott:2014rca}, respectively.}. 
Subsequently, it was shown~\cite{Borsato:2014exa,Borsato:2014hja,Lloyd:2014bsa,Borsato:2015mma} 
that these can be included in a novel integrable all-loop world-sheet 
S matrix, which was determined, up to dressing phases, for 
$\AdS_3\times \Sphere^3\times \Torus^4$ and 
$\AdS_3\times \Sphere^3\times \Sphere^3\times \Sphere^1$ supported 
by R-R flux and mixed R-R/NS-NS flux. While the dressing phases are 
not fixed by the symmetries of the theory,  they satisfy crossing 
equations which were also found.

This progress in exact results was accompanied by a large 
body of perturbative world-sheet computations and a number of successful 
comparisons between the two was performed in the massive sector, 
see for example~\cite{Rughoonauth:2012qd,Abbott:2012dd,Beccaria:2012kb,Beccaria:2012pm, 
Sundin:2013ypa,Sundin:2013uca,Hoare:2013pma,Bianchi:2013nra,Hoare:2013ida,
Engelund:2013fja,Abbott:2013ixa,Hoare:2013lja,Sundin:2014sfa,Bianchi:2014rfa,
Hernandez:2014eta,Stepanchuk:2014kza,Roiban:2014cia}.~\footnote{Further papers on integrable $\AdS_3/\CFT_2$ holographic results include~\cite{Abbott:2015mla,Abbott:2014pia,Wulff:2015mwa,
Wulff:2014kja,Wulff:2013kga,Chervonyi:2016ajp,Kluson:2015lia,
Hernandez:2015nba,Ahn:2014tua,David:2014qta,Kluson:2016dca,
Banerjee:2016avv,Prinsloo:2015apa}.} In addition, 
these comparisons have also resulted in two unresolved issues. Firstly, 
the massless dispersion relation,  determined via a (super-)symmetry 
argument in~\cite{Sundin:2014ema}, does not agree with a two-loop 
perturbative computation~\cite{Borsato:2014exa,Borsato:2014hja}. 
Secondly, the massive dressing factor obtained by solving the crossing 
equations~\cite{Borsato:2013hoa} was found to be slightly different 
from the one calculated using perturbative world-sheet 
calculations~\cite{Beccaria:2012kb,Beccaria:2012pm}. The first 
issue's resolution might come about by employing a more symmetric 
regularisation scheme, or a modified definition of asymptotic states. 
Given that the all-loop massless dispersion relation follows from a supersymmetric 
shortening condition, this issue deserves to be better understood. 
A possible explanation for the second issue was recently proposed 
in~\cite{Abbott:2015pps}, where it was argued that a proper 
incorporation of the wrapping corrections of massless modes is 
likely to correct the perturbative world-sheet calculations in 
a way that would make them consistent with crossing.

Beside these issues, a more complete understanding of how the massless 
modes enter the integrable construction was still lacking. While the 
S-matrix and the crossing equations for processes involving massless 
modes had been found~\cite{Borsato:2014exa,Borsato:2014hja,Lloyd:2014bsa,Borsato:2015mma}, 
the analytic properties of massless modes, and their dressing factors 
remained to be determined. In this paper we address these outstanding 
problems and find the minimal solutions to the crossing equations, 
giving a detailed exposition of the results announced in~\cite{Borsato:2016kbm}. 
On general grounds dressing phases have an expansion in the coupling 
constant, with the leading and next-to-leading orders on the string 
theory side conventionally referred to as the Arutyunov-Frolov-Staudacher 
(AFS)~\cite{Arutyunov:2004vx} and Hern\'andez-L\'opez 
(HL)~\cite{Hernandez:2006tk} orders. The minimal solution for the 
massless dressing factor is non-trivial only at HL order, while the 
mixed-mass dressing factor minimal solution is non-trivial at AFS 
and HL order. Our solutions have a very natural interpretation as 
coming from a ``massless limit'' of the corresponding massive phases. 
By considering this limit for the non-perturbative Beisert-Eden-Staudacher (BES) phase~\cite{Beisert:2006ez} we argue that no 
natural candidate for homogeneous solutions exists at higher orders, 
while a homogeneous AFS order term might be natural. Further, we 
investigate the possibility of massless bound states in the spectrum. 
We argue that such states are not allowed kinematically and
confirm their absence by an explicit analysis of the all-loop S matrix, 
including our dressing factors. It is worth pointing out that the 
absence of massless bound states is also a well-known feature of {\em relativistic} massless integrable models~\cite{Zamolodchikov:1992zr}.

We then turn to the derivation of the Bethe equations for the 
complete closed string spectrum from the S matrix. We impose 
periodicity and employ the nesting procedure to find the Bethe 
equations. The structure of the Bethe equations depends on a 
choice of grading for the underlying super-algebra. We write 
them in the two inequivalent gradings, showing how these are related through a 
fermionic duality. We show that these equations reduce to the 
Bethe equations for the massive modes~\cite{Borsato:2013qpa} 
when no massless excitations are present. We demonstrate that the 
spectrum has degeneracies that follow from the global 
$\algPSU(1,1|2)^2$ symmetry of the theory, as well as from translations along the four directions of the torus. 
We determine the weak-coupling or spin-chain limit of the Bethe 
equations and show these latter equations can also be obtained 
directly from the weakly-coupled limit of the S matrix. Further 
we write down the Bethe equations for the mixed NS-NS and R-R 
flux supported $\AdS_3\times \Sphere^3\times \Torus^4$ background, 
though in this case solving the crossing equations remains an open problem.

Subsequently, We turn to the Berenstein-Maldacena-Nastase (BMN)
limit~\cite{Berenstein:2002jq} of the S matrix. In this near-relativistic limit 
the massless excitations become either left- or right- moving on 
the worldsheet and the massless S matrix degenerates into two S matrices, 
depending on whether massless particles of same or different worldsheet 
chiralities are scattered. For the scattering of same-chirality particles 
the S matrix is difficult to interpret within a perturbative 
worldsheet framework. On the other hand the S matrix for scattering 
particles of opposite worldsheet chiralities has a good perturbative 
expansion. Both these features are entirely in agreement with the 
general results for relativistic massless integrable theories~\cite
{Zamolodchikov:1992zr,Fendley:1993wq,Fendley:1993xa} (see also~\cite
{Polyakov:1983tt,Polyakov:1984et,Fioravanti:1996rz}). We compare 
the expansion of the mixed worldsheet chirality S matrix with the 
recent perturbative world-sheet results~\cite{Sundin:2016gqe}. For 
the most part we find exact agreement. When comparing certain terms 
that depend on the dressing factors, some of the perturbative calculations 
suffer from infrared ambiguities making a comparison less well-defined.

Finally, we study the Yangian symmetry underlying the massless-massless 
S-matrix. We find that the scattering problem in this sector is 
controlled by a Yangian algebra of the same general type as the 
massive-massive one. We obtain the specific evaluation representation 
and the corresponding crossing-symmetry conditions, consistent with 
the traditional Hopf-algebra framework. We display the Yangian 
hypercharge generator, and give the rules for its coproduct and 
charge-conjugation at the zeroth and the first Yangian level. 
One advantage of small-rank algebras is that it becomes rather elementary to 
prove a host of determinantal identities. The existence of such identities, 
and of the associated Yangian central elements, is connected to general 
principles of integrability. It is however often difficult to get a 
hold of them in higher-dimensional AdS/CFT situations. In this respect, the massless 
sector of $\AdS_3$ in particular reveals itself as a favoured playground for 
testing a variety of exact algebraic methods~\cite
{Borsato:2013qpa,Pittelli:2014ria,Stromwall:2016dyw}.

This paper is organised as follows. In Section~\ref{sec:crossing} 
we investigate the analytic structure of the massless modes and 
solve the crossing equations for the massless and mixed-mass 
dressing factors. We discuss possible homogeneous solutions of the crossing
equations as well as the absence of massless bound states. In Section~\ref{sec:BEs} 
we  determine the Bethe equations for the closed string spectrum, 
show that these have the expected symmetries and find their 
weak-coupling limit. We also comment on how the Bethe equations generalise to the background 
supported by mixed NS-NS and R-R flux. In Section~\ref{sec:pert-comp} 
we compare the near-BMN limit of our S-matrix to the 
perturbative calculations of~\cite{Sundin:2016gqe}, while in 
Section~\ref{sec:yangian} we investigate the underlying Yangian 
symmetry in the massless mode sector. Following the conclusion, 
we present a number of appendices where some of the more 
technical results are contained. In Appendix~\ref{app:Zamolo} 
we give a short review of massless scattering in relativistic 
integrable systems which we hope might furnish an easy access-point to 
this classic material.

\section{Solving crossing}
\label{sec:crossing}

Symmetries severely constrain the two-body worldsheet S matrix of strings on $\AdS_3\times S^3\times \Torus^4$, determining it up to four independent dressing factors.
Scattering of purely massive excitations involves $\sigma^{\bullet\bullet}$ and  ${\tilde \sigma}^{\bullet\bullet}$, while massless-massless and mixed-mass scattering involve $\sigma^{\circ\circ}$ and $\sigma^{\circ\bullet},\sigma^{\bullet\circ}$, respectively.\footnote{The dressing factors $\sigma^{\circ\bullet}$ and $\sigma^{\bullet\circ}$ are related by unitarity.} The dressing factors satisfy crossing equations which severely restrict their form. Solutions to the crossing equations for 
$\sigma^{\bullet\bullet}$, ${\tilde \sigma}^{\bullet\bullet}$, have been found some time ago~\cite{Borsato:2013hoa} and agree, modulo a small discrepancy discussed in the introduction, with a number of direct calculations~\cite{Beccaria:2012kb,Beccaria:2012pm}.
In this section we solve the crossing equations for the massless and mixed mass dressing phases $\sigma^{\circ\circ}$ and $\sigma^{\circ\bullet},\sigma^{\bullet\circ}$.

\subsection{The crossing transformation}
Let us start by describing the crossing transformation for massive and massless excitations.
Recall that massive excitations have a dispersion relation
\begin{equation}
E(p)=\sqrt{m^2+4h^2\sin\big(\frac{p}{2}\big)^2},\qquad m=\pm1.
\end{equation}
It is then useful to introduce Zhukovski variables $x^\pm$
\begin{equation}
\label{eq:massive-Zhukovski}
x^\pm(p)=\frac{|m|+E(p)}{2 h\, \sin\tfrac{p}{2}}\,e^{\pm i\frac{p}{2}},
\end{equation}
so that%
\footnote{In what follows, we will often indicate the arguments of functions as subscripts where convenient, \textit{e.g.}\ $x^\pm_p\equiv x^\pm(p)$.}
\begin{equation}
\label{eq:shortening-cond}
E(p)=\frac{ih}{2}\big(x^-_p-\frac{1}{x^-_p}-x^+_p+\frac{1}{x^+_p}\big),\quad
\frac{2i\,|m|}{h}=x^+_p+\frac{1}{x^+_p}-x^-_p-\frac{1}{x^-_p}, \quad
e^{ip}=\frac{x^+_p}{x^-_p}.
\end{equation}

\begin{figure}[t]
  \centering
  \includegraphics{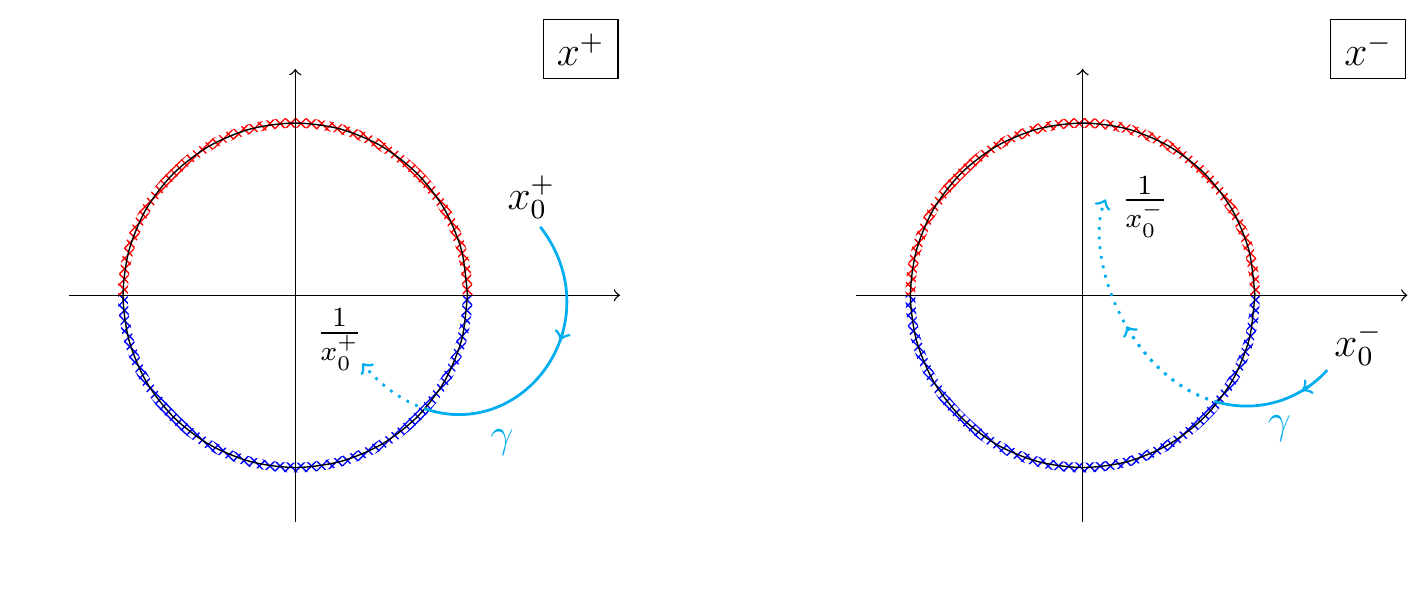}

\caption{The crossing transformation for massive variables in the $x^\pm$ planes. Note that, following the conventions of~\cite{Borsato:2013hoa}, the path $\gamma$ crosses the unit circle \textit{below} the real line. The red and blue zig-zag patterns depict the branch cuts of the massive dressing factors~\cite{Borsato:2013hoa}.}
\label{fig:crossing-massive}
\end{figure}

Massless excitations have a dispersion relation of the form
\begin{equation}
E(p)=2h\,\left|\sin \frac{p}{2}\right|,
\end{equation}
which, just like the massive one, is $2\pi$-periodic. 
The massless Zhukovski variables
\begin{equation}
\label{eq:massless-zhukovski}
x^{\pm}_p=e^{\pm\frac{i}{2}p}\ \text{sgn}\left(\sin\frac{p}{2}\right)\,
\end{equation}
can be thought of as the $m\to0$ limit of equation~\eqref{eq:massive-Zhukovski}, and have the same $2\pi$-periodicity.
Finally, as can be seen from the shortening condition~\eqref{eq:shortening-cond}, massless variables satisfy $x^+_p\,x^-_p=1$ for any $p$, so that it is useful to define
\begin{equation}
\label{eq:massless-condition}
x_p\equiv x^{+}_p=\frac{1}{x^{-}_p}\qquad \text{for massless modes}.
\end{equation}
Then their dispersion is simply $E(p)=-ih(x_p-1/x_p)$.

\begin{figure}[t]
  \centering
  \includegraphics{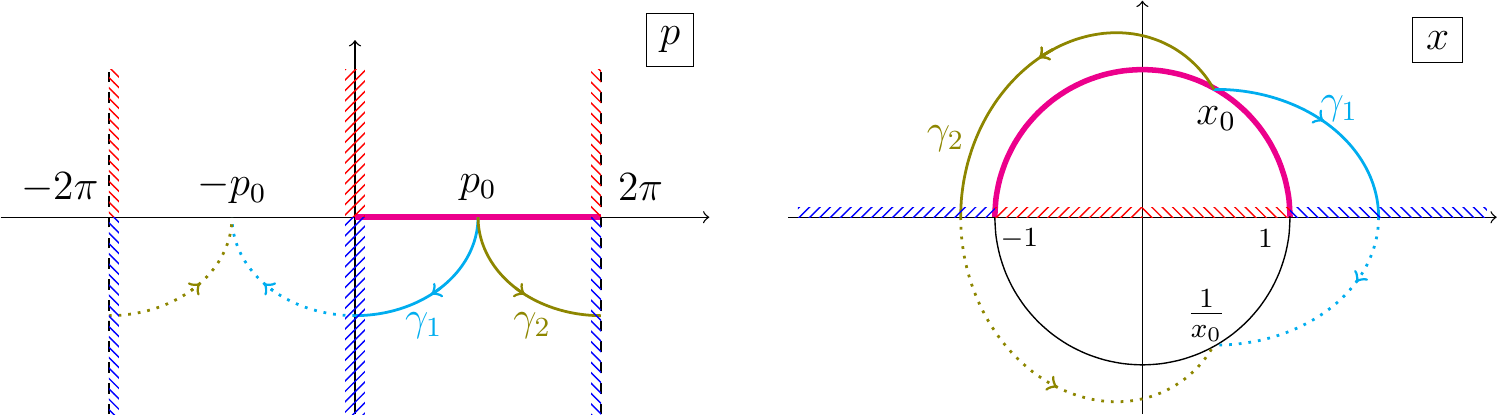}
  \caption{The $p$- and $x$-planes for massless particles. The thick magenta line indicates the domain of real physical momenta. The zig-zag patterns denote where the energy changes sign. Given a point $p_0$ or $x_0$ in corresponding to the real momentum, we depict curves $\gamma_1,\gamma_2$ for the crossing transformation. On the $p$-plane, these send $p_0\to-p_0$ and cross a cut of the energy. On the $x$-plane, we have $x_0\to 1/x_0$ while crossing the real line with $|x|>1$.}
  \label{fig:crossingPU}
\end{figure}

In analogy with relativistic massless particles that have $E_p\sim |p|$, we may consider a fundamental region for $p$ to be $-\pi\leq p<\pi$ and introduce a notion of left- and right-movers \textit{on the world-sheet}. Right-movers would then have $0< p<\pi$, and lie in the upper-right quadrant of the $x$-plane, while left-movers with $-\pi<p<0$ would lie in the upper-left quadrant owing to the $\text{sgn}(\sin p/2)$ in equation~\eqref{eq:massless-zhukovski},  see also Fig.~\ref{fig:crossingPU}.
However, it is inconvenient to use a discontinuous map for the massless Zhukovski variables. Taking advantage of the periodicity of~\eqref{eq:massless-zhukovski}, we therefore define the fundamental region to be
\begin{equation}
0\leq p < 2\pi.
\end{equation}
This somewhat obscures the parallel with the relativistic case, but has the advantage that all the discontinuities of the kinematics lie at the boundary of the fundamental region. Left-movers now have $\pi<p<2\pi$ and are still mapped to the upper-left quadrant in Fig.~\ref{fig:crossingPU}.
Note that all singularities of $x(p)$ and hence of $E(p)$ lie at the boundaries of our fundamental region. 

Under the crossing transformation, energy and momentum must change sign,
\begin{equation}
p\to \bar{p}= -p,\qquad
E(p)\to E(\bar{p})=-E(p).
\end{equation}
In terms of the Zhukovski variables, crossing takes a similar form for massive and massless modes
\begin{equation}
x^\pm_{p}\to x^\pm_{\bar{p}}=\frac{1}{x^{\pm}_p},\qquad
x_{p}\to x_{\bar{p}}=\frac{1}{x_p}.
\end{equation}
However, the crossing transformation looks different in the two cases.
The physical region for massive modes is $|x^\pm_p|>1$, with the imaginary part being positive for $x^+_p$ and negative for $x^-_p$. Then the crossing transformation takes us \textit{inside} the unit circle, see Fig.~\ref{fig:crossing-massive}. If the dressing factor has branch cuts, it is natural to define them on the circle, where $E(p)$ changes sign.

For massless modes, again we want to send $p\to-p$ through the branch cut of the energy in the $p$-plane, see Fig.~\ref{fig:crossingPU}. In the $x$-plane, real momenta live on the upper-half circle owing to the definition~\eqref{eq:massless-condition}. Crossing takes us to the lower half-circle by crossing the real line, where $E(p)$ changes sign. Comparing Fig.~\ref{fig:crossingPU} with Fig.~\ref{fig:crossing-massive} we can think of the path $\gamma_1$ as coming from the limit of~$\gamma$ for $x^+$ in the massive case, when its endpoints tend to the unit circle---which is what happens as we take $m\to0$.

\begin{figure}[t]
  \centering
  \includegraphics{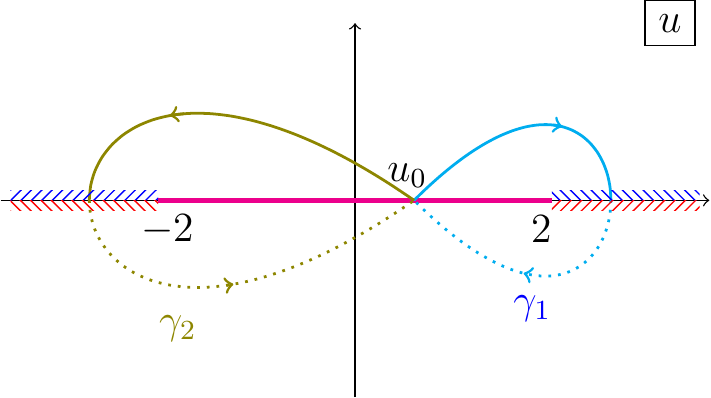}
  \caption{The $u$-plane for massless variable.}
  \label{fig:crossingU}
\end{figure}

It is also convenient to introduce the massless rapidity
\begin{equation}
u=x+\frac{1}{x}=2\cos\frac{p}{2},\qquad
\text{so that}\quad E(u)=h\,\sqrt{4-u^2}.
\end{equation}
Clearly the physical region is $-2\leq u\leq2$. The energy has cuts for real $u$  with $|u|>2$, \textit{cf.}\ Fig.~\ref{fig:crossingU}. 
The crossing transformation takes $u$ to itself through the cuts.

\subsection{The crossing equations}
The S~matrix of $\AdS_3\times\Sphere^3\times\Torus^4$ strings contains four dressing factors. For massive modes, $\sigma^{\bullet\bullet}$ and $\widetilde{\sigma}^{\bullet\bullet}$ correspond to the scattering of particle of equal or opposite $m$, respectively. Scattering of two massless modes gives  $\sigma^{\circ\circ}$, while mixed-mass scattering gives $\sigma^{\circ\bullet}$. They must obey the crossing equations
\begin{equation}
\label{eq:crossingeqs}
\begin{aligned}
\left(\sigma^{\bullet\bullet}_{pq}\right)^2 \ \left(\widetilde{\sigma}^{\bullet\bullet}_{\bar{p}q}\right)^2 &= \left( \frac{x^-_q}{x^+_q} \right)^2 \frac{(x^-_p-x^+_q)^2}{(x^-_p-x^-_q)(x^+_p-x^+_q)} \frac{1-\frac{1}{x^-_px^+_q}}{1-\frac{1}{x^+_px^-_q}}, \\
\left(\sigma^{\bullet\bullet}_{\bar{p}q}\right)^2 \ \left(\widetilde{\sigma}^{\bullet\bullet}_{pq}\right)^2 &= \left( \frac{x^-_q}{x^+_q} \right)^2 \frac{\left(1-\frac{1}{x^+_px^+_q}\right)\left(1-\frac{1}{x^-_px^-_q}\right)}{\left(1-\frac{1}{x^+_px^-_q}\right)^2} \frac{x^-_p-x^+_q}{x^+_p-x^-_q},\\
\left(\sigma^{\circ\circ}_{\bar{p}q}\right)^2\ \left(\sigma^{\circ\circ}_{pq}\right)^2 &= F(w_p-w_q)\,f(x_p,x_q)^2,\\
\left(\sigma^{\circ\bullet}_{\bar{p}q}\right)^2\ \left(\sigma^{\circ\bullet}_{pq}\right)^2 &= \frac{f(x_p,x_q^+)}{f(x_p,x_q^-)},\qquad
\left(\sigma^{\bullet\circ}_{\bar{p}q}\right)^2\ \left(\sigma^{\bullet\circ}_{pq}\right)^2 = x^{-2}_q \frac{f(x_p^+,x_q)}{f(x_p^-,x_q)},
\end{aligned}
\end{equation}
where 
\begin{equation}
F(w)=\frac{w+i}{w},
\qquad
f(x,y)=
\left(\frac{xy-1}{x-y}\right).
\end{equation}
These equations are supplemented by the constraints due to unitarity which requires%
\footnote{%
Braiding unitarity relates $\sigma^{\circ\bullet}_{pq}$ to the massive-massless phase $\sigma^{\bullet\circ}_{qp}=1/\sigma^{\circ\bullet}_{pq}$~\cite{Borsato:2014hja}. In fact, owing to that relation, we need only consider $\sigma^{\circ\bullet}_{pq}$.
}
\begin{equation}
\label{eq:unitarity}
\sigma^{\circ\circ}_{pq}\,\sigma^{\circ\circ}_{qp}=|\sigma^{\circ\circ}_{pq}|=1,\qquad
\sigma^{\circ\bullet}_{pq}\sigma^{\bullet\circ}_{qp}=1,\qquad
|\sigma^{\circ\bullet}_{pq}|=|\sigma^{\bullet\circ}_{pq}|=1.
\end{equation}
Introducing%
\footnote{The discussion below is written for any dressing factor $\sigma$ and associated phase $\theta$. It applies equally to each of the four dressing factors $\sigma^{\bullet\bullet}$,  ${\widetilde \sigma}^{\bullet\bullet}$,  $\sigma^{\circ\bullet}$ and  $\sigma^{\circ\circ}$.}
\begin{equation}
\sigma=e^{i\theta},
\end{equation}
we will often write $\theta(x^\pm,y^\pm)\equiv \theta(x^+,x^-,y^+,y^-)$ when no ambiguity can arise. For physical rapidities the phases have an expansion~\cite{Arutyunov:2004vx,Hernandez:2006tk,Beisert:2005wv}  
\begin{equation}
\label{eq:phase-exp}
\theta(x^\pm,y^\pm)=\sum_{r,s}c_{r,s}(h)\bigl(q_r(x^\pm)q_s(y^\pm)-q_r(y^\pm)q_s(x^\pm)\bigr),
\end{equation}
in terms of the local charges $q_r$  of the integrable system~\cite{Beisert:2004hm}. Above, the $c_{r,s}$ are $h$-dependent coefficients and the local charges are given by
\begin{equation}
\label{eq:charges-definition}
q_r(x^\pm)=\frac{i}{r-1}\left(\frac{1}{(x^+)^{r-1}}-\frac{1}{(x^-)^{r-1}}\right),
\qquad
q_1(x^\pm)=-i\log \frac{x^+}{x^-}.
\end{equation}
In $\AdS_3$ backgrounds the momentum $q_1(x)\equiv p$ may enter the 
expansion~\eqref{eq:phase-exp}~\cite{Beccaria:2012kb}. It is convenient
to express $\theta$ in terms of a simpler function $\chi$
\begin{equation}
\label{eq:def-of-theta-as-chis}
\theta(x^\pm,y^\pm)=\chi(x^+,y^+)+\chi(x^-,y^-)-\chi(x^+,y^-)-\chi(x^-,y^+),
\end{equation}
in the case of massive scattering. For mixed-mass scattering one can use~\eqref{eq:massless-condition} to reduce the decomposition to two terms, while for massless scattering no decomposition is necessary.
The expansion for $\chi$ analogous to~\eqref{eq:phase-exp} takes the form
\begin{equation}
\label{eq:chi-exp}
\chi(x,y)=-\sum_{r,s}\frac{c_{r,s}(h)}{(r-1)(s-1)x^{r-1}y^{s-1}}.
\end{equation}
At strong coupling the coefficients $c_{r,s}$ have the expansion
\begin{equation}
c_{r,s}(h)=\sum_{n=0}^\infty c^{(n)}_{r,s}h^{1-n}.
\end{equation}
The $n=0,1$ terms are known as AFS~\cite{Arutyunov:2004vx} and HL~\cite{Hernandez:2006tk} orders, respectively. 

\subsection{\texorpdfstring{Finding $\sigma^{\circ\circ}$}{Finding sigma\^{}oo}}

In this subsection we will find the all-loop expression for $\sigma^{\circ\circ}$. The crossing equation for $\sigma^{\circ\circ}$ can be decomposed into two auxiliary problems
\begin{equation}
\sigma_F^{\circ\circ}(\bar{p},q)^2\sigma_F^{\circ\circ}(p,q)^2=F(w_p-w_q),
\qquad
\sigma_f^{\circ\circ}(\bar{p},q)^2\sigma_f^{\circ\circ}(p,q)^2=f(x_p,x_q)^2,
\end{equation}
which we will solve separately. 

\subsubsection{\texorpdfstring{Finding the massless dressing factor $\sigma^{\circ\circ}_F$}{Finding sigma\^{}oo\string_F}}

The crossing equation for $\sigma^{\circ\circ}_F$ involves an auxiliary rapidity~$w(p)$. While its dependence on $p$ is undetermined, we know that it arises from an $\algSU(2)$ invariance and that under crossing we have~\cite{Borsato:2014exa,Borsato:2014hja}
\begin{equation}
w(\bar{p})=w(p)+i.
\end{equation}
Using this, and the fact that~$F(w)$ is of difference form in~$w$, we can write down a familiar minimal solution for~$\sigma_F$ \emph{of difference form}~\cite{Reshetikhin:1990jn}
\begin{equation}
\sigma_F^{\circ\circ}(p,q)=\Psi(w_p-w_q),
\qquad
\Psi(w)=i\frac{\Gamma(1-\frac{i}{2}w)}{\Gamma(1+\frac{i}{2}w)}\frac{\Gamma(\frac{1}{2}+\frac{i}{2}w)}{\Gamma(\frac{1}{2}-\frac{i}{2}w)}.
\end{equation}
It is also straightforward to write homogeneous solutions to the $\sigma_F$ crossing equation. However, as we will argue in Section~\ref{sec:pert-comp} no such solutions are compatible with perturbative string theory computations of~\cite{Sundin:2016gqe} and we are led to take  $w\rightarrow\infty$. This trivializes the complete $\algSU(2)_\circ$ S matrix. In particular the dressing factor $\sigma^{\circ\circ}_F\rightarrow 1$, which is consistent with crossing in this limit, since $F(w)\rightarrow 1$. The massless dressing factor then reduces to
\begin{equation}
\sigma^{\circ\circ}=\sigma^{\circ\circ}_f.
\end{equation}
In terms of $\theta^{\circ\circ}$ the crossing equation takes the form
\begin{equation}
\label{eq:crossignchimm}
\theta^{\circ\circ}(\bar{x},y)+\theta^{\circ\circ}(x,y)=-\frac{i}{2}\log f(x,y)^2.
\end{equation}

\subsubsection{\texorpdfstring{The Riemann-Hilbert problem for $\theta^{\circ\circ}$}{The Riemann-Hilbert problem for theta\^{}oo}}

It is useful to express the crossing equation for $\sigma^{\circ\circ}$ in terms of~$u$. For example~\eqref{eq:crossignchimm} takes the form
\begin{equation}
\theta^{\circ\circ}(\bar{u},v)+\theta^{\circ\circ}(u,v)=-\frac{i}{2}\log f(x_u,x_v)^2.
\end{equation}
This equation is defined when~$|\Re u|<2$, see Fig.~\ref{fig:crossingPU}, and involves the value of~$\sigma_f$ on two sheets. If we analytically continue the point~$u$ to somewhere very close to the cut of~$E(u)$ (and let's say above it), the equation reads
\begin{equation}
\label{eq:riemann-hilbert}
\theta^{\circ\circ}(u+i0,v)+\theta^{\circ\circ}(u-i0,v)=-\frac{i}{2}\log f(x_u,x_v)^2,\qquad
\Re u>2.
\end{equation}
This is a Riemann-Hilbert problem that can be solved by the Sochocki-Plemelj theorem~\cite{Sochocki:1873a,Plemelj:1908a}, see also \textit{e.g.}~\cite{Volin:2009uv}. The resulting solution is \textit{minimal} in the sense that by construction it only allows for the singularities necessary to solve~\eqref{eq:riemann-hilbert}, and needs to be appropriately anti-symmetrised to account for unitarity~\eqref{eq:unitarity}. While it is useful to characterize the solution in this way, it is more convenient to work directly in the $x$-plane, and to make anti-symmetry more manifest from the beginning. To this end, it is helpful to briefly recall some properties of the massive HL phase $\chi^{\mbox{\scriptsize HL}}$.

The massless \emph{all-loop} crossing equation for~$\theta^{\circ\circ}$ has the same form as that of the massive HL phase $\chi^{\mbox{\scriptsize HL}}$
\begin{equation}
\chi^{\mbox{\scriptsize HL}}(\bar{x},y)+
\chi^{\mbox{\scriptsize HL}}(x,y)=-\frac{i}{2}\log f(x,y),
\label{eq:massive-hl-cross}
\end{equation}
where now $x,y$ are interpreted as massive variables.
The massive HL phase has the integral representation
\begin{equation}
\begin{aligned}
\chi^{\mbox{\scriptsize HL}}(x^\pm,y^+)
&=\left(\inturl - \intdlr\right)\frac{dz}{4\pi}\frac{1}{x^\pm-z}G_-(z,y^+),\\
\chi^{\mbox{\scriptsize HL}}(x^\pm,y^-)
&=\left(\inturl - \intdlr\right)\frac{dz}{4\pi}\frac{1}{x^\pm-z}G_+(z,y^-),
\end{aligned}
\label{eq:def-chi-hl}
\end{equation}
valid in the physical region $|x^\pm|>1$, $|y^\pm|>1$, with
\begin{equation}
G_\pm(z,y)=\log\left(\pm i(y-z)\right)-\log\left(\pm i(y-\tfrac{1}{z})\right).
\end{equation}
This definition deviates slightly from the one known in the literature for the choice of the branch-cuts of the logarithms in $G_\pm(z,y)$.%
\footnote{Our choice here ensures that such branch-cuts do not intersect the upper-half disc (lower-half disk) in the $z$-plane in the case of $G_+(z,y)$ ($G_-(z,y)$). As we will see in the next subsection this choice simplifies the analysis of crossing for $\chi^{\mbox{\scriptsize HL}}$.}
In Appendix~\ref{app:chi-asym} we show that~\eqref{eq:def-chi-hl} defines an anti-symmetric function. The expression in~\eqref{eq:def-chi-hl} has discontinuities across the unit circle; defining  $\chi^{\HL}(x,y)$ \emph{in the crossed region} $|x|<1$ through analytic continuation, one may check that it satisfies~\eqref{eq:massive-hl-cross}.
Is is also worth mentioning that these expressions can be integrated to an expression involving dilogarithms that has appeared in the $\AdS_5$ literature~\cite{Beisert:2006ib,Arutyunov:2006iu}.

\subsubsection{Deforming the contour for the HL phase}
\label{sec:def-cont}
While the massive HL phase satisfies the same crossing equation as $\theta^{\circ\circ}$,  the position of its branch cuts differs from those of the massless kinematics. Massless dressing factors should have branch cuts for real~$x$, as illustrated in Fig.~\ref{fig:crossingPU}. Instead~\eqref{eq:def-chi-hl} has a branch-cut on the half-circle, which is exactly where \textit{physical} massless particles should live.
Thankfully, we can analytically extend the integral~\eqref{eq:def-chi-hl} by deforming its integration contour, in such a  way as to move the branch cut to the unit segment, as showin in Fig.~\ref{fig:move-cnt}. What is more, as we show in Appendix~\ref{app:proof-cr-hl}, the resulting function continues to satisfy the crossing equation that follows from~\eqref{eq:massive-hl-cross}. 
Relegating the technical details to Appendix~\ref{app:deform-hl}, after the contour shift we find
\begin{equation}\label{eq:hl-generic}
\begin{aligned}
\theta^{\HL} (x^\pm,y^\pm) &=\int\limits_{-1+i\epsilon}^{1+i\epsilon} \frac{dz}{4\pi}G_-(z,y^+)\bigl( g(z,x^+)-g(z,x^-) \bigr)
\\ &\,\,
-\int\limits_{-1-i\epsilon}^{1-i\epsilon}  \frac{dz}{4\pi}G_+(z,y^-) \bigl( g(z,x^+)-g(z,x^-) \bigr) \\
& \,\,-\frac{i}{2} \left( G_-(\tfrac{1}{x^-},y^+)-G_+(\tfrac{1}{x^+},y^-) \right),
\end{aligned}
\end{equation}
where\footnote{The function $g$ does \emph{not} depend on the choice of sign $\pm$
that enters $G_\pm$.}
\begin{equation}
g(z,x)\equiv\frac{\partial}{\partial z}G_{\pm}(z,x)=\frac{1}{z-x}-\frac{1}{z-\tfrac{1}{x}}+\frac{1}{z}.
\end{equation}
When shifting the contour, we regulate with an $\epsilon$-prescription to avoid collision with the branch-cuts of the functions $G_\pm$ (see Fig.~\ref{fig:move-cnt}).
The contributions appearing in the third line of~\eqref{eq:hl-generic} do not involve integrals and come from poles that we pick when moving the contours of integration, as depicted in Fig.~\ref{fig:move-cnt}.
The representation~\eqref{eq:hl-generic} is valid for $\Im x^+>0, \Im x^-<0$ and similarly for $y^\pm$, with no restrictions on the masses of the excitations. 
\begin{figure}[t]
  \centering
  \includegraphics{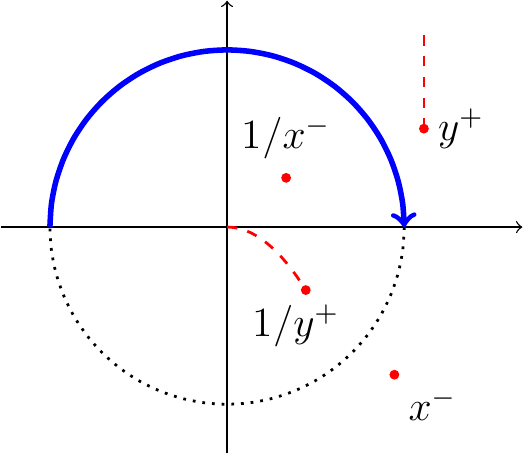}
  %
  \hspace{0.5cm}
  \raisebox{2.2cm}{$=$}
  \hspace{0.5cm}
  \includegraphics{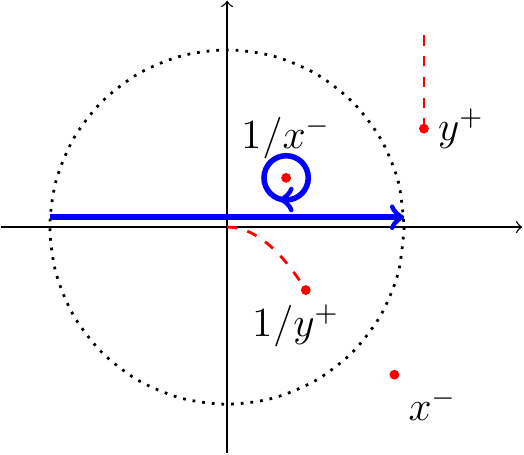}

  \caption{Red dashed lines correspond to the branch cuts of the integrand, while red dots to the poles. In this specific example we move the contour of integration (blue line) from the upper semicircle to the real interval $[-1,+1]$ and we pick up a pole.}
  \label{fig:move-cnt}
\end{figure}

The phase in equation~\eqref{eq:hl-generic} has discontinuities on the real line, as we wanted. As proven in Appendix~\ref{app:proof-cr-hl}, we therefore have constructed an anti-symmetric function that satisfies the crossing equation for~$\sigma^{\circ\circ}$ and is compatible with massless kinematics. Furthermore, this functions still satisfies massive crossing~\eqref{eq:massive-hl-cross} too, as long as the crossing path is taken as in Fig.~\ref{fig:crossing-massive}. Hence it can be employed in the massive, massless, or mixed-mass cases.
A similar result could have been found starting from the dilogarithm expression of~\cite{Beisert:2006ib,Arutyunov:2006iu}. As all these functions coincide in the region $|x^\pm|>1$, $|y^\pm|>1$, they give rise to the same analytic extension once we move the branch cut to the real line.

\subsubsection{Expansion coefficients \texorpdfstring{for $\theta^{\mbox{\scriptsize HL}}$}{for theta\^{}HL} with arbitrary masses}
It is often useful to represent the dressing factors as a sum over conserved charges, \cf equation~\eqref{eq:phase-exp}. Such a representation is valid in the physical region, which depends on the particular kinematics of the excitations under consideration. Specifically, for massive kinematics, the expansion is well-defined for $|x|,|y|>1$, while for massless modes the variable $x$ or $y$ lie on the upper-half-circle. In this latter case it is worth noting that the expansion is in fact a Fourier series.  In Appendix~\ref{app:four-and-crs} we show that the $c_{r,s}$ coefficients for $\theta^{\text{\scriptsize HL}}|_{m_x=m_y=0}$  are the same as the HL coefficients~\cite{Hernandez:2006tk}, \textit{i.e.}
\begin{equation}
c^{\mbox{\scriptsize HL}}_{r,s}=\frac{1-(-1)^{r+s}}{\pi}
\frac{(r-1)(s-1)}{(r-s)(r+s-2)},\qquad\text{for arbitrary masses}.
\label{eq:massless-crs}
\end{equation}
This means that, just as the expression~\eqref{eq:hl-generic} is valid for massive, massless and mixed-mass kinematics, so is the double-series representation~\eqref{eq:phase-exp} with the coefficients as in~\eqref{eq:massless-crs}.

\subsubsection{Minimal solution of the massless crossing equation}
\label{sec:mass-phase-crossing-soln}
The expression~\eqref{eq:hl-generic} is well defined in the massless kinematics, and as shown in Appendix~\ref{app:proof-cr-hl}, $\tfrac{1}{2}\theta^{\HL}$,
satisfies the massless crossing equation~\eqref{eq:crossignchimm} where the crossing path is taken as in Fig.~\ref{fig:crossingPU}. Evaluating~\eqref{eq:hl-generic} in the massless kinematics~\eqref{eq:massless-condition} we conclude that the minimal solution for $\theta^{\circ\circ}$ is
\begin{equation}\label{eq:hl-massless}
\begin{aligned}
\theta^{\circ\circ}_{\mbox{\scriptsize min}}(x,y)=\frac{1}{2}
\theta^{\mbox{\scriptsize HL}} (x^\pm,y^\pm)\Big|_{m_x=m_y=0}
 &= \int\limits_{-1+i\epsilon}^{1+i\epsilon} \frac{dz}{4\pi}G_-(z,y)
\left( \frac{1}{z-x}-\frac{1}{z-\tfrac{1}{x}} \right)
\\ &\,\,\,\,\,\,
- \int\limits_{-1-i\epsilon}^{1-i\epsilon} \frac{dz}{4\pi}G_+(z,\tfrac{1}{y})
\left( \frac{1}{z-x}-\frac{1}{z-\tfrac{1}{x}} \right)\\
& \,\,\,\,\,\,-\frac{i}{4} \left( G_-(x,y)-G_+(\tfrac{1}{x},\tfrac{1}{y}) \right).
\end{aligned}
\end{equation}
In Appendix~\ref{app:chi-asym} we show that  the above expression for $\theta_{\mbox{\scriptsize min}}^{\circ\circ}$ is anti-symmetric under $x\leftrightarrow y$. Therefore, up to possible CDD factors',  we have solved the crossing equation for $\theta^{\circ\circ}$.

\subsection{Solution of homogeneous crossing}
As is well known~\cite{Castillejo:1955ed}, solutions of crossing equations are only defined up to CDD factors, \ie, solutions of the homogeneous crossing equation. There is a huge degeneracy of such solutions in the massless case. Evaluating~\eqref{eq:charges-definition} on the massless kinematics, we find $q_{r+1}(x)=\frac{i}{r}(1/x^r-x^r)$. Therefore any finite linear combination of functions of the form
\begin{equation}
\vartheta^{(\text{hom})}_{r,s}(x,y)= c_{r,s} \big(q_r(x)q_s(y)-q_r(y)q_s(x)\big),
\end{equation}
solves the homogeneous crossing equation. While it is very hard to exclude all such possible solutions, they do not appear to have any particular physical significance. Therefore, we shall focus on a particular class of solution of homogeneous crossing that emerge as massless limits of the massive kinematics.

Note that in the massless limit where $x^+=1/x^-$, the right-hand-side crossing equation for massive modes~\eqref{eq:crossingeqs} becomes the same as the one of the massless crossing equation~\eqref{eq:crossingeqs}.\footnote{Similarly in this limit the crossing equation satisfied by the BES phase~\eqref{eq:bes-crossing} also reduces to the massless crossing equation.} Therefore, one might wonder whether, in the massless limit, the phases $\sigma^{\bullet\bullet}$ and $\widetilde{\sigma}^{\bullet\bullet}$ might suggest a natural homogeneous solution, and whether such a solution should be included in $\sigma^{\circ\circ}$. Our minimal solution $\sigma^{\circ\circ}_{\mbox{\scriptsize min}}$ was constructed out of the massless limit of the HL-order term in the BES phase. Below we investigate the (leading) AFS term and the higher order terms in the BES phase in the massless limit. As we will show, in this limit the higher-order terms 
become trivial, while the AFS term will provide a hint that a homogeneous term at this order might naturally be expected.

\subsubsection{The BES phase}
An important ingredient of integrable holography is the non-perturbative BES phase~\cite{Beisert:2006ez} which, can be written as a double contour integral over unit circles~\cite{Dorey:2007xn}
\begin{equation}
\label{eq:bes-phase}
\chi^{\BES}(x,y)=-i\oint\frac{dz}{2\pi i}\oint\frac{dz'}{2\pi i}\frac{1}{x-z}
\frac{1}{x-z'}\log\frac{\Gamma\left(1+i h(z+\tfrac{1}{z}-z'-\tfrac{1}{z'})\right)}
{\Gamma\left(1-i h(z+\tfrac{1}{z}-z'-\tfrac{1}{z'})\right)}.
\end{equation}
The resulting phase satisfies the crossing equation
\begin{equation}
\theta^{\BES}(x^\pm,y^\pm)+\theta^{\BES}(\bar{x}^\pm,y^\pm)
=\frac{y^-}{y^+}\frac{x^--y^+}{x^--y^-}\frac{1-\frac{1}{x^+y^+}}{1-\frac{1}{x^+y^-}}.
\label{eq:bes-crossing}
\end{equation}
The BES phase~\eqref{eq:bes-phase} can be expanded asymptotically using that~\cite{Borsato:2013hoa}
\begin{equation}
\label{eq:bes-expansion}
i\log\frac{\Gamma(1+i x)}{\Gamma(1-i x)}=-x\log\frac{x^2}{e^2}- \frac{\pi}{2}\text{sign}(\text{Re}\,x)-2\sum_{n=0}^\infty\frac{\zeta(-2n-1)}{2n+1}\frac{(-1)^n}{x^{2n+1}},
\end{equation}
where the first term gives the AFS phase and the second one gives the HL one~\cite{Vieira:2010kb}.

\subsubsection{Sub-leading orders}
We start by considering the terms arising from the series in~\eqref{eq:bes-expansion}, \ie, these beyond HL order which can be written as
\begin{equation}
I_n(x_1,x_2)=\oint\frac{dz_1}{2\pi i}\frac{1}{z_1-x_1}\oint\frac{dz_2}{2\pi i}\frac{1}{z_2-x_2}\frac{1}{(z_1-z_2)^{2n+1}}\frac{1}{\big(1-\frac{1}{z_1z_2}\big)^{2n+1}}.
\end{equation}
The above expression is ill-defined due to the poles at $z_1=z_2$ and $z_1 z_2=1$. We can regularize this integral by giving a principal-value prescription in \textit{e.g.}~$z_1$. In fact, to preserve antisymmetry we should sum over the case where the principal value is in $z_1$ and in $z_2$.%
\footnote{Equivalently, we may think of taking the radii of the integration circles to be \textit{e.g.}\ $r_1=1$ and $r_2=1+\epsilon$ with $\epsilon>0$ and small. Anti-symmetry then requires us to sum over this case and the one with $r_1\leftrightarrow r_2$.} 
In this way, one recovers for any $n$ the expressions of ref.~\cite{Beisert:2006ib}.

In the massless case we have additional poles on the integration contour.
Since $|x_i|=1$, the poles at $z_i=x_i$ need to be regularized too. By applying the same prescription, we now find that the integral in \textit{e.g.}\ $z_1$ vanishes as all poles for $z_1$ are on the unit circle.
We conclude that the regularization of $I_n(x_1,x_2)$ vanishes when $|x_i|=1$, and therefore we do not expect the sub-leading pieces of the BES phase to play a role in the massless kinematics.

\subsubsection{HL order}
In the previous sub-section, we discussed how $\theta^{\HL}_m$ can be deformed to solve crossing for~$\sigma^{\circ\circ}$.  Recall that  $\sigma^{\bullet\bullet}$ and $\widetilde{\sigma}^{\bullet\bullet}$ differ from one another by a phase $\sigma^-$~\cite{Borsato:2013hoa}
\begin{equation}
\sigma^-=\frac{\sigma^{\bullet\bullet}}{\widetilde{\sigma}^{\bullet\bullet}}.
\end{equation}
Further, $\sigma^-$ is non-trivial precisely at HL order and the corresponding $\chi^-$ is
\begin{equation}
\chi^-(x,y)=\left(\inturl - \intdlr\right)\frac{dz}{4\pi}\left(\frac{H(z,y)}{x-z}-\frac{H(z,x)}{y-z}\right),
\end{equation}
where
\begin{equation}
H(z,x)=\log\left(x-z\right)\left(1-\tfrac{1}{xz}\right).
\end{equation}
For massless kinematics, the crossing equation for $\sigma^-$ becomes homogeneous and so one may wondered whether, in this limit, $\sigma^-$ gives rise to a natural non-trivial homogeneous solution at the HL order. In Appendix~\ref{app:sigma-minus} we show that by deforming the contour and restricting to massless 
kinematics, $\sigma^-$ becomes trivial.

\subsubsection{AFS order}
At the leading order of the strong-coupling expansion~\eqref{eq:bes-expansion} one recovers the AFS phase, which can be concisely expressed as a series
\begin{equation}
\chi^{\AFS}(x,y)=-\sum_{r=2}^\infty\sum_{s=r+1}^\infty 
\frac{c^{\AFS}_{r,s}}{(r-1)(s-1)}\frac{1}{x^{r-1}y^{s-1}},
\qquad
c^{\AFS}_{r,s}=\delta_{s+1,r}-\delta_{r+1,s},
\end{equation}
which is valid for arbitrary masses $m_x$ and $m_y$.
Performing the sum we find
\begin{equation}
\label{eq:afs-summed}
\chi^{\AFS}(x,y)=\tfrac{1}{x} - \tfrac{1}{y} + (y+\tfrac{1}{y}-x-\tfrac{1}{x})
\log\left(1 - \tfrac{1}{x y}\right),
\end{equation}
In contrast to $\sigma^-$ and the sub-leading orders of the BES phase, the above expression for the AFS phase does not vanish in the massless limit.%
\footnote{%
We could have reached the same conclusion working in terms of the integral obtained from the expansion of~\eqref{eq:bes-expansion}.}
Furthermore, it is straightforward to
check that for two massless excitations under crossing
\begin{equation}
0=\Big[\theta^{\AFS}(x,y)+\theta^{\AFS}(\bar{x},y)\Big]_{m_x=m_y=0}.
\end{equation}
In fact, as is clear from equation~\eqref{eq:afs-summed}, taking the massless limit and performing the crossing transformation are two commuting operations.
Therefore, while the AFS phase constructed as the $m\rightarrow 0$ limit of the massive AFS phase is non-zero, it satisfies the homogeneous crossing equation. This observation suggests that a homogeneous solution at AFS order may be present in $\theta^{\circ\circ}$.

\subsubsection{Solution of the massless crossing equation}
On the basis of the above arguments, we propose the following solution for the massless crossing equation
\begin{equation}
\theta^{\circ\circ}(x,y)=\Big[
\theta^{\text{AFS}}(x,y)+\tfrac{1}{2}\theta^{\text{HL}}(x,y)\Big]_{m_x=m_y=0},
\end{equation}
which differs from the minimal one by the addition of the AFS term that plays the role of a CDD factor.

\subsection{Determining the mixed-mass dressing factor \texorpdfstring{$\sigma^{\circ\bullet}$}{sigma\^{}ob}}

As we have discussed above, the deformation of the integration contour for the HL-order phase can be carried out independently of the mass of the excitations. Further, the AFS-order phase is also well behaved when the mass is varied. As a result, it is straightforward to write down a solution for the mixed-mass crossing equation~\eqref{eq:crossingeqs}. By defining 
\begin{equation}
\theta^{\circ\bullet}(x^\pm,y)=\Big[
\theta^{\text{AFS}}(x,y^\pm)+\tfrac{1}{2}\theta^{\text{HL}}(x,y^\pm)\Big]_{m_x=0,m_y=1},
\end{equation}
in terms of~(\ref{eq:hl-generic},~\ref{eq:afs-summed}) we can check that the crossing equation is satisfied. More specifically, the AFS part of the phase satisfies the crossing equation
\begin{equation}
\begin{aligned}
&\Big[\theta^{\AFS}(x^\pm,y)+\theta^{\AFS}(\bar{x}^\pm,y)\Big]_{{m_x=1,m_y=0}}=i\log y,\\
&\Big[\theta^{\AFS}(x,y^\pm)+\theta^{\AFS}(\bar{x},y^\pm)\Big]_{{m_x=0,m_y=1}}=0.
\end{aligned}
\end{equation}
in the mixed-mass case, while the HL part satisfies
\begin{equation}
\begin{aligned}
\Big[\theta^{\text{HL}}(x^\pm,y)+\theta^{\text{HL}}(\bar{x}^\pm,y)\Big]_{{m_x=1,m_y=0}}=-i \log \frac{f(x^+,y)}{f(x^-,y)},\\
\Big[\theta^{\text{HL}}(x,y^\pm)+\theta^{\text{HL}}(\bar{x},y^\pm)\Big]_{{m_x=0,m_y=1}}=-i \log\frac{f(x,y^+)}{f(x,y^-)},
\end{aligned}
\end{equation}
As for the solution of the homogeneous crossing equation, it is worth noticing that in this case the massless limit of the crossing equation~\eqref{eq:bes-crossing} does not give our mixed-mass equation. Therefore, it appears that no natural candidate exists for a  physically relevant class of solutions to homogeneous crossing.

\subsection{Absence of bound states}
\label{sec:boundstates}
The massless S~matrix is given explicitly in Appendix~M of~\cite{Borsato:2014hja}, supplemented by the dressing factors we just constructed. We are interested in studying its poles, and discussing whether they may be interpreted as arising from bound states.
Several entries of  the S~matrix have a simple pole when
\begin{equation}
\label{eq:bound-state}
x^-(p_1) = x^+(p_2). 
\end{equation}
This is the familiar bound-state condition for the scattering of massive particles with $m_1m_2>0$, \textit{cf.}~\cite{Dorey:2006dq}. In the massless kinematics~\eqref{eq:massless-condition}, this reads
\begin{equation}
\label{eq:massless-bound-state}
\frac{1}{x(p_1)} = x(p_2).
\end{equation}
It follows that the total energy and momentum of any putative bound state would be
\begin{equation}
E(p_1)+E(p_2)=-ih\big(x_1-\frac{1}{x_1}+x_2-\frac{1}{x_2}\big)=0,
\qquad
e^{i(p_1+p_2)}=(x_1x_2)^2=1.
\end{equation}
Hence such a bound state would have to be a singlet of the symmetry algebra. Further, solving~\eqref{eq:massless-bound-state} in terms of momenta we find
\begin{equation}
e^{\frac{i}{2}(p_1+p_2)}=1.
\end{equation}
For $p_i$ in the physical strip (\ie, when $\Re p_i\in [0,2 \pi)$), 
 the only solutions are $\Re p_1=\Re p_2=0$. Therefore, this putative bound state would have to be a singlet constructed out of two particles with purely imaginary and opposite momenta, and is therefore unphysical. As discussed in 
Appendix~\ref{app:singularities-dressing}, this structure is not modified by the dressing factors.

The above conclusions can be further confirmed by considering the residues of the $\algPSU(1|1)^2_{\text{c.e.}}$-invariant S~matrix of~\cite{Borsato:2014hja} when the momenta satisfy equation~\eqref{eq:massless-bound-state}. One finds non-vanishing residues from four out of the six S-matrix elements, resulting in a $2\times2$ residue matrix, just like in the massive case. However, when imposing the massless condition~\eqref{eq:massless-condition}, the residue matrix has rank one, again indicating that only one mode could be potentially propagating. This leads us to conclude that any putative bound-state would have to be a singlet and hence unphysical.

Let us now consider the case of one massive and one massless particle. Since massive particles have $|x^\pm|>1$ while massless ones have $|x|=1$, it is easy to check that it is impossible to construct a bound state satisfying~\eqref{eq:bound-state} for a pair of particle with complex and conjugate momenta. Interestingly, looking at the total energy and momentum, we find
\begin{equation}
\begin{aligned}
&E_1+E_2=-\frac{ih}{2}\big(x_1^+-\frac{1}{x_1^+}-x_1^-+\frac{1}{x_1^-}+2(x_2-\frac{1}{x_2})\big)=
-\frac{ih}{2}\big(x_1^+-\frac{1}{x_1^+}+x_1^--\frac{1}{x_1^-}\big),
\\
&e^{i(p_1+p_2)}=\frac{x_1^+}{x_1^-}(x_2)^2=x_1^+ x_1^-.
\end{aligned}
\end{equation}
Therefore, such a two-particle configurations has the same dispersion relation as a single particle in the ``mirror'' kinematics~\cite{Arutyunov:2007tc}.
We conclude that there are no physical bound states between massless and massive particles.

\section{Bethe equations}
\label{sec:BEs}
In this section we write down the all-loop nested Bethe equations for the full worldsheet theory, including the massless modes. 
They are found by imposing periodicity of the wave-function on the worldsheet, and their solutions give quantisation conditions for the momenta of the excitations.
Our analysis will be restricted to states which carry zero momentum and winding along the $\Torus^4$. Given the complexity of the S-matrix we need to appeal to the ``nesting procedure''~\cite{Yang:1967bm,Beisert:2005tm} to diagonalise it, which introduces new auxiliary roots. In Sections~\ref{sec:bos-grad-Bethe} and~\ref{sec:fermionic-duality}, we write down the Bethe equations in bosonic and fermionic gradings respectively, and discuss some of their properties and symmetries. In Section~\ref{sec:mass-zero-modes} we comment on extra symmetries of the Bethe equations associated to the presence of the massless modes, while in Section~\ref{sec:weak-coupling-limit} we present the weak coupling limit of the equations. In Section~\ref{sec:mixed-flux} we discuss the Bethe equations for the $\AdS_3 \times \Sphere^3 \times \Torus^4$ background supported by mixed NS-NS and R-R three-form flux.

\subsection{Bethe equations in the bosonic grading}\label{sec:bos-grad-Bethe}
The derivation of the Bethe equations closely follows the one discussed in~\cite{Borsato:2013qpa} for the massive sector. 
Here we include also the massless excitations, and rather than giving the details of the procedure, which can be found in~\cite{Borsato:2016hud}, we outline the principal points of the method.

\begin{figure}
  \centering
  \includegraphics{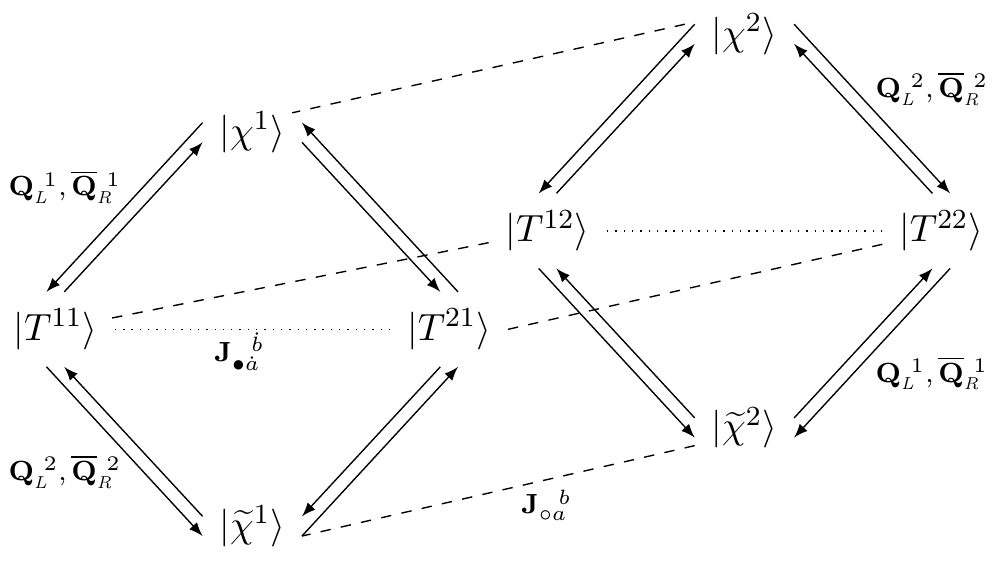}
  \caption{The massless module. Dashed lines denote the action of $\algSU(2)_\circ$ generators.}
  \label{fig:massless}
\end{figure}

We begin by choosing a maximal set of excitations above the BMN vacuum that scatter diagonally with each other, which will make up the level I of the Bethe equations. 
There are several possible sets of such excitations, corresponding to different choices of gradings of the $\algPSU(1,1|2)^2$ symmetry algebra. One possible choice, compatible with that of~\cite{Borsato:2013qpa}, is
\begin{equation*}
  Y^{\smallL}, \quad Z^{\smallR}, \quad \chi^1, \quad \chi^2 ,
\end{equation*}
\begin{figure}[t]
  \centering
  \includegraphics{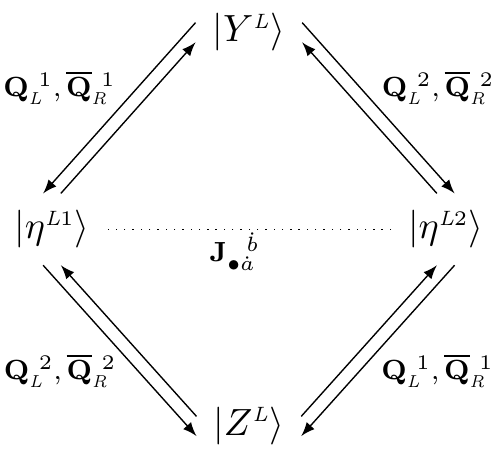}
\hspace{2cm}
  \includegraphics{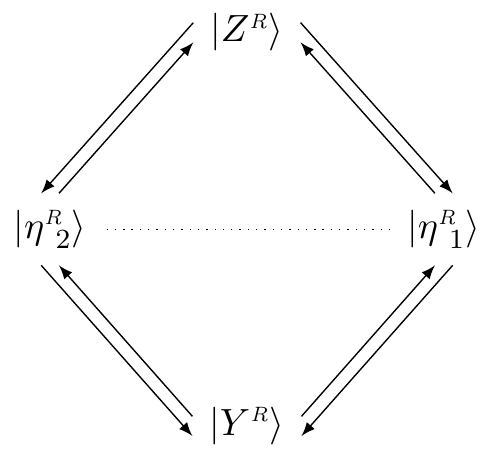}
  \caption{The Left and Right massive modules. We indicate explicitly only the lowering supercharges, corresponding to the arrows pointing downwards. In each module the fermions transform in a doublet of $\algSU(2)_\bullet$.}
  \label{fig:massive}
\end{figure}
The first two excitations are massive and belong to the representations shown in Fig.~\ref{fig:massive}, while the latter two are from the massless  module depicted in Fig.~\ref{fig:massless}.
Note that we have assumed that the $\algSU(2)_{\circ}$ symmetry acting on the massless excitations trivially commutes with the S matrix, as implied by the perturbative computations of~\cite{Sundin:2015uva,Sundin:2016gqe}. If this were not true, only one of the massless fermions would appear at level I, and we would have an additional auxiliary root corresponding to the action of the lowering operator of $\algSU(2)_{\circ}$. For completness, we have written down such a set of Bethe equations in Appendix~\ref{sec:nest-su2}.

For each level-I excitation we introduce a set of momentum-carrying Bethe roots. Since the two massless fermions carry exactly the same charges except for the $\algSU(2)_{\circ}$ spin, we will only use a single type of root to describe both at the same time.  The Bethe roots and excitation numbers corresponding to each type of level-I excitation are summarised in the table below.
\begin{center}
  \begin{tabular}{lccc}
    \toprule
    & Left massive & Right massive & Massless \\
    \midrule
    Level-I excitation & $Y^{\smallL}$ & $Z^{\smallR}$ & $\chi^1$, $\chi^2$ \\
    Bethe root & $x^{\pm}$ & $\bar{x}^{\pm}$ & $z^{\pm}$ \\
    Excitation number & $N_2$ & $N_{\bar{2}}$ & $N_0$ \\
    \bottomrule
  \end{tabular}
\end{center}
Since scattering among these excitations is diagonal, it is trivial to write down an eigenstate of the S-matrix and the corresponding Bethe equations as long as no other excitations are present.

When we include other excitations, non-diagonal processes arise.
For these we use the nesting procedure, see Appendix~\ref{app:nesting}, and introduce auxiliary Bethe roots, which correspond to the action of the supercharges on the level-I excitations. 
Acting once on a level-I state we generate level-II excitations, namely the massive fermions $\eta^{\smallL a},\eta^{\smallR a}$ of Fig.~\ref{fig:massive} and the massless bosons $T^{a\alpha}$ of Fig.~\ref{fig:massless}.
As suggested by these figures it is enough to introduce two sets of auxiliary roots
\begin{center}
  \begin{tabular}{lcc}
    \toprule
    & $\gen{Q}^{\smallL 1}$, $\overline{\gen{Q}}^{\smallR 1}$ & \qquad $\gen{Q}^{\smallL 2}$, $\overline{\gen{Q}}^{\smallR 2}$ \\
    \midrule
    Bethe root & $y_1$ & $y_3$ \\
    Excitation number & $N_1$ & $N_3$ \\
    \bottomrule
  \end{tabular}
\end{center}
Below we will discuss how two additional sets of auxiliary roots can be introduced, so that each type of root corresponds to the action of one supercharge.

The massive momentum-carrying roots satisfy the equations 
\begin{equation}
  \begin{split}
    \label{eq:BA-2}
    \left(\frac{x_k^+}{x_k^-}\right)^L 
    &=
    \prod_{\substack{j = 1\\j \neq k}}^{N_2} \nu_k^{-1}\nu_j \frac{x_k^+ - x_j^-}{x_k^- - x_j^+} \frac{1- \frac{1}{x_k^+ x_j^-}}{1- \frac{1}{x_k^- x_j^+}} (\sigma^{\bullet\bullet}_{kj})^2
    \prod_{j=1}^{N_1}  \nu_k^{\frac{1}{2}} \, \frac{x_k^- - y_{1,j}}{x_k^+ - y_{1,j}}
    \prod_{j=1}^{N_3} \nu_k^{\frac{1}{2}} \, \frac{x_k^- - y_{3,j}}{x_k^+ - y_{3,j}}
    \\ &\phantom{= }\times
    \prod_{j=1}^{N_{\bar{2}}} \nu_j  \frac{1- \frac{1}{x_k^+ \bar{x}_j^+}}{1- \frac{1}{x_k^- \bar{x}_j^-}} \frac{1- \frac{1}{x_k^+ \bar{x}_j^-}}{1- \frac{1}{x_k^- \bar{x}_j^+}} (\widetilde{\sigma}^{\bullet\bullet}_{kj})^2
    \\ &\phantom{= }\times
    \prod_{j=1}^{N_{0}} 
    \nu_k^{-1/2} \nu_j 
    \frac{x_k^+ - z_j^-}{x_k^- - z_j^+}
    \left(\frac{1 - \frac{1}{x_k^- z_j^-}}{1 - \frac{1}{x_k^+ z_j^+}}\right)^{\!\!\frac{1}{2}}
    \left(\frac{1 - \frac{1}{x_k^+ z_j^-}}{1 - \frac{1}{x_k^- z_j^+}}\right)^{\!\!\frac{1}{2}}
    (\sigma^{\bullet\circ}_{kj})^2
    ,
  \end{split}
\end{equation}
and
\begin{equation}
  \begin{split}
    \label{eq:BA-2b}
    \left(\frac{\bar{x}_k^+}{\bar{x}_k^-}\right)^L 
    &=
    \prod_{\substack{j = 1\\j \neq k}}^{N_{\bar{2}}} \frac{\bar{x}_k^- - \bar{x}_j^+}{\bar{x}_k^+ - \bar{x}_j^-} 
    \frac{1- \frac{1}{\bar{x}_k^+ \bar{x}_j^-}}{1- \frac{1}{\bar{x}_k^- \bar{x}_j^+}}  (\sigma^{\bullet\bullet}_{kj})^2
    \prod_{j=1}^{N_{1}} \nu_k^{\frac{1}{2}} \frac{1 - \frac{1}{\bar{x}_k^+ y_{1,j}}}{1- \frac{1}{\bar{x}_k^- y_{1,j}}}
    \prod_{j=1}^{N_{3}} \nu_k^{\frac{1}{2}}\frac{1 - \frac{1}{\bar{x}_k^+ y_{3,j}}}{1- \frac{1}{\bar{x}_k^- y_{3,j}}} 
    \\ &\phantom{= }\times
    \prod_{j=1}^{N_2} \nu_k^{-1} \frac{1- \frac{1}{\bar{x}_k^- x_j^-}}{1- \frac{1}{\bar{x}_k^+ x_j^+}} \frac{1- \frac{1}{\bar{x}_k^+ x_j^-}}{1- \frac{1}{\bar{x}_k^- x_j^+}}(\widetilde{\sigma}^{\bullet\bullet}_{kj})^2
    \\ &\phantom{= }\times
    \prod_{j=1}^{N_{0}} 
    \nu_k^{-1/2} 
    \left(\frac{1- \frac{1}{\bar{x}_k^- z_j^-}}{1- \frac{1}{\bar{x}_k^+ z_j^+}}\right)^{\!\!\frac{3}{2}} 
    \left(\frac{1- \frac{1}{\bar{x}_k^+ z_j^-}}{1- \frac{1}{\bar{x}_k^- z_j^+}}\right)^{\!\!\frac{1}{2}} 
    (\sigma^{\bullet\circ}_{kj})^2
    .
  \end{split}
\end{equation}
Above, the factors of $\nu_k$ are either equal to $1$ or to $e^{ip_k}$ depending on whether the equations are written in the \emph{spin-chain frame} or the \emph{string frame}.\footnote{
  The difference between the spin-chain  and string frames is in the normalisation of the two-particle states. The S matrix of~\cite{Borsato:2014exa,Borsato:2014hja} was written in the string frame, while the Bethe equations for the massive sector of~\cite{Borsato:2013qpa} were written in the spin-chain frame.
}
The massless momentum-carrying roots satisfy the equation
\begin{equation}
  \begin{split}
    \label{eq:BA-0}
    \left(\frac{z_k^+}{z_k^-}\right)^L 
    &=
    \prod_{\substack{j = 1\\j \neq k}}^{N_0} 
    \nu_k^{-1/2} \nu_j^{+1/2}
    \frac{z_k^+ - z_j^-}{z_k^- - z_j^+}(\sigma^{\circ\circ}_{kj})^2
    \prod_{j=1}^{N_1}\nu_k^{\frac{1}{2}} \frac{z_k^- - y_{1,j}}{z_k^+ - y_{1,j}}
    \prod_{j=1}^{N_3} \nu_k^{\frac{1}{2}}  \frac{z_k^- - y_{3,j}}{z_k^+ - y_{3,j}}
    \\ &\phantom{= }\times
    \prod_{j=1}^{N_{2}}
    \nu_k^{-1} \nu_j^{1/2}
    \frac{z_k^+ - x_j^-}{z_k^- - x_j^+}
    \left(\frac{1-\frac{1}{z_k^+x_j^+}}{1-\frac{1}{z_k^-x_j^-}}\right)^{\!\!\frac{1}{2}}
    \left(\frac{1-\frac{1}{z_k^+x_j^-}}{1-\frac{1}{z_k^-x_j^+}}\right)^{\!\!\frac{1}{2}}
    (\sigma^{\circ\bullet}_{kj})^2
    \\ &\phantom{= }\times
    \prod_{j=1}^{N_{\bar{2}}} 
    \nu_j^{1/2}
    \left(\frac{1-\frac{1}{z_k^+\bar{x}_j^+}}{1-\frac{1}{z_k^-\bar{x}_j^-}}\right)^{\!\!\frac{3}{2}}
    \left(\frac{1-\frac{1}{z_k^+\bar{x}_j^-}}{1-\frac{1}{z_k^-\bar{x}_j^+}}\right)^{\!\!\frac{1}{2}}
    (\sigma^{\circ\bullet}_{kj})^2
    .
  \end{split} 
\end{equation}
In addition, the auxiliary roots satisfy
\begin{equation}
  \label{eq:BA-1-3}
  1 
  =
  \prod_{j=1}^{N_2} \frac{y_{I,k} - x_j^+}{y_{I,k} - x_j^-} \nu_j^{-\frac{1}{2}}
  \prod_{j=1}^{N_{\bar{2}}} \frac{1 - \frac{1}{y_{I,k} \bar{x}_j^-}}{1- \frac{1}{y_{I,k} \bar{x}_j^+}} \nu_j^{-\frac{1}{2}}
  \prod_{j=1}^{N_0} \frac{y_{I,k} - z_j^+ }{y_{I,k} - z_j^- } \nu_j^{-\frac{1}{2}} , 
\end{equation}
where $I = 1,3$. A physical solution further satisfies the level-matching condition
\begin{equation}\label{eq:level-matching}
  1 = \prod_{j=1}^{N_2} \frac{x_j^+}{x_j^-} \prod_{j=1}^{N_{\bar{2}}} \frac{\bar{x}_j^+}{\bar{x}_j^-} \prod_{j=1}^{N_0} \frac{z_j^+}{z_j^-}.
\end{equation}
The energy of a state is given by
\begin{equation}\label{eq:energy}
  D - J = N_2 + N_{\bar{2}} 
  + ih \sum_{k=1}^{N_2} \Bigl( \frac{1}{x_k^+} - \frac{1}{x_k^-} \Bigr)
  + ih \sum_{k=1}^{N_{\bar{2}}} \Bigl( \frac{1}{\bar{x}_k^+} - \frac{1}{\bar{x}_k^-} \Bigr)
  + ih \sum_{k=1}^{N_0} \Bigl( \frac{1}{z_k^+} - \frac{1}{z_k^-} \Bigr) .
\end{equation}
We shall refer to the above Bethe equations as being in the {\em bosonic grading}.

Let us now consider some basic properties of these equations. In~\eqref{eq:BA-2}--~\eqref{eq:BA-1-3} we have written the Bethe equations using two auxiliary roots. On the other hand, in~\cite{Borsato:2013qpa} the massive sector of the equations was written in terms of four auxiliary roots. To see how the two expressions are related consider $N_1$ roots of type~1. We can split these into two groups of $M_1$ and $M_{\bar{1}}$ roots, and introduce a new set of roots $y_{\bar{1},k}$ defined by
\begin{equation}
  \label{eq:1-to-1bar}
  y_{\bar{1},k} = \frac{1}{y_{1,k+M_1}} .
\end{equation}
We then find that the couplings between the roots of type~1 and the momentum carrying roots of type~2 and~$\bar{2}$ can be written as
\begin{equation}
  \begin{aligned}
    \prod_{j=1}^{N_1} \frac{x_k^- - y_{1,j}}{x_k^+ - y_{1,j}} 
    &= 
    \Bigl( \frac{x_k^-}{x_k^+} \Bigr)^{M_{\bar{1}}}
    \prod_{j=1}^{M_1} \frac{x_k^- - y_{1,j}}{x_k^+ - y_{1,j}}
    \prod_{j=1}^{M_{\bar{1}}} \frac{1 - \frac{1}{x_k^- y_{\bar{1},j}}}{1 - \frac{1}{x_k^+ y_{\bar{1},j}}} ,
    \\
    \prod_{j=1}^{N_1} \frac{1 - \frac{1}{\bar{x}^+_k y_{1,j}}}{1 - \frac{1}{\bar{x}^+_k y_{1,j}}}
    &=
    \Bigl( \frac{\bar{x}^-_k}{\bar{x}^+_k} \Bigr)^{M_{\bar{1}}}
    \prod_{j=1}^{M_1} \frac{1 - \frac{1}{\bar{x}^+_k y_{1,j}}}{1 - \frac{1}{\bar{x}^-_k y_{1,j}}}
    \prod_{j=1}^{M_{\bar{1}}} \frac{\bar{x}^+_k - y_{\bar{1},j}}{\bar{x}^-_k - y_{\bar{1},j}} .
  \end{aligned}
\end{equation}
In the same way we can group the roots of type~3 into two groups of $M_3$ and $M_{\bar{3}}$ roots, respectively. In this way we obtain one set of auxiliary roots for each supercharge. If we then go to the spin-chain frame and shift the length $L$ by
\begin{equation}
  L \to L - M_{\bar{1}} - M_{\bar{3}} ,
\end{equation}
and only consider the massive sector by setting $K_0=0$ we exactly reproduce the Bethe equations of~\cite{Borsato:2013qpa}.

\begin{figure}
  \centering
  \includegraphics{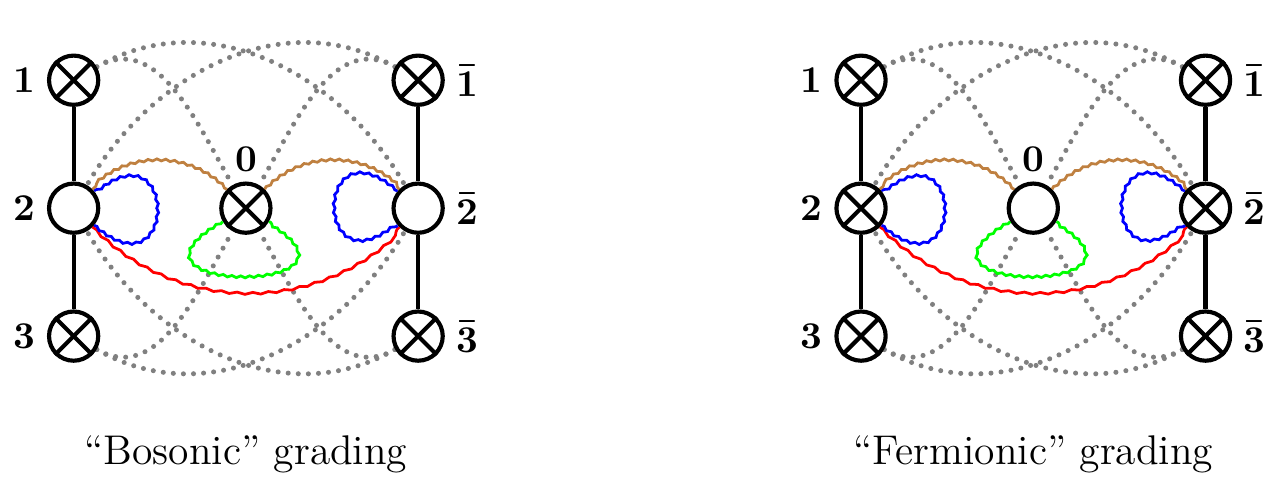}

  \caption{The Bethe equations can be represented with two copies of the Dynkin diagram for $\algPSU(1,1|2)$ supplemented by one node for massless fermions. This can be done in two different ways (left and right panel), corresponding to different choices of the superalgebra grading. We use solid lines for Dynkin links, and dotted lines for the other interactions between auxiliary nodes and momentum carrying ones. Blue and red wavy links are used for dressing phases of the massive sector $\sigma^{\bullet\bullet}$ and $\widetilde{\sigma}^{\bullet\bullet}$ respectively. Brown wavy links represent the dressing phase $\sigma^{\circ\bullet}$, while the green one represents $\sigma^{\circ\circ}$. Note that we here indicate four types of auxiliary roots. As explained in the main text we can get rid of two sets of auxiliary roots by changing roots of type $\bar{1}$ and $\bar{3}$ to type $1$ and $3$ by the inverse of the transformation in~\eqref{eq:1-to-1bar}. The map between the two gradings will be discussed in Section~\ref{sec:fermionic-duality}.}
  \label{fig:bethe-equations}
\end{figure}

A graphical representation of the Bethe equations is given in Fig.~\ref{fig:bethe-equations}.

\paragraph{Frame independence.}

The frame factors $\nu_j$ describe the difference between the string frame and spin-chain frame S matrices. The spectrum must be independent of the chosen frame. Since each state in the spectrum corresponds to a (level-matched) solution of the Bethe equations, the equations should have the same solutions in either frame. To check this, let us collect the factors of $\nu$ coming from the various equations. For the auxiliary root equation~\eqref{eq:BA-1-3} we get a factor of the form
\begin{equation}
  \prod_j^{N_2} \nu_j^{-\frac{1}{2}} \prod_j^{N_{\bar{2}}} \nu_j^{-\frac{1}{2}} \prod_j^{N_0} \nu_j^{-\frac{1}{2}}
\end{equation}
which is equal to identity in the string frame once we impose level matching.

For the momentum carrying nodes we get, after again imposing level matching, that the right hand side of each equation contains an overall phase
\begin{equation}
  \nu_k^{-N_2 + \frac{1}{2} ( N_1 + N_3 - N_0 ) } 
\end{equation}
This factor is trivial in the spin-chain frame. In the string frame it can be absorbed in a change of the length $L$
\begin{equation}
  L \to L - N_2 + \frac{1}{2} ( N_1 + N_3 - N_0 ) .
\end{equation}
Hence the Bethe equations in the two frames are identical up to the interpretation of the unphysical parameter $L$.

\paragraph{Global $\algPSU(1,1|2)^2$  symmetry.}

To understand how the global symmetry algebra acts on a solution of the Bethe equations it is useful to work in the spin-chain frame by setting $\nu_k = 1$.
Given a solution to the Bethe equations we can then construct a new solution by adding an additional root of type~$1$, $2$ or~$3$ with zero momentum, in other words with $y_{I,k}$ and $x_k^{\pm}$ sitting at infinity. This is a manifestation of the $\algPSU(1,1|2)_L$ symmetry of the spectrum. Finding the $\algPSU(1,1|2)_R$ symmetry is a bit more subtle. Firstly, we can add a root of type~$\bar{2}$ at infinity which encodes the $\algSU(1,1)_R$ symmetry. To also see the corresponding supersymmetries we need to add a root of type~$1$ or type~$3$ at $y = 0$. Doing this we see right away that the equation for the new auxiliary root is satisfied provided the original solution satisfies the level matching condition~\eqref{eq:level-matching}. Adding a root at $y=0$ in the equations for the momentum-carrying roots we find that we indeed get a new solution to the equations provided we at the same time shift the length $L$ by
\begin{equation}
  L \to L - 1 .
\end{equation}
Hence, the supersymmetries of the $\algPSU(1,1|2)_R$ algebra are \emph{dynamical} and relate spin-chain states of different lengths. 
In the string frame, where $\nu_k = e^{ip_k}$, adding a root at infinity (or zero in the case of roots of type~$1$ and~$3$) is again a symmetry provided we shift the length $L$ in an appropriate way.

By expanding around a root at infinity we can read off the global charges of the corresponding state expressed in terms of the excitation numbers. 
The $\algPSU(1,1|2)_L \times \algPSU(1,1|2)_R$ symmetry algebra has four Cartan elements: two $\algSU(1,1)$ charges $\Delta_L$ and $\Delta_R$ and two angular momenta $J_L$ and $J_R$. It is convenient to combine them into
\begin{equation}
  \Delta = \Delta_L + \Delta_R , \quad
  J = J_L + J_R, \quad
  S = \Delta_L - \Delta_R , \quad
  K = J_L - J_R.
\end{equation}
We then find\footnote{
  To find these expressions it is useful to note that in the bosonic grading the Dynkin labels of the $\algPSU(1,1|2)_{\smallL}$ algebra correspond to the Cartan elements $(\Delta_{\smallL} - J_{\smallL}, 2J_{\smallL}, \Delta_{\smallL} - J_{\smallL})$, while for the algebra $\algPSU(1,1|2)_{\smallR}$ we have $(J_{\smallR}-\Delta_{\smallR},2\Delta_{\smallR},J_{\smallR}-\Delta_{\smallR})$.
}
\begin{equation}
  \label{eq:charges-bos}
  \begin{aligned}
    \Delta &= L + N_{\bar{2}} + \tfrac{1}{2} \bigl( N_1 + N_3 - N_0 \bigr) + \delta D , \\
    J &= L - N_2 + \tfrac{1}{2} \bigl( N_1 + N_3 - N_0 \bigr) , \\
    S &= \phantom{L} - N_{\bar{2}} + \tfrac{1}{2} \bigl( N_1 + N_3 - N_0 \bigr) , \\
    K &= \phantom{L} - N_2 + \tfrac{1}{2} \bigl( N_1 + N_3 - N_0 \bigr) ,
  \end{aligned}
\end{equation}
where
\begin{equation}
  \delta D = + ih \sum_{k=1}^{N_2} \Bigl( \frac{1}{x_k^+} - \frac{1}{x_k^-} \Bigr)
  + ih \sum_{k=1}^{N_{\bar{2}}} \Bigl( \frac{1}{\bar{x}_k^+} - \frac{1}{\bar{x}_k^-} \Bigr)
  + ih \sum_{k=1}^{N_0} \Bigl( \frac{1}{z_k^+} - \frac{1}{z_k^-} \Bigr) 
\end{equation}
is the anomalous dimension of the state.

There are also global symmetries corresponding to insertions of massless momentum carrying roots. However, these are more complicated and will be further discussed below.

\paragraph{Global $\algSU(2)_{\circ}$ symmetry.}

In the Bethe equations we have included a single type of massless momentum-carrying root, $N_0$. In the world-sheet theory there are two $\algPSU(1|1)^4_{\text{c.e.}}$ multiplets of massless excitations, which form a doublet under the $\algSU(2)_{\circ}$ symmetry. In order to keep track of the $\algSU(2)_{\circ}$ quantum numbers, we could have introduced two different types of massless roots. However, the S matrix acts trivially on the $\algSU(2)_{\circ}$ indices\footnote{%
  In~\cite{Borsato:2014exa,Borsato:2014hja} a non-trivial $\algSU(2)_{\circ}$ was included in the massless sector. However, in order to match with perturbative results this part of the massless S matrix has to be trivial~\cite{Sundin:2015uva,Sundin:2016gqe}.%
} %
the two types of massless excitations enter the Bethe equations in an identical way. Note that this indicates that a solution to the Bethe ansatz equations which includes $N_0$ massless modes represents $2^{N_0}$ degenerate states. It would be very interesting to investigate these degeneracies further. For completeness, we present the Bethe equations in the case of a non-trivial $\algSU(2)_{\circ}$ symmetry in Appendix~\ref{sec:nest-su2}

\paragraph{Further degeneracies.}

In addition to the symmetries described above, the spectrum obtained from the Bethe equations~\eqref{eq:BA-2}--\eqref{eq:BA-1-3} contains another big set of degeneracies. To see this we note that, \eg, the roots of type~$1$ and type~$3$ enter the Bethe equations in a completely symmetric way. Hence, for a given solution we can obtain a new solution by removing a root of type~$1$ and inserting a new root of type~$3$ at the same position
\begin{equation}
  N_1 \to N_1 - 1 , \qquad
  N_3 \to N_3 + 1 , \qquad
  y_{1,N_1} \to y_{3,N_3+1} .
\end{equation}
This situation is very similar to the $\algPSU(1,1|2)$ sector of $\superN=4$ SYM~\cite{Beisert:2005fw}. A part of this degeneracy is explained by the $\algSU(2)_{\bullet}$ outer automorphism that acts on $\algPSU(1,1|2)^2$. However, that symmetry only explains the degeneracy of states that live in the same irreducible $\algSU(2)_{\bullet}$ representation, but the Bethe equations show a degeneracy also between different such representations. In~\cite{Beisert:2007sk} this degeneracy was explained by the existence of an infinite set of bilocal operators that commute with the $\algPSU(1,1|2)$ symmetry. It would be interesting to further study these operators in the $\AdS_3$ case and to see how the degeneracies noted above manifests itself in the string theory spectrum.

\subsection{Fermionic duality and Bethe equations in the fermionic grading}
\label{sec:fermionic-duality}

The Bethe equations~\eqref{eq:BA-2}--\eqref{eq:BA-1-3} are written in a particular grading of the $\algPSU(1,1|2)^2$ algebra, where the simple positive roots correspond to the generators
\begin{equation}
  \gen{Q}_{+--}^{\smallL} , \qquad
  \gen{L}_+^{\smallL} , \qquad  
  \gen{Q}_{+-+}^{\smallL} , \qquad
  \gen{Q}_{-++}^{\smallR} , \qquad
  \gen{S}_+^{\smallR} , \qquad
  \gen{Q}_{-+-}^{\smallR} .
\end{equation}
By performing a fermionic duality transformation on the nodes corresponding to auxiliary roots we obtain equations in a different grading~\cite{Essler:1992nk,Beisert:2005fw,Borsato:2013qpa}. Here we consider the case where the simple roots are all odd and correspond to
\begin{equation}
  \gen{Q}_{-++}^{\smallL} , \qquad
  \gen{Q}_{++-}^{\smallL} , \qquad  
  \gen{Q}_{+-+}^{\smallL} , \qquad
  \gen{Q}_{+--}^{\smallR} , \qquad
  \gen{Q}_{+++}^{\smallR} , \qquad
  \gen{Q}_{-+-}^{\smallR} .
\end{equation}
Let us define the following polynomial of degree $N_2 + N_{\bar{2}} + N_0 - 1$
\begin{equation}\label{eq:P-x}
  \begin{split}
    P(\zeta)
    &= 
    \prod_{j=1}^{N_2} ( \zeta - x_j^+ ) \nu_j^{-\frac{1}{2}}
    \prod_{j=1}^{N_{\bar{2}}} \bigl( \zeta - \frac{1}{\bar{x}_j^-} \bigr) \nu_j^{-\frac{1}{2}}
    \prod_{j=1}^{N_0} ( \zeta - z_j^+ ) \nu_j^{-\frac{1}{2}}
    \\ &\qquad
    -
    \prod_{j=1}^{N_2} ( \zeta - x_j^- )
    \prod_{j=1}^{N_{\bar{2}}} \bigl( \zeta - \frac{1}{\bar{x}_j^+} \bigr)
    \prod_{j=1}^{N_0} ( \zeta - z_j^- ) .
  \end{split}
\end{equation}
Using equation~\eqref{eq:BA-1-3} and the level-matching condition~\eqref{eq:level-matching} we find that $P(\zeta)$ has $N_1 + 1$ roots
\begin{equation}\label{eq:P-y}
  P(y_{1,j}) = 0 , \qquad
  P(0) = 0 .
\end{equation}
The polynomial can therefore also be written as
\begin{equation}
  P(\zeta) = 
  \zeta 
  \prod_{j=1}^{N_1} ( \zeta - y_{1,j} ) 
  \prod_{j=1}^{\tilde{N}_1} ( \zeta - \tilde{y}_{1,j} ) ,
\end{equation}
where we have introduced the additional $\tilde{N}_1$ roots $\tilde{y}_{1,j}$. In order to match the degrees of the polynomial we find
\begin{equation}
  \tilde{N}_1 = N_2 + N_{\bar{2}} + N_0 - N_1 - 2 .
\end{equation}
By evaluating the ratios
\begin{equation}
  \frac{P(x_k^+)}{P(x_k^-)} , \qquad
  \frac{P(1/\bar{x}_k^-)}{P(1/\bar{x}_k^+)} , \qquad
  \frac{P(z_k^+)}{P(z_k^-)} ,
\end{equation}
using the two different expressions for $P(\zeta)$ we obtain a set of relations that can be used to rewrite the Bethe equations. The resulting equations for the momentum-carrying roots read
\begin{align}
  \begin{split}
    \label{eq:BA-2-dual}
    \left(\frac{x_k^+}{x_k^-}\right)^{\tilde{L}}
    &=
    \prod_{\substack{j = 1\\j \neq k}}^{N_2} \nu_k^{-1}\nu_j 
    \frac{1- \frac{1}{x_k^+ x_j^-}}{1- \frac{1}{x_k^- x_j^+}} 
    (\sigma^{\bullet\bullet}_{kj})^2
    \prod_{j=1}^{N_{\bar{2}}} \nu_j  
    \frac{1- \frac{1}{x_k^+ \bar{x}_j^-}}{1- \frac{1}{x_k^- \bar{x}_j^+}} 
    (\widetilde{\sigma}^{\bullet\bullet}_{kj})^2
    \\ &\phantom{= }\times
    \prod_{j=1}^{N_{0}} 
    \nu_k^{-1/2} \nu_j 
    \left(\frac{1 - \frac{1}{x_k^- z_j^-}}{1 - \frac{1}{x_k^+ z_j^+}}\right)^{\!\!\frac{1}{2}}
    \left(\frac{1 - \frac{1}{x_k^+ z_j^-}}{1 - \frac{1}{x_k^- z_j^+}}\right)^{\!\!\frac{1}{2}}
    (\sigma^{\bullet\circ}_{kj})^2
    \\ &\phantom{= }\times
    \prod_{j=1}^{\tilde{N}_1}  \nu_k^{\frac{1}{2}} \, \frac{x_k^+ - \tilde{y}_{1,j}}{x_k^- - \tilde{y}_{1,j}}
    \prod_{j=1}^{N_3} \nu_k^{\frac{1}{2}} \, \frac{x_k^- - y_{3,j}}{x_k^+ - y_{3,j}}
    ,
  \end{split} \\
  \begin{split}
    \label{eq:BA-2b-dual}
    \left(\frac{\bar{x}_k^+}{\bar{x}_k^-}\right)^{\tilde{L}} 
    &=
    \prod_{\substack{j = 1\\j \neq k}}^{N_{\bar{2}}} 
    \frac{1- \frac{1}{\bar{x}_k^+ \bar{x}_j^-}}{1- \frac{1}{\bar{x}_k^- \bar{x}_j^+}}  
    (\sigma^{\bullet\bullet}_{kj})^2
    \prod_{j=1}^{N_2} \nu_k^{-1} 
    \frac{1- \frac{1}{\bar{x}_k^+ x_j^-}}{1- \frac{1}{\bar{x}_k^- x_j^+}}
    (\widetilde{\sigma}^{\bullet\bullet}_{kj})^2
    \\ &\phantom{= }\times
    \prod_{j=1}^{N_{0}} 
    \nu_k^{-1/2} 
    \left(\frac{1- \frac{1}{\bar{x}_k^- z_j^-}}{1- \frac{1}{\bar{x}_k^+ z_j^+}}\right)^{\!\!\frac{1}{2}} 
    \left(\frac{1- \frac{1}{\bar{x}_k^+ z_j^-}}{1- \frac{1}{\bar{x}_k^- z_j^+}}\right)^{\!\!\frac{1}{2}} 
    (\sigma^{\bullet\circ}_{kj})^2
    \\ &\phantom{= }\times
    \prod_{j=1}^{\tilde{N}_{1}} \nu_k^{\frac{1}{2}} \frac{1 - \frac{1}{\bar{x}_k^- \tilde{y}_{1,j}}}{1- \frac{1}{\bar{x}_k^+ \tilde{y}_{1,j}}}
    \prod_{j=1}^{N_{3}} \nu_k^{\frac{1}{2}}\frac{1 - \frac{1}{\bar{x}_k^+ y_{3,j}}}{1- \frac{1}{\bar{x}_k^- y_{3,j}}} 
    .
  \end{split}
  \\
  \begin{split}
    \label{eq:BA-0-dual}
    \left(\frac{z_k^+}{z_k^-}\right)^{\tilde{L}} 
    &=
    \prod_{\substack{j = 1\\j \neq k}}^{N_0} 
    \nu_k^{-1/2} \nu_j^{+1/2}
    (\sigma^{\circ\circ}_{kj})^2
    \prod_{j=1}^{\tilde{N}_1}\nu_k^{\frac{1}{2}} \frac{z_k^+ - y_{1,j}}{z_k^- - y_{1,j}}
    \prod_{j=1}^{N_3} \nu_k^{\frac{1}{2}}  \frac{z_k^- - y_{3,j}}{z_k^+ - y_{3,j}} 
    \\ &\phantom{= }\times
    \prod_{j=1}^{N_{2}}
    \nu_k^{-1} \nu_j^{1/2}
    \left(\frac{1-\frac{1}{z_k^+x_j^+}}{1-\frac{1}{z_k^-x_j^-}}\right)^{\!\!\frac{1}{2}}
    \left(\frac{1-\frac{1}{z_k^+x_j^-}}{1-\frac{1}{z_k^-x_j^+}}\right)^{\!\!\frac{1}{2}}
    (\sigma^{\circ\bullet}_{kj})^2
    \\ &\phantom{= }\times
    \prod_{j=1}^{N_{\bar{2}}} 
    \nu_j^{1/2}
    \left(\frac{1-\frac{1}{z_k^+\bar{x}_j^+}}{1-\frac{1}{z_k^-\bar{x}_j^-}}\right)^{\!\!\frac{1}{2}}
    \left(\frac{1-\frac{1}{z_k^+\bar{x}_j^-}}{1-\frac{1}{z_k^-\bar{x}_j^+}}\right)^{\!\!\frac{1}{2}}
    (\sigma^{\circ\bullet}_{kj})^2
    ,
  \end{split} 
\end{align}
where the new length $\tilde{L}$ is given by
\begin{equation}
  \tilde{L} = L + N_{\bar{2}} - 1.
\end{equation}
The new roots of type~$\tilde{1}$ satisfy
\begin{equation}
  \label{eq:BA-1-dual}
  1
  = 
  \prod_{j=1}^{N_2} \frac{\tilde{y}_{1,k} - x_j^-}{\tilde{y}_{1,k} - x_j^+} \nu_j^{+\frac{1}{2}}
  \prod_{j=1}^{N_{\bar{2}}} \frac{1 - \frac{1}{\tilde{y}_{1,k} \bar{x}_j^+}}{1- \frac{1}{\tilde{y}_{1,k} \bar{x}_j^-}} \nu_j^{+\frac{1}{2}}
  \prod_{j=1}^{N_0} \frac{\tilde{y}_{1,k} - z_j^- }{\tilde{y}_{1,k} - z_j^+} \nu_j^{+\frac{1}{2}} ,
\end{equation}
while the equations for the roots of type~$3$ and~$4$ are unchanged. These equations can be represented pictorially as in the right panel of Fig.~\ref{fig:bethe-equations}.

By changing the grading of the superalgebra we also change the definition of highest weight states. Let $\ket{\text{h.w.}}$ be a highest weight state $\ket{\text{h.w.}}$ in the original ($\algSU(2) \times \algSL(2)$) grading, and hence correspond to a solution of the Bethe equations~\eqref{eq:BA-2}--\eqref{eq:BA-1-3} with no roots at infinity. Using the fermionic duality transformation we can find the corresponding solution in the dual equations~\eqref{eq:BA-2-dual}--\eqref{eq:BA-1-dual}. However, the state $\ket{\text{h.w.}}$ is \emph{not} a highest weight state in the fermionic grading. In this case the solution with no roots at infinity instead correspond to the state $\gen{Q}_{-++}^{\smallL} \gen{Q}_{+--}^{\smallR} \ket{\text{h.w.}}$, which is a descendent in the original grading, but a primary in the fermionic one.

The global $\algPSU(1,1|2)^2$ charges of a solution to the Bethe equations in the fermionic grading are given by\footnote{
  In the fermionic grading the Dynkin labels of $\algPSU(1,1|2)_{\smallL}$ and $\algPSU(1,1|2)_{\smallR}$ correspond to $(J_{\smallL}-\Delta_{\smallL},\Delta_{\smallL} +J_{\smallL},\Delta_{\smallL}-J_{\smallL})$, and $(\Delta_{\smallR}-J_{\smallR},\Delta_{\smallR} +J_{\smallR},J_{\smallR}-\Delta_{\smallR})$, respectively.
}
\begin{equation}
  \label{eq:charges-ferm}
  \begin{aligned}
    \Delta &= \tilde{L} + \frac{1}{2} \bigl( N_2 + N_{\bar{2}} - \tilde{N}_1 + N_3 \bigr) + \delta\Delta , \\
    J &= \tilde{L} - \frac{1}{2} \bigl( N_2 + N_{\bar{2}} + \tilde{N}_1 - N_3 \bigr) , \\
    S &= \phantom{L} + \frac{1}{2} \bigl( N_2 - N_{\bar{2}} - \tilde{N}_1 + N_3 \bigr) , \\
    K &= \phantom{L} - \frac{1}{2} \bigl( N_2 - N_{\bar{2}} + \tilde{N}_1 - N_3 \bigr) .
  \end{aligned}
\end{equation}
Plugging back the expressions for $\tilde{L}$ and $\tilde{N}_1$ into these expressions and comparing the charges in the bosonic grading from equation~\eqref{eq:charges-bos} we see that that they match up to a shift
\begin{equation}
  L \to L - 1 , \qquad N_1 \to N_1 + 2 ,
\end{equation}
which exactly corresponds to the application of the two supercharges discussed above.

Note that in the fermionic grading the degeneracy discussed above, where a root of type~$1$ can be made into a root of type~$3$, instead takes the form of adding both a root of type~$\tilde{1}$ and of type~$3$ at the same position.

\subsection{Massless zero modes}
\label{sec:mass-zero-modes}

As discussed above, descendants of a primary operator under the global $\algPSU(1,1|2)^2$ symmetry can be constructed in the Bethe equations by adding additional Bethe roots at infinity. A momentum carrying root of type~$2$ or~$\bar{2}$ sitting at infinity corresponds to a massive world-sheet excitation with zero momentum. The massless momentum carrying roots of type~$0$ are constrained to live on the unit circle, and can therefore never sit at infinity. For these roots a zero momentum excitation instead sits at $z^+ = z^- = z_0^{(\pm)} = \pm 1$. As a shorthand we will denote either of those points by $z_0$.

It is straightforward to check that if we insert an extra root of type~$0$ at $z^{\pm} = z_0$ in the dualised Bethe equations of Section~\ref{sec:fermionic-duality} we obtain a new solution. In the ``fermionic'' grading the massless momentum carrying roots correspond to bosonic excitations on $\Torus^4$, and the symmetry may be naturally interpreted as giving rise to infinitesimal shifts along the torus. 

If we instead work in the ``bosonic'' grading, adding a massless root at $z^{\pm} = z_0$ does \emph{not}, in general, give a solution to the Bethe equations. At first sight this might seem strange -- after all the symmetries of the system should not depend on the choice of grading. However, it is important to remember that equivalent solutions to the Bethe equations in different gradings do not correspond to exactly the same operator but to two operators in the same supermultiplet. To understand what happens with the root at $z^{\pm} = z_0$ when we dualise back to the original grading we go back to the polynomial $P(\zeta)$ defined in~\eqref{eq:P-x} and~\eqref{eq:P-y}. Let us consider a particular solution to the Bethe equations in the fermionic grading such that $\tilde{y}_1 = z_0$ is not an auxiliary Bethe root of the solution. To transform this solution to the bosonic grading we construct a polynomial $P(\zeta)$
\begin{equation}
  \begin{split}
    P(\zeta)
    &= 
    \prod_{j=1}^{N_2} ( \zeta - x_j^+ ) \nu_j^{-\frac{1}{2}}
    \prod_{j=1}^{N_{\bar{2}}} \bigl( \zeta - \frac{1}{\bar{x}_j^-} \bigr) \nu_j^{-\frac{1}{2}}
    \prod_{j=1}^{N_0} ( \zeta - z_j^+ ) \nu_j^{-\frac{1}{2}}
    \\ &\qquad
    -
    \prod_{j=1}^{N_2} ( \zeta - x_j^- )
    \prod_{j=1}^{N_{\bar{2}}} \bigl( \zeta - \frac{1}{\bar{x}_j^+} \bigr)
    \prod_{j=1}^{N_0} ( \zeta - z_j^- ) .
  \end{split}
\end{equation}
If we now add a root $z_{N_0+1}^{\pm} = z_0$ we obtain a new polynomial $\hat{P}(\zeta)$ with a degree one higher than that of $P(\zeta)$, 
\begin{equation}
  \begin{split}
    \hat{P}(\zeta)
    &= 
    (\zeta - z_{N_0+1}^+) \prod_{j=1}^{N_2} ( \zeta - x_j^+ ) \nu_j^{-\frac{1}{2}}
    \prod_{j=1}^{N_{\bar{2}}} \bigl( \zeta - \frac{1}{\bar{x}_j^-} \bigr) \nu_j^{-\frac{1}{2}}
    \prod_{j=1}^{N_0} ( \zeta - z_j^+ ) \nu_j^{-\frac{1}{2}}
    \\ &\qquad
    -
    (\zeta - z_{N_0+1}^-) \prod_{j=1}^{N_2} ( \zeta - x_j^- )
    \prod_{j=1}^{N_{\bar{2}}} \bigl( \zeta - \frac{1}{\bar{x}_j^+} \bigr)
    \prod_{j=1}^{N_0} ( \zeta - z_j^- ) 
    \\
    &= (\zeta - z_0) P(\zeta) .
  \end{split}
\end{equation}
This polynomial has a root at $\zeta = z_0$. Since we have assumed that there is no auxiliary roots at this position in the fermionic grading this root has to be an auxiliary root in the bosonic grading, and we will denote it by $y_{1,N_1+1} = z_0$. Hence, we expect that if we add not only a massless root at $z^{\pm} = z_0$ but also an auxiliary root at $y_1 = z_0$ we will get a symmetry of the bosonic Bethe equations. It is straightforward to check that his works in the original equations. Consider, \eg, the equation for a momentum carrying root $x_k^{\pm}$ of type~$2$. The coupling of this root to the new massless root gives a factor of the form\footnote{Note that the massive--massless dressing phase is trivial when one of the excitations have vanishing momentum.}
\begin{equation}
  \nu_k^{-\frac{1}{2}} \frac{x_k^+ - z_0}{x_k^- - z_0}
\end{equation}
on the right hand side of~\eqref{eq:BA-2}. At the same time the extra root of type~$1$ at $y_1 = z_0$ gives a factor
\begin{equation}
  \nu_k^{+\frac{1}{2}}  \frac{x_k^- - z_0}{x_k^+ - z_0} ,
\end{equation}
which exactly cancels the contribution from the massless root. Similar cancelation occur in all the other original Bethe equations. However, we also need to consider the equations satisfied by the two new roots $z_{N_0+1}^{\pm}$ and $y_{1,N_1+1}$. Here we run into a problem since both of these equations contain factors of the form
\begin{equation}
  \frac{z_{N_0+1}^- - y_{1,N_1+1}}{z_{N_0+1}^+ - y_{1,N_1+1}} .
\end{equation}
When we plug in the values of the two roots this takes the indeterminate form $0/0$. Hence we need to be a bit careful. Let us therefore write out the equations for the two roots and plug in the values $z_{N_0+1}^{\pm} = y_{1,N_1+1} = z_0$ everywhere except in the factor written above. For the $y_{1,N_1+1}$ root we get (here we have assumed a physical state with vanishing total momentum)
\begin{equation}
  1
  = 
  \frac{y_{1,N_1+1} - z_{N_0+1}^+}{y_{1,N_1+1} - z_{N_0+1}^-} 
  \prod_{j=1}^{N_2} \frac{z_0 - x_j^+}{z_0 - x_j^-} 
  \prod_{j=1}^{N_{\bar{2}}} \frac{1 - \frac{1}{z_0 \bar{x}_j^-}}{1 - \frac{1}{z_0 \bar{x}_j^+}} 
  \prod_{j=1}^{N_0} \frac{z_0 - z_j^+}{z_0 - z_j^-} 
\end{equation}
and for $z_{N_0+1}^{\pm}$ we get
\begin{equation}
  1
  =
  \frac{z_{N_0+1}^- - y_{1,N_1+1}}{z_{N_0+1}^+ - y_{1,N_1+1}}
  \prod_{j=1}^{N_0} \frac{z_0 - z_j^-}{z_0 - z_j^+}
  \prod_{j=1}^{N_2} \frac{z_0 - x_j^-}{z_0 - x_j^+}
  \prod_{j=1}^{N_{\bar{2}}} \frac{1 - \frac{1}{z_0 \bar{x}_j^+}}{1 - \frac{1}{z_0 \bar{x}_j^-}} .
\end{equation}
Hence the two remaining Bethe equations are actually identical. We introduce the notation\footnote{For a physical configuration $\Upsilon$ is real.}
\begin{equation}
  e^{i \Upsilon}
  =
  \prod_{j=1}^{N_0} \frac{z_0 - z_j^-}{z_0 - z_j^+}
  \prod_{j=1}^{N_2} \frac{z_0 - x_j^-}{z_0 - x_j^+}
  \prod_{j=1}^{N_{\bar{2}}} \frac{1 - \frac{1}{z_0 \bar{x}_j^+}}{1 - \frac{1}{z_0 \bar{x}_j^-}} .
\end{equation}
The full Bethe ansatz equation is then solved if we can find a limit such that
\begin{equation}
  z_{N_0+1}^{\pm} \to z_0 , \qquad
  y_{1,N_1+1} \to z_0 , \qquad
  \frac{z_{N_0+1}^+ - y_{1,N_1+1}}{z_{N_0+1}^- - y_{1,N_1+1}} \to e^{i\Upsilon} .
\end{equation}
This sort of limiting procedure is well-known in the Bethe ansatz literature, see for example the exceptional states in $\AdS_5 \times \Sphere^5$~\cite{Arutyunov:2012tx}. We can solve the equation by writing
\begin{equation}
  z_{N_0+1}^{\pm} \approx z_0 \pm i \epsilon \tan\frac{\Upsilon}{2} , \qquad
  y_{1,N_1+1} \approx z_0 - \epsilon , \qquad
  \epsilon \to 0 .
\end{equation}
Hence, there is a symmetry in the Bethe equations in the bosonic grading, where we add a massless momentum carrying root and a root of type~$1$. As discussed in the previous section, a root of type~$1$ can be freely replaced by a root of type~$3$ at the same position. This gives rise to a second massless zero mode. In the fermionic grading this solution now has three additional roots, one massless, one of type~$\tilde{1}$ and one of type~$3$. Again we need to take a particular limit to get a solution of the full equation
\begin{equation}
  z_{N_0+1}^{\pm} \approx z_0 \pm i \epsilon \tan\frac{\Upsilon}{2} , \qquad
  \tilde{y}_{1,\tilde{N}_1+1} \approx z_0 - \epsilon , \qquad
  y_{3,N_3+1} \approx z_0 - \epsilon , \qquad
  \epsilon \to 0 .
\end{equation}
The two massless zero modes found above differ only in the $\algSU(2)_{\bullet}$ quantum numbers. Since the massless momentum-carrying roots can carry two different $\algSU(2)_{\circ}$ spins, we thus have the four expected shift symmetries along $\Torus^4$ corresponding to the addition of massless zero-modes. We have interpreted them as being related to changes in the position of the state along the torus. 

\subsection{The weak-coupling limit}
\label{sec:weak-coupling-limit}

Let us consider the weak coupling limit of the bosonic grading Bethe equations. The weak coupling limit is defined by sending $h \to 0$ while keeping the momentum fixed. To understand how the Bethe roots scale in this limit we remind the reader that the massive momentum-carrying roots $x^{\pm}$ can be written in terms of the momentum as
\begin{equation}
  x^{\pm} = \frac{1 + \sqrt{1 + 4h^2\sin^2\frac{p}{2}}}{2h\sin\frac{p}{2}} e^{\pm \frac{ip}{2}} ,
\end{equation}
with an identical expression for $\bar{x}^{\pm}$. The massless momentum-carrying roots $z^{\pm}$, on the other hand, take the form
\begin{equation}
\label{eq:massless-rapidity-no-h}
  z^{\pm} = e^{\pm\frac{ip}{2}} \text{sgn}\left(\sin\tfrac{p}{2}\right).
\end{equation}
In the small $h$ limit it is then useful to introduce, for the massive roots
\begin{equation}
  x^{\pm} \approx \frac{u \pm \frac{i}{2}}{h} , \qquad \bar{x}^{\pm} \approx \frac{\bar{u} \pm \frac{i}{2}}{h} ,
\label{eq:weak-massive-raps}
\end{equation}
while the massless roots $z^{\pm}$ remain fixed, since equation~\eqref{eq:massless-rapidity-no-h} has no dependence on $h$.

For the auxiliary roots of type~$1$ and~$3$ we need to be a bit more careful. When we send the coupling $h$ to zero there are in general three possible behaviours for such a root: the position of the roots goes to infinity, it goes to zero or it remains finite. 
That the auxiliary roots split into such three cases can also be seen by looking directly at the nesting procedure applied to the weak-coupling S matrix. We refer the reader to Appendix~\ref{sec:nest-proc-weak} for a discussion of this.
We will therefore split the $N_1$ auxiliary roots of type~$1$ into three different groups of $M_1 + M_{\bar{1}} + R_1$ roots  depending on their scaling with $h$.
\begin{equation}\label{eq:aux-roots-weak}
  y_{1,i} \approx \begin{cases}
    \frac{v_{1,i}}{h} , & i = 1, \dotsc, M_1 , \\
    \frac{h}{v_{\bar{1},i-M_1}} , & i = M_1 + 1, \dotsc, M_1 + M_{\bar{1}} , \\
    r_{1,i-M_1-M_{\bar{1}}} , & i = M_1 + M_{\bar{1}} + 1 , \dotsc, N_1 = M_1 + M_{\bar{1}} + R_1 ,
  \end{cases}
\end{equation}
and similarly for the roots of type~$3$.

The weak coupling Bethe equations in the spin-chain frame then take the form
\begin{align}
  1 &= \prod_{j=1}^{N_2} \frac{v_{1,k} - u_j - \frac{i}{2}}{v_{1,k} - u_j + \frac{i}{2}} ,
  \\
\label{eq:Bethe-weak-2}
  \left(\frac{u_k + \frac{i}{2}}{u_k - \frac{i}{2}}\right)^{\!\!\tilde{L}-N_0+R_1+R_3}
    &= \prod_{\substack{j = 1\\j \neq k}}^{N_2} \frac{u_k - u_j + i}{u_k - u_j - i}
  \prod_{j=1}^{M_1} \frac{u_k - v_{1,j} - \frac{i}{2}}{u_k - v_{1,j} + \frac{i}{2}}
  \prod_{j=1}^{M_3} \frac{u_k - v_{3,j} + \frac{i}{2}}{u_k - v_{3,j} - \frac{i}{2}}
  ,
  \\
  1 &= \prod_{j=1}^{N_2} \frac{v_{3,k} - u_j - \frac{i}{2}}{v_{3,k} - u_j + \frac{i}{2}} ,
  \\
  1 &= \prod_{j=1}^{N_{\bar{2}}} \frac{v_{\bar{1},k} - \bar{u}_j + \frac{i}{2}}{v_{\bar{1},k} - \bar{u}_j - \frac{i}{2}} ,
  \\
  \left(\frac{\bar{u}_k + \frac{i}{2}}{\bar{u}_k - \frac{i}{2}}\right)^{\!\!\tilde{L}}
    &= \prod_{\substack{j = 1\\j \neq k}}^{N_2} \frac{u_k - u_j - i}{u_k - u_j + i}
  \prod_{j=1}^{M_{\bar{1}}} \frac{\bar{u}_k - v_{\bar{1},j} + \frac{i}{2}}{\bar{u}_k - v_{\bar{1},j} - \frac{i}{2}}
  \prod_{j=1}^{M_{\bar{3}}} \frac{\bar{u}_k - v_{\bar{3},j} + \frac{i}{2}}{\bar{u}_k - v_{\bar{3},j} - \frac{i}{2}}
  ,
  \\
  1 &= \prod_{j=1}^{N_{\bar{2}}} \frac{v_{\bar{3},k} - \bar{u}_j + \frac{i}{2}}{v_{\bar{3},k} - \bar{u}_j - \frac{i}{2}} ,
  \\
  \left(\frac{z_k^+}{z_k^-}\right)^{\!\!\tilde{L}}
    &= \prod_{\substack{j = 1\\j \neq k}}^{N_0} \frac{z_k^+ - z_j^-}{z_k^- - z_j^+} ( \sigma_{kj}^{\circ\circ} )^2
  \\ & \quad \times
  \prod_{j=1}^{N_2} \frac{x_j^-}{x_j^+} 
  \prod_{j=1}^{R_1} \frac{z_k^- - r_{1,j}}{z_k^+ - r_{1,j}}
  \prod_{j=1}^{R_3} \frac{z_k^- - r_{3,j}}{z_k^+ - r_{3,j}}
  ,
  \\
  1 &= \prod_{j=1}^{N_0} \frac{r_{1,k} - z_k^+}{r_{1,k} - z_k^-}   \prod_{j=1}^{N_2} \frac{x_j^-}{x_j^+} ,
  \\
  1 &= \prod_{j=1}^{N_0} \frac{r_{3,k} - z_k^+}{r_{3,k} - z_k^-}   \prod_{j=1}^{N_2} \frac{x_j^-}{x_j^+} ,
\end{align}
where $\tilde{L}=L+M_{\bar{1}}+M_{\bar{3}}$. Above, we have imposed the zero-momentum condition. We have also used the fact that at weak-coupling the massive dressing factors and $\sigma^{\bullet\circ}$ are all sub-leading.\footnote{In the weak-coupling limit the massive phases are $\order(h)$~\cite{Borsato:2013hoa}. Since the massless rapidities do not depend on $h$, the AFS- and  HL-order contributions to 
$\sigma^{\circ\circ}$ will be $\order(h)$ and $\order(h^0)$, respectively. As a result, the HL-order massless dressing factor contributes to the weak-coupling Bethe equations. On the other hand, using the expansion~\eqref{eq:weak-massive-raps} for massive rapidities, one finds that the mixed-mass
AFS- and HL-order phases are $\order(h^2)$ and $\order(h)$, respectively. Hence, in the weak-coupling limit, $\sigma^{\bullet\circ}$ is sub-leading.} For $N_0 = 0$ we obtain exactly the expected standard Bethe equations for two decoupled $\algPSU(1,1|2)$ spin-chains in the relevant gradings~\cite{OhlssonSax:2011ms}. When we add some massless excitations things get more interesting. Firstly we note that the length of left and right massive spin chains is not the same unless $N_0-R_1-R_3=0$. Hence, it is natural to interpret the massless fermions as \emph{chiral} states. The fundamental massless fermion in this grading is charged only under the right copy of $\algPSU(1,1|2)$. Furthermore, the massless fermions feel a twist corresponding to the total momentum of the excitations of type~$2$.

From the above set of equations it is easier to understand the massless zero modes. Let us for simplicity assume we have a solution involving only massive modes. If we now try to add a single massless root $z^{\pm}$ we find that we need to satisfy the Bethe equation
\begin{equation}
  \left(\frac{z^+}{z^-}\right)^{\!\!\tilde{L}} = \prod_{j=1}^{N_2} \frac{x_j^+}{x_j^-} .
\end{equation}
Clearly $z^{\pm} = 1$ is not a solution unless the roots of type~$2$ carry total momentum zero. Even in that case, however, the spin-chain length felt by the roots of type~$2$ decreases by one, since adding the root $z$ increases $N_0$. Hence the Bethe equation for roots of type~$2$ is no longer satisfied.\footnote{%
The only exception seems to be when we only have zero-momentum roots of type~$2$. In this case we have a descendant of the ground state in the left-moving sector and the state is short or semi-short.%
} %
If we, on the other hand, also add a root of type~$R_1$ or~$R_3$ we can solve the equations for the new roots using a limiting procedure similar to what we discussed above. In this case the length felt by roots of type~$2$ does \emph{not} change, so this is precisely the weak-coupling limit of the four $\grpU(1)$ symmetries discussed in Section~\ref{sec:mass-zero-modes}.

\subsection{Bethe equations for the mixed RR and NSNS background}
\label{sec:mixed-flux}

Throughout this paper we focus on strings in an $\AdS_3 \times \Sphere^3 \times \Torus^4$ background supported by pure R-R three-form flux. However, this background can be generalised to a one-parameter family of backgrounds containing a mix of R-R and NS-NS flux, in such a way that the worldsheet sigma model remains integrable~\cite{Cagnazzo:2012se,Babichenko:2014yaa}. In the bosonic sigma model the NS-NS flux gives rise to a Wess-Zumino-Witten (WZW) term whose integer level $k$ parametrises the amount of NS-NS flux. The mixed flux S matrix was constructed in the massive sector in~\cite{Hoare:2013pma,Hoare:2013ida}, and for the full model in~\cite{Lloyd:2014bsa}.

The Bethe equations constructed above can be straightforwardly generalised to the mixed flux case. In fact, we have normalised the S matrix in such a way that the equations~(\ref{eq:BA-2}--\ref{eq:BA-1-3}) are valid for any value of the flux, provided the dressing phase is adjusted in the appropriate manner. However, the conditions satisfied by spectral parameters $x^{\pm}$, $\bar{x}^{\pm}$ and $z^{\pm}$ explicitly depend on the WZW level $k$, so we have to be a bit careful when deriving the global charges corresponding to a solution of the Bethe equations are slightly adjusted~\cite{Babichenko:2014yaa}. In the bosonic grading we find
\begin{equation}
  \begin{aligned}
    \Delta &= L + N_{\bar{2}} + \tfrac{1}{2} \bigl( N_1 + N_3 - N_0 \bigr) + \delta D , \\
    J &= L - N_2 + \tfrac{1}{2} \bigl( N_1 + N_3 - N_0 \bigr) , \\
    S &= \phantom{L} - N_{\bar{2}} + \tfrac{1}{2} \bigl( N_1 + N_3 - N_0 \bigr) , \\
    K &= \phantom{L} - N_2 + \tfrac{1}{2} \bigl( N_1 + N_3 - N_0 \bigr) - \frac{k}{2\pi} P ,
  \end{aligned}
\end{equation}
where $P$ is the total momentum of all momentum-carrying excitations. Note that $K$ is quantised, since the total momentum satisfy the level-matching constraint $P \in 2\pi\Integers$ and the WZW level $k$ is integer-valued.

The anomalous dimension $\delta D$ is now given by
\begin{equation}
  \delta D =
  ih\sum_{k=1}^{N_2} \Bigl( \frac{1}{x_k^+} - \frac{1}{x_k^-} \Bigr)
  + ih\sum_{k=1}^{N_{\bar{2}}} \Bigl( \frac{1}{\bar{x}_k^+} - \frac{1}{\bar{x}_k^-} \Bigr)
  + ih\sum_{k=1}^{N_0} \Bigl( \frac{1}{z_k^+} - \frac{1}{z_k^-} \Bigr)
  + \frac{k}{2\pi} \bigl( P_2 - P_{\bar{2}} + P_0 \bigr) ,
\end{equation}
where $P_2$, $P_{\bar{2}}$ and $P_0$ denote the total momentum of each type of excitation,
\begin{equation}
  e^{iP_2} = \prod_{k=1}^{N_2} \frac{x_k^+}{x_k^-} , 
  \qquad
  e^{iP_{\bar{2}}} = \prod_{k=1}^{N_{\bar{2}}} \frac{\bar{x}_k^+}{\bar{x}_k^-} , 
  \qquad
  e^{iP_0} = \prod_{k=1}^{N_0} \frac{z_k^+}{z_k^-} .
\end{equation}
With these adjustments the properties of the Bethe equations discussed earlier in this section remain valid also in the mixed flux case.

\section{Near-BMN expansion of S~matrix}
\label{sec:pert-comp}
In this section we will consider our S~matrix, including the phases, in the near-BMN limit. While the expansion of the S~matrix is straightforward, the scalar factors require some additional care. In the next subsection, we collect some useful expressions for them. In the next, we proceed to compare with the results found recently by Sundin and Wulff~\cite{Sundin:2016gqe}.

\subsection{BMN expansion of S matrix and dressing factors}

\newcommand{\mom}[1]{\mathrm{#1}}

To expand the all-loop S~matrix in the near-BMN limit we rescale the momentum $p$ to be small, which for massive modes amounts to $|x_p^\pm|$ to be large. In the massless kinematics we should distinguish between two cases
\begin{equation}
p\gtrsim 0 \quad\Rightarrow\quad x_p\sim +1,
\qquad\text{and}\qquad
p\lesssim 0 \quad\Rightarrow\quad x_p\sim -1.
\end{equation}
With our choice of the fundamental region $0\leq p<2\pi$, we can equivalently get the latter case $x_p\sim -1$ by taking $p$ to be close to and smaller than $2\pi$. With this in mind, we introduce the rescaled variables
\begin{equation}
\label{eq:momrescaling}
  p \to \mom{p}/h , \qquad q \to \mom{q}/h
\end{equation}
and take the coupling $h$ to be large. The dispersion relation for the massive excitations then takes the form
\begin{equation}
  \omega_{\mom{p}} = \sqrt{1 + \mom{p}^2} + \order(1/h^2),
\end{equation}
while for the massless excitations we have
\begin{equation}
  \omega_{\mom{p}} = |\mom{p}|  + \order(1/h^2).
\end{equation}
We see from this equation that we recover the relativistic massless dispersion relation. In particular this means that, in this limit, there is a clear notion of right- and left-movers on the worldsheet. Our non-relativistic S~matrix is well-defined for any value of the momenta $p,q$, and can in principle be expanded at any arbitrary point. In particular, nothing stops us from expanding it when both momenta have the same worldsheet-chirality. An easy way to see that this regime is incompatible with perturbation theory is to expand the massless  crossing equation~\eqref{eq:crossingeqs}  for \textit{e.g.}\ $0<\mom{p}<\mom{q}$. The right-hand-side becomes
\begin{equation}
f(x_p,x_q)^2 =  \frac{(\mom{p}-\mom{q})^2}{(\mom{p}+\mom{q})^2}+\order(1/h^2),
\qquad \text{for}\ 0<\mom{p}<\mom{q}.
\end{equation}
The first non-trivial order appears at $\order(h^0)$, corresponding to where in perturbation theory is the scattering is free! To satisfy the crossing equation at this order, the dressing factor at  $\order(h^0)$ should have a non-trivial analytic structure, and indeed one can check that this is the case.
It is not surprising that this kinematics does not make sense from the point of view of near-BMN perturbation theory. In the near-relativistic limit the both particles would have the same speed and direction, which results in divergences signifying that the scattering is ill-defined.
Since here we want to compare with perturbation theory, we will restrict to the case where the massless particles have opposite chirality, taking
\begin{equation}
x_p\sim+1, \quad x_q\sim-1, \qquad\text{\ie}\qquad\mom{q}<0<\mom{p}.
\end{equation}

The uniform light-cone gauge was not fixed in the same way in the perturbative calculation of~\cite{Sundin:2016gqe} as in the all-loop S matrix of~\cite{Borsato:2014exa,Borsato:2014hja}. In order to write the exact S matrix in a general $a$-gauge the S matrix should be multiplied by an additional overall phase $\sigma_{\text{gauge}}^{-a}$, where~\cite{Arutyunov:2005hd,Arutyunov:2006gs}
\begin{equation}
  \sigma_{\text{gauge}}(p,q) = \exp\bigl( i( p E_q - q E_p) \bigr) = 1 + \tfrac{i}{h} ( \mom{p} \omega_{\mom{q}} - \mom{q} \omega_{\mom{p}} ) + \order(1/h^3) ,
\end{equation}
where $E_p$ is the all-loop dispersion relation. If we assume that $q < 0 < p$ we can simplify this for various values of the masses
\begin{equation}
  \begin{aligned}
    \sigma_{\text{gauge}}^{\bullet\bullet} &= 1 + \tfrac{i}{h} (\mom{p} \sqrt{1+\mom{q}^2} - \mom{q} \sqrt{1+\mom{p}^2} ) + \order(1/h^3), \\
    \sigma_{\text{gauge}}^{\circ\bullet} &= 1 - \tfrac{i}{h} (\mom{p} \mom{q} + \mom{q} \sqrt{1+\mom{p}^2} ) + \order(1/h^3) , \\
    \sigma_{\text{gauge}}^{\circ\circ} &= 1 - \tfrac{i}{h} 2 \mom{p} \mom{q} ) + \order(1/h^3) .
  \end{aligned}
\end{equation}

We also need to expand the dressing phases to one-loop order. Recall that the massless and mixed-mass phases contained an AFS order and an HL term. We start by expanding the AFS phase (\cf equation~\eqref{eq:afs-summed}) for general values of the masses
\begin{equation}
  \log\sigma_{\text{AFS}}^2 = 
  \frac{i}{2h} \biggl( \frac{(m_{\mom{p}} \mom{p} - m_{\mom{q}} \mom{q})^2}{\mom{q}\omega_{\mom{p}} - \mom{p}\omega_{\mom{q}}} + (\mom{q}\omega_{\mom{p}} - \mom{p}\omega_{\mom{q}}) + 2(m_{\mom{q}}\mom{p}-m_{\mom{p}}\mom{q}) \biggr) + \order(1/h^3) .
\end{equation} 
We can easily take the masses to $1$ or $0$ in the above expression%
\footnote{This is true here due to the simple form of the AFS phase. In general, one can expect an order-of-limits issue when taking $h\to\infty$ and $m\to0$, \cf the next section.}
\begin{equation}
  \begin{aligned}
    \log(\sigma_{\text{AFS}}^{\bullet\bullet})^2 &= 
    \frac{i}{2h} \biggl( \frac{(\mom{p} - \mom{q})^2}{\mom{q}\omega_{\mom{p}} - \mom{p}\omega_{\mom{q}}} + (\mom{q}\omega_{\mom{p}} - \mom{p}\omega_{\mom{q}}) + 2(\mom{p}-\mom{q}) \biggr) + \order(1/h^3),
    \\
    \log(\sigma_{\text{AFS}}^{\bullet\circ})^2 &= 
    \frac{i}{2h} \biggl( \frac{\mom{q}}{\omega_{\mom{p}} + \mom{p}} + (\mom{q}\omega_{\mom{p}} - \mom{p}\omega_{\mom{q}}) - 2\mom{q} \biggr) + \order(1/h^3) ,
    \\
    \log(\sigma_{\text{AFS}}^{\circ\circ})^2 &= 
    \frac{i}{2h} \biggl( \mom{q}\omega_{\mom{p}} - \mom{p}\omega_{\mom{q}} \biggr) + \order(1/h^3).
  \end{aligned}
\end{equation}

The one-loop terms get corrections from the HL part of the phases. For the purpose of computing this expansion we can equivalently work in terms of the integral~\eqref{eq:hl-massless} or the dilogarithm expression. Before proceeding, it is worth remarking an order-of-limits issue. Here we want to first take the mass of one or both particles to vanish, which imposes the $x^+_p x^-_p=1$, and only afterwards take the momenta to be small as well. Proceeding in the opposite order would result in ambiguities akin to IR divergences when sending $m\to0$, see also~\cite{Sundin:2015uva}. With this in mind, we find that the HL phase~\eqref{eq:hl-generic} in the massless-massless kinematics is
\begin{equation}
\label{eq:phase-massless-1loop}
\theta^{\circ\circ}_{\text{HL}}=
\theta^{\text{HL}}(\mom{p},\mom{q})\Big|_{m_p=m_q=0}  = -\frac{1}{8\pi h^2}\,\mom{p}\mom{q}\,\Big(1-\log \frac{-\mom{p}\mom{q}}{16h^2}\Big)+\order(1/h^4),
\end{equation}
and in the massless-massive kinematics is
\begin{equation}
\label{eq:phase-mixed-1loop}
\theta^{\bullet\circ}_{\text{HL}}=
\theta^{\text{HL}}(\mom{p},\mom{q})\Big|_{m_p=1,m_q=0} =
\frac{1}{4\pi h^2}\,\frac{\mom{p}^2\mom{q}}{\mom{p}+\omega_\mom{p}}
\log \Big[\frac{-\mom{q}}{4h(\mom{p}+\omega_\mom{p})}\Big]+\order(1/h^4).
\end{equation}
It may appear bizarre to have a perturbative expansion involving \textit{logarithms} of the coupling constant. However, here this simply follows from homogeneity requirements. The massless Zhukovski variable $x_p$ is only a function of the momentum~\eqref{eq:massless-zhukovski}, and does not involve any scale. Therefore, the only scale comes from introducing the coupling in~\eqref{eq:momrescaling} and the expansion of the dressing factors in the massless kinematics will depend on $\mom{p}/h$.

There are many processes in the full S matrix, and we will not write them all down here. For the scattering between the highest weight state in each representation we find, at one loop
\begin{equation}
\label{eq:prediction-scattering}
  \begin{aligned}
    S\ket{Y^{\smallL} Y^{\smallL}} &=
    +\bigl( 1 - \gamma_1 + \tfrac{1}{2} \gamma_1^2 - 2i\theta^{\bullet\bullet}_{\text{HL}} \bigr) \ket{Y^{\smallL} Y^{\smallL}} , 
    \\
    S\ket{Y^{\smallL} Z^{\smallR}} &= 
    +\bigl( 1 - \gamma_2 + \tfrac{1}{2} \gamma_2^2 - 2i\tilde{\theta}^{\bullet\bullet}_{\text{HL}} \bigr) \ket{Z^{\smallR} Y^{\smallL}} , 
    \\
    S\ket{Y^{\smallL} \chi^{a}} &= 
    +\bigl( 1 - \gamma_3 + \tfrac{1}{2} \gamma_3^2 - 2i\theta^{\bullet\circ}_{\text{HL}} \bigr) \ket{\chi^a Y^{\smallL}} ,
    \\
    S\ket{Z^{\smallR} \chi^{a}} &= 
    +\bigl( 1 - \gamma_4 + \tfrac{1}{2} \gamma_4^2 - 2i\theta^{\bullet\circ}_{\text{HL}}\bigr) \ket{\chi^a Z^{\smallR}} ,
    \\
    S\ket{\chi^a \chi^b} &=
    -\bigl(1 - \gamma_5 + \tfrac{1}{2} \gamma_5^2 - 2i\theta^{\circ\circ}_{\text{HL}}\bigr) \ket{\chi^b \chi^a} ,
  \end{aligned}
\end{equation}
where the last term on each line gives the contribution from the one-loop dressing phase,%
\footnote{%
  The one-loop expansions of the massive dressing factors can be found in~\cite{Borsato:2013hoa}.
}
 and
\begin{equation}
  \begin{aligned}
    \gamma_1 &= -\frac{i}{2h} \frac{\mom{p}+\mom{q}}{\mom{p}-\mom{q}} ( \mom{p}\omega_{\mom{q}} + \mom{q} \omega_{\mom{p}} ) + \frac{i(a-\tfrac{1}{2})}{h} ( \mom{p}\omega_{\mom{q}} - \mom{q} \omega_{\mom{p}} ) + \order(1/h^3), \\
    \gamma_2 &= +\frac{i}{2h} ( \mom{p}\omega_{\mom{q}} + \mom{q} \omega_{\mom{p}} )  + \frac{i(a-\tfrac{1}{2})}{h} ( \mom{p}\omega_{\mom{q}} - \mom{q} \omega_{\mom{p}} ) + \order(1/h^3), \\
    \gamma_3 &= +\frac{i}{2h} (\mom{p}+\mom{q}) ( \omega_\mom{p} - \mom{p}) + \frac{i(a-\tfrac{1}{2})}{h} ( \mom{p}\omega_{\mom{q}} - \mom{q} \omega_{\mom{p}} ) + \order(1/h^3), \\
    \gamma_4 &= -\frac{i}{2h} (\mom{p}+\mom{q}) ( \omega_\mom{p} - \mom{p}) + \frac{i(a-\tfrac{1}{2})}{h} ( \mom{p}\omega_{\mom{q}} - \mom{q} \omega_{\mom{p}} ) + \order(1/h^3), \\
    \gamma_5 &= + \frac{i(a-\tfrac{1}{2})}{h} ( \mom{p}\omega_{\mom{q}} - \mom{q} \omega_{\mom{p}} ) + \order(1/h^3) .
  \end{aligned}
\end{equation}
The expansion of the rational part of the all-loop S matrix in the gauge $a=1/2$ exactly agrees with all the perturbative results of~\cite{Sundin:2016gqe}.

\subsection{Comparison with perturbative results}
Sundin and Wulff have recently computed worldsheet S~matrix for $\AdS_3\times\Sphere^3\times\Torus^4$ strings at one loop, and obtained the imaginary part of the massless-massless scattering at two loops~\cite{Sundin:2016gqe}. It is straightforward to check that the matrix part of the S~matrix, given by ratios of S-matrix elements, coincides with the result found by symmetry arguments~\cite{Borsato:2014exa,Borsato:2014hja}. In particular, we see, as anticipated above, that the $\algSU(2)_{\circ}$ S~matrix trivializes. Recall that~\cite{Borsato:2014exa,Borsato:2014hja}
\begin{equation}
  \Smat_{\algSU(2)}(p,q)=\left(1-\frac{1}{1+i(w_p-w_q)}\right)\mathbf{1}+\frac{1}{1+i(w_p-w_q)}\Pi.
\end{equation}
Up to one loop in perturbation theory, we find that the coefficient of the permutation matrix $\Pi$ vanishes and the S-matrix trivializes. Note also that this together with the crossing equation~\eqref{eq:crossing} forces $w_p$ to be very singular at small momentum,%
\footnote{It is easy to see that $w_p$ should go at least like $w_p\sim h^4/\mom{p}^4$ in the near-BMN limit.}
something with no clear interpretation in perturbation theory. Therefore we will set $w_p = \infty$, trivialising the $\algSU(2)_{\circ}$ S matrix.

As for the dressing factors, it is straightforward to check that they perfectly fit the prediction~\eqref{eq:prediction-scattering} at tree level. At one loop, the expressions found by Sundin and Wulff are
\begin{equation}
\theta_{\text{SW}}^{\circ\circ}(\mom{p},\mom{q})=
-\frac{1}{8\pi h^2} \mom{pq}\Big(1-\log(-4\mom{p}\mom{q})\Big)
+\order(g^{-3}),
\end{equation}
and
\begin{equation}
\label{eq:SW-mixed}
\theta_{\text{SW}}^{\bullet\circ}(\mom{p},\mom{q})=
-\frac{1}{4\pi h^2}\frac{\mom{p}^2\mom{q}}{\mom{p}+\omega_{\mom{p}}}\Big(1-\log\frac{-2\mom{q}}{\omega_{\mom{p}}-\mom{p}}\Big)+\order(g^{-3}).
\end{equation}
These should be compared to equations \eqref{eq:phase-massless-1loop}--\eqref{eq:phase-mixed-1loop}. It is clear that there are several differences between the two sets of formul\ae. Let us start from the massless phase. We find that the discrepacy is 
\begin{equation}
\theta_{\text{SW}}^{\circ\circ}(\mom{p},\mom{q})-\theta_{\text{HL}}^{\circ\circ}(\mom{p},\mom{q})=
\frac{1}{8\pi h^2}\,\mom{pq}\,\log \frac{1}{64h^2}.
\end{equation}
The mismatch takes the form of a scale in the logarithm, and can be seen as arising from the ambiguity in regularising an infra-red divergence~\cite{Sundin:2016gqe}. Note also that this term is a solution of the homogeneous crossing equation, and hence hard to rule out on symmetry grounds. Similar ambiguities may justify the discrepancy in the mixed-mass phase. In that case, it should also be noted that the dependence on the light-cone momentum $\mom{p}_{\pm}=\omega_{\mom{p}} \pm \mom{p}$ in the logarithm is also different; in fact, using $\mom{p}_+\mom{p}_-=1$ the powers of $\mom{p}_+$ appearing in $\theta_{\text{HL}}^{\bullet\circ}$ and $\theta_{\text{SW}}^{\bullet\circ}$ are different. Since this discrepancy occurs in a non-rational term, it will in principle affect the crossing equations. However, the two expressions are related by picking opposite conventions for the path of the crossing transformation.

At two loops, Sundin and Wulff predict that, in the massless-massless sector, the dressing factor should vanish for their particular choice of $a$-gauge. While as discussed we also do not expect any novel contribution to the phase at this order, we would get a sub-leading rational term from the expansion of the AFS phase at next-to-leading order. On the one hand, one may argue that this discrepancy can be removed by picking a different CDD factor,%
\footnote{%
One way to pick a different solution  of the homogeneous crossing equation is to replace the AFS phase with a term proportional to the ``gauge'' phase, $\theta_{\text{AFS}}^{\circ\circ}\to-\frac{1}{2}\theta_{\text{gauge}}^{\circ\circ}$.
} as it is anyway a solution of homogeneous crossing. On the other hand, we know that at this order perturbation theory and symmetry arguments predict a different dispersion relation, so that a mismatch is in a sense not unexpected. At this stage it is not clear whether one should add a (somewhat artificial) CDD factor, or whether the issue will be properly accounted for by better understanding the near-BMN limit of massless modes.

It therefore appears that there are several ambiguities in giving an interpretation of our results in the near-BMN limit: the presence of $\log h$ terms is mirrored in perturbation theory by ambiguities related to IR divergences. This can also be seen as an order-of-limits issue. One may take the massive-massive HL phase in the near-BMN limit, and send one or both masses to zero~\cite{Sundin:2016gqe}, or send $m_p\to0$ in~\eqref{eq:SW-mixed}. Then, even if qualitatively one recovers these results, singularities of the form $\log m$ and $1/m$ emerge. It would be very interesting to understand better how to remove these IR ambiguities and more precisely relate perturbative calculations to a large-$h$ expansion of the all-loop phase.

\section{Yangian symmetry of the massless S-matrix}
\label{sec:yangian}

In this section we discuss the Yangian symmetry underlying the massless-sector S-matrix, focusing on the massless-massless scattering. Large part of the structure which we will uncover descends from the massive Yangian~\cite{Borsato:2013qpa,Regelskis:2015xxa} in the natural limit. Nevertheless, the limit itself is rather subtle, and we find it useful to confirm the algebraic construction starting from scratch in the strict massless representation. Moreover, we find it beneficial to reproduce the salient details for the convenience of the reader, which altogether justifies the following separate treatment. We shall then refer to Appendix~\ref{app:Zamolo} for the implementation in the S-matrix crossing-unitarity problem.  

\subsection{Yangian generators, evaluation representation and crossing}

We shall work in this section with the following symmetry generators on the elementary \emph{massless} excitations, forming the centrally-extended algebra associated to $\algSU(1|1)_L \oplus \algSU(1|1)_R $. The non-vanishing (anti-)commutation relations read
\begin{equation}
  \label{eq:alge}
  \{\gen{Q}_{\smallL}, \overline{\gen{Q}}_{\smallL}\} = \gen{H}_{\smallL}, \quad 
  \{\gen{Q}_{\smallR}, \overline{\gen{Q}}_{\smallR}\} = \gen{H}_{\smallR}, \quad 
  \{\gen{Q}_{\smallL}, \gen{Q}_{\smallR}\} = \gen{C}, \quad  
  \{\overline{\gen{Q}}_{\smallL}, \overline{\gen{Q}}_{\smallR}\} = \overline{\gen{C}}. 
\end{equation}
The representation we consider is given by the following action:
\begin{equation}\label{eq:leftrep}
  \begin{aligned}
    \overline{\gen{Q}}_{\smallR} &:= - \sqrt{h \sin \frac{p}{2}}\begin{pmatrix}0&0\\1&0\end{pmatrix}, &
    \gen{Q}_{\smallR} &:= - \sqrt{h \sin \frac{p}{2}}\begin{pmatrix}0&1\\0&0\end{pmatrix}, \\
    \gen{Q}_{\smallL} &:= \phantom{+}\sqrt{h \sin \frac{p}{2}}\begin{pmatrix}0&0\\1&0\end{pmatrix}, &
    \overline{\gen{Q}}_{\smallL} &:= \phantom{+}\sqrt{h \sin \frac{p}{2}}\begin{pmatrix}0&1\\0&0\end{pmatrix}, 
  \end{aligned}
\end{equation}
with central-charge eigenvalues\footnote{We shall concentrate on the \emph{left} representation and on the positive-momentum worldsheet-movers for this algebraic discussion, as it will be sufficient for the purposes of illustration.}  
\begin{equation}\label{eq:defofex}
  H_{\smallL} = H_{\smallR} = - C = - \bar{C} = h \, \left|\sin \frac{p}{2}\right|.
\end{equation}
The coupling constant $h$ is taken to be real and positive. The coproduct we equip the generators with is given by
\begin{equation}
  \label{eq:coprod}
  \begin{aligned}
    \Delta(\gen{H}_{\smallL}) &:= \gen{H}_{\smallL} \otimes \matId + \matId \otimes \gen{H}_{\smallL}, \quad &
    \Delta(\gen{Q}_{\smallL}) &:= \gen{Q}_{\smallL} \otimes e^{-i\frac{p}{4}} + e^{i\frac{p}{4}} \otimes \gen{Q}_{\smallL}, \\
    \Delta(\gen{H}_{\smallR}) &:= \gen{H}_{\smallR} \otimes \matId + \matId \otimes \gen{H}_{\smallR}, \quad &
    \Delta(\overline{\gen{Q}}_{\smallL}) &:= \overline{\gen{Q}}_{\smallL} \otimes e^{i\frac{p}{4}} + e^{- i\frac{p}{4}} \otimes \overline{\gen{Q}}_{\smallL},\\
    \Delta(\gen{C}) &:= \gen{C} \otimes e^{- i\frac{p}{2}} + e^{i\frac{p}{2}} \otimes \gen{C}, \quad &
    \Delta(\gen{Q}_{\smallR}) &:= \gen{Q}_{\smallR} \otimes e^{-i\frac{p}{4}} + e^{i\frac{p}{4}} \otimes \gen{Q}_{\smallR}, \\
    \Delta(\overline{\gen{C}}) &:= \overline{\gen{C}} \otimes e^{i\frac{p}{2}} + e^{- i\frac{p}{2}} \otimes \overline{\gen{C}}, \quad &
    \Delta(\overline{\gen{Q}}_{\smallR}) &:= \overline{\gen{Q}}_{\smallR} \otimes e^{i\frac{p}{4}} + e^{- i\frac{p}{4}} \otimes \overline{\gen{Q}}_{\smallR},
  \end{aligned}
\end{equation}
where $p$ is the momentum, which is a central element in $\algSU(1|1)_L \oplus \algSU(1|1)_R $ with coproduct $\Delta(p) = p \otimes \matId + \matId \otimes p$.  We will use the notation $p_1$ ($p_2$) for the momentum in the first (second) factor of the tensor product.
The $R$-matrix (stripped-off of a scalar factor) is obtained as a limit from the massive one as follows:
\begin{align}\label{eq:RLL}
  R \ket{\phi} \otimes \ket{\phi} &=+\ket{\phi} \otimes \ket{\phi},\nonumber \\
  R \ket{\phi} \otimes \ket{\psi} &=-\csc \frac{p_1 + p_2}{4} \, \sin \frac{p_1 - p_2}{4} \ket{\phi} \otimes \ket{\psi} + {{}} \, \csc \frac{p_1 + p_2}{4} \, \sqrt{\sin \frac{p_1}{2} \sin \frac{p_2}{2}} \ket{\psi} \otimes \ket{\phi}, \nonumber \\
  R \ket{\psi} \otimes \ket{\phi} &=+\csc \frac{p_1 + p_2}{4} \, \sin \frac{p_1 - p_2}{4}  \ket{\psi} \otimes \ket{\phi} + {{}} \, \csc \frac{p_1 + p_2}{4} \, \sqrt{\sin \frac{p_1}{2} \sin \frac{p_2}{2}} \ket{\phi} \otimes \ket{\psi}, \nonumber \\
  R \ket{\psi} \otimes \ket{\psi} &=-\ket{\psi} \otimes \ket{\psi},
\end{align}
and it satisfies 

\begin{equation}\label{eq:definiLL}
  \Delta^{\text{op}} (\gen{a})\,  R = R\, \Delta (\gen{a}) \qquad\forall \, \, \gen{a} \in \algSU(1|1)_{\smallL}\oplus \algSU(1|1)_{\smallR},
\end{equation}
where $\Delta^{op} = \Pi \circ \Delta$, and $\Pi$ is the graded permutation.

Crossing symmetry is implemented as follows. By imposing
\begin{equation}
  \label{eq:axiom}
  \mu (\Sigma \otimes \matId) \Delta = \eta \, \epsilon
\end{equation}
where $\mu$ is the multiplication in the algebra, and $\eta$ and $\epsilon$ are the unit and counit,\footnote{
  The co-unit map $\epsilon$ takes values in $\Complex$ and gives $0$ for the generators of $\algSU(1|1)_L \oplus \algSU(1|1)_R$ and $1$ for the identity element of the algebra. The unit $\eta$ is a map from $\Complex$ to the algebra that sends $1$ to the identity element of the algebra.
}
respectively, 
one can check from (\ref{eq:coprod}) that the antipode on all the Lie superalgebra generators is simply 
\begin{equation}
  \Sigma(\gen{J}) = - \gen{J},
\end{equation}
from which one can verify the crossing relation
\begin{equation}
  \label{eq:relo}
  \Sigma(\gen{J}(p)) = C^{-1} \, \underline{\gen{J}}(-p)^{str} \, C
\end{equation}
with
\begin{equation}
  C = \begin{pmatrix}1&0\\0&i\end{pmatrix}
\end{equation}
the charge-conjugation matrix, 
\begin{equation}
  \label{eq:crossa}
  \underline{\gen{Q}}_{\smallL}=\gen{Q}_{\smallR}, \qquad \underline{\overline{\gen{Q}}}_{\smallL}=\overline{\gen{Q}}_{\smallR}
\end{equation}
and ${}^{str}$ denoting supertransposition. Although the L and R charges form isomorphic representations, (\ref{eq:crossa}) motivates us to still formally distinguish them from the point of view of crossing.

Let us focus on the left-sector generators for the remainder of this section (the right sector being effectively isomorphic to the left one). The R-matrix~\eqref{eq:RLL} is invariant under the following Yangian \emph{level-one} supercharges:
\begin{align}
  \label{eq:compl}
  \Delta(\gen{Q}_{\smallL}^{(1)})  &:= \gen{Q}^{(1)}_{\smallL} \otimes e^{-i\frac{p}{4}} + e^{i\frac{p}{4}} \otimes \gen{Q}^{(1)}_{\smallL} +  e^{i\frac{p}{4}} \gen{H}_{\smallL} \otimes \gen{Q}_{\smallL},\nonumber \\
  \Delta(\overline{\gen{Q}}_{\smallL}^{(1)}) &:= \overline{\gen{Q}}^{(1)}_{\smallL} \otimes e^{i\frac{p}{4}} + e^{- i\frac{p}{4}} \otimes \overline{\gen{Q}}^{(1)}_{\smallL}+\overline{\gen{Q}}_{\smallL} \otimes e^{i\frac{p}{4}} \gen{H}_{\smallL},
\end{align}
with the \emph{evaluation representation} understood as
\begin{equation}
  \label{eq:eval}
  \gen{Q}_{\smallL}^{(1)} = u_{\smallL} \, \gen{Q}_{\smallL}, \qquad \overline{\gen{Q}}_{\smallL}^{(1)} = u_{\smallL} \, \overline{\gen{Q}}_{\smallL}, \qquad u_{\smallL} = \frac{h}{2 i} \, e^{i\frac{p}{2}}.
\end{equation}
Commuting these charges with the Lie superalgebra (\emph{level-zero}) supercharges, we can generate the level-one central charges, which are automatically symmetries of the R-matrix. In this matrix representation, all these level-one charges will then be provided by multiplying the level-zero ones by the evaluation parameter $u_{\smallL}$. On the other hand the coproduct will be rather more complicated following equation~\eqref{eq:compl}. 

If we supplement this algebra with the hypercharge-generator $\Ygen{B} = (-)^F$, $F$ being the fermion-number generator, we see that the entire symmetry algebra contains the Yangian $Y(\algGL(1|1))$: in the so-called Drinfeld's second realization~\cite{Drinfeld:1987sy}, this reads
\begin{equation}
  \label{eq:Lie}
  \begin{gathered}
    \comm{\Ygen{B}_0}{\Ygen{e}_n} = -2\Ygen{e}_n , \qquad
    \comm{\Ygen{B}_0}{\Ygen{f}_n} = 2\Ygen{f}_n, \qquad
    \acomm{\Ygen{e}_m}{\Ygen{f}_n} = \Ygen{k}_{m+n} , \\
    \comm{\Ygen{B}_m}{\Ygen{h}_n} = 
    \comm{\Ygen{B}_m }{\Ygen{k}_n} =
    \comm{\Ygen{k}_m }{\Ygen{k}_n} = 
    \comm{\Ygen{k}_m }{\Ygen{e}_n} =
    \comm{\Ygen{k}_m }{\Ygen{f}_n} =
    \acomm{\Ygen{e}_m}{\Ygen{e}_n} = 
    \acomm{\Ygen{f}_m}{\Ygen{f}_n} = 0 , \\
    \comm{\Ygen{B}_{m+1}}{\Ygen{e}_n} - \comm{\Ygen{B}_m}{\Ygen{e}_{n+1}} + \acomm{\Ygen{B}_m}{\Ygen{e}_n} = 0, \qquad
    \comm{\Ygen{B}_{m+1}}{\Ygen{f}_n} - \comm{\Ygen{B}_m}{\Ygen{f}_{n+1}} - \acomm{\Ygen{B}_m}{\Ygen{f}_n} = 0.
  \end{gathered}
\end{equation}
The following representation satisfies all the relations~\eqref{eq:Lie}
\begin{equation}
  \Ygen{e}_n = u_{\smallL}^n \, \gen{Q}_{\smallL}, \qquad  
  \Ygen{f}_n = u_{\smallL}^n \, \overline{\gen{Q}}_{\smallL}, \qquad
  \Ygen{k}_n = u_{\smallL}^n \, \gen{H}_{\smallL}, \qquad 
  \Ygen{B}_n = u_{\smallL}^n (-)^F.
\end{equation}
One can check that the R-matrix is also invariant under~\cite{Pittelli:2014ria}
\begin{align}
  \label{eq:complb}
  \Delta(B_0) &= B_0 \otimes \matId + \matId \otimes B_0 \nonumber \\
  \Delta(B_1) &= B_1 \otimes \matId + \matId \otimes B_1 + 2 \, e^{i\frac{p}{4}} \overline{\gen{Q}}_{\smallL} \otimes e^{i\frac{p}{4}} \gen{Q}_{\smallL}.
\end{align}
By a simple map, it is possible to obtain a new level-one hypercharge generator,
\begin{equation}
  \gen{b}_1 = B_1 + \frac{1}{2} \gen{Q}_{\smallL} \overline{\gen{Q}}_{\smallL} - \frac{1}{2} \overline{\gen{Q}}_{\smallL} \gen{Q}_{\smallL}, 
\end{equation}
with a more symmetric form for its coproduct\cite{Pittelli:2014ria}
\begin{align}
  \label{eq:complbb}
  \Delta(\gen{b}_1)  := \gen{b}_1 \otimes \matId + \matId \otimes \gen{b}_1 +  e^{- i\frac{p}{4}} \gen{Q}_{\smallL} \otimes e^{- i\frac{p}{4}} \overline{\gen{Q}}_{\smallL} +  e^{i\frac{p}{4}} \overline{\gen{Q}}_{\smallL} \otimes e^{i\frac{p}{4}} \gen{Q}_{\smallL},
\end{align}
where
\begin{equation}
  \gen{b}_1 = - \frac{i h}{2} \cos \frac{p}{2} \, (-)^F.
\end{equation}

Let us now show how crossing symmetry extends to the Yangian. It is straighforward to show, by applying (\ref{eq:axiom})
this time with $\epsilon (\gen{J}^{(1)})=0$, that 
\begin{equation}
  \Sigma(\gen{Q}_{\smallL}^{(1)}) = - \gen{Q}_{\smallL}^{(1)} + \gen{H}_{\smallL} \, \gen{Q}_{\smallL}, \qquad \Sigma(\overline{\gen{Q}}_{\smallL}^{(1)}) = - \overline{\gen{Q}}_{\smallL}^{(1)} + \gen{H}_{\smallL} \, \overline{\gen{Q}}_{\smallL}.
\end{equation}
It is then an easy exercise to verify that, on the positive branch of the dispersion relation, the following crossing relations hold also at level one:
\begin{equation}
  \label{eq:combo}
  \Sigma(\gen{Q}_{\smallL}^{(1)}(p)) = C^{-1} \, \underline{\gen{Q}}_{\smallL}^{(1)}(-p)^{str} C, \qquad \Sigma(\overline{\gen{Q}}_{\smallL}^{(1)}(p)) = C^{-1} \, \underline{\overline{\gen{Q}}}_{\smallL}^{(1)}(-p)^{str} C, 
\end{equation}
where
\begin{equation}
  \underline{\gen{Q}}_{\smallL}^{(1)}=u_{\smallL} \, \gen{Q}_{\smallR}, \qquad \underline{\overline{\gen{Q}}}_{\smallL}^{(1)}=u_{\smallL} \, \overline{\gen{Q}}_{\smallR}.
\end{equation}
In fact, since the relation (\ref{eq:relo}) holds and the evaluation representation connects the level-one generators with the level-zero in the very simple fashion (\ref{eq:eval}), crossing symmetry combined with (\ref{eq:combo}) just amounts to the following relation:
\begin{equation}
  u_{\smallL}(-p) = u_{\smallL}(p) - H_{\smallL},
\end{equation}
which follows from equation~(\ref{eq:defofex}). On the other hand, in order for the generators $B_0$ and $B_1$ to beconsistent with crossing symmetry one needs to choose
\begin{equation}
  \underline{B}_0 = - B_0, \qquad \underline{\gen{b}}_1 = - \gen{b}_1 + H_{\smallL}.
\end{equation} 

This then extends, simply by the homomorphism property, to the whole Yangian, and it implies the crossing symmetry for the R-matrix:
\begin{equation}
  (C^{-1} \otimes \matId) \, R_{\smallR,\smallL}(-p_1,p_2)^{str_1} (C \otimes \matId) \, R_{\smallL,\smallL}(p_1,p_2) = \matId \otimes \matId.
\end{equation}

\subsection{Determinantal identites and Yangian centre}

Finally, we would like to point out some curious features of the massless S-matrix, related to determinants and the centre of the Yangian agebra.

Let us provide an explanation of how to adapt the RTT formulation of Yangians to the present case. The massless R-matrix $R$, which we have shown to be invariant under the Yangian $Y$, can be written as a matrix depending on two \emph{spectral parameters} $(p_1, p_2)$:
\begin{equation}
\label{erre}
R = R_{ijkl}(p_1,p_2) E_{ij} \otimes E_{kl}.
\end{equation}
We further introduce a set of Yangian elements (\emph{generating functions}) $T_{ij}(p)$ such that
\begin{equation}
T_{ij}(p) \in Y, \qquad p \in \Complex,
\end{equation}
and define
\begin{equation}
\label{spaces}
\mathcal T(p):= E_{ij}\otimes T_{ij}(p),
\end{equation}
where repeated indices are summed over. We will define the Yangian in the RTT presentation to be the algebra generated by the elements $T_{ij}(p)$, subject to the RTT relations:
\begin{equation}
\label{above}
R_{12}(p_1,p_2)\mathcal T_{13}(p_1)\mathcal T_{23}(p_2)=\mathcal T_{23}(p_2)\mathcal T_{13}(p_1)R_{12}(p_1,p_2).
\end{equation}
The subscripts appearing in $\mathcal T$ and $R$ in (\ref{above}) indicate now the spaces the respective tensor products acts upon, cf. (\ref{spaces}) and (\ref{erre}). The Laurent coefficients $T_{ij}^n$, obtained by expanding $T_{ij}(p)$ around a suitable point $p^{(0)}$ (such that the expansion is well-defined), namely
\begin{equation}
T_{ij}(p):=\sum_n T_{ij}^n \Big(p - p^{(0)}\Big)^n,
\end{equation}
are supposed to generate all the levels of the Yangian. In turn, the RTT relations should reproduce, by the very same Laurent expansion, all the relations at each Yangian level, and one expects to be able to recast them, after highly non-trivial manipulations, into the relations of Drinfeld's second realisation of the Yangian.

One then uses the fact that the R-matrix itself gives a representation of the element $\mathcal T$. This is because, by construction, the R-matrix satisfies the RTT relations, since they reduce to the Yang-Baxter equation in this case. This means that we can define
\begin{equation}
\mathcal{T}(p_1) = E_{ij}\otimes T_{ij}(p_1) = (\pi_1 \otimes \mathbbmss{1}) \, \mathcal{R}
\end{equation}
where $\mathcal{R}$ is the universal R-matrix\footnote{We will assume its existence for the purposes of the arguments in this section, although a mathematical proof of this statement is still lacking for $AdS/CFT$ (see however the recent progress obtained in~\cite{Beisert:2016qei} in the context of $q$-deformations of the $AdS_5$ integrable system).}, and $\pi_i \equiv \pi_{p_i}$ projects abstract algebra elements into the representation characterised by the spectral parameter $p_i$. Therefore, we obtain
\begin{equation}
E_{ij}\otimes \pi_2 \Big[T_{ij}(p_1)\Big]  =  (\pi_1 \otimes \pi_2) \, \mathcal{R} = R(p_1,p_2),
\end{equation}
which is the matrix we started with. This means that, knowing the R-matrix, we can construct a representation of the Yangian generating functions, where one of the spectral parameters is the generating parameter, while the other one is the representation parameter. Notice that we could have of course repeated the discussion swapping the role of $p_1$ and $p_2$ in the discussion, and interchanging the spaces we were projecting onto. Straightforward comparison produces
\begin{equation}
R_{ijkl} (p_1,p_2) E_{kl} = \pi_2 \Big[T_{ij}(p_1)\Big].
\end{equation} 
The elements $T_{ij}^n$ are supposed to generate the same Yangian we have described in the previous subsection, this time in the so-called RTT presentation~\cite{Faddeev:1987ih,Beisert:2014hya}.

We now calculate the following element of the Yangian\footnote{This discussion is partly inspired by~\cite{Nazarov:1991a}, where bosonic $\algGL$-type algebras and central element of the corresponding Yangians were studied.}
\begin{equation}
  Z(p_1) = \tr \otimes \tr \, T(p_1) \otimes T^\dagger(p_1), \qquad T(p_1) = T_{ij}(p_1) E_{ij}, 
\end{equation}
where the ${}^\dagger$ involves super-transposition. $T(p)$ is now a matrix with entries taking values in $Y$. We can verify that, in the representation $\pi_2$, this corresponds to
\begin{equation}
  \label{eq:RE11-RE22}
  \Big| R_{1111}(p_1,p_2) + R_{2211}(p_1,p_2) \Big|^2 \, E_{11}+ \Big| R_{2222}(p_1,p_2) + R_{1122}(p_1,p_2) \Big|^2 \, E_{22} ,
\end{equation} 
which is central (\ie, proportional to the identity matrix $\matId=E_{11} + E_{22}$) by direct calculation, since \begin{equation}
  R_{1111}(p_1,p_2) + R_{2211}(p_1,p_2) = -\Big(R_{2222}(p_1,p_2) + R_{1122}(p_1,p_2)\Big).
\end{equation} 

Moreover, we notice that a determinantal identity, which was introduced in~\cite{Hoare:2014kma}, holds here as well. If we focus on the submatrices formed by 
\begin{equation}
  \Omega_1 = \{|\phi\rangle \otimes |\phi \rangle, |\psi\rangle \otimes |\psi \rangle\}, \qquad \mbox{and} \qquad \Omega_2 = \{|\phi\rangle \otimes |\psi \rangle, |\psi\rangle \otimes |\phi \rangle\},
\end{equation} 
respectively, then the determinants
\begin{equation}
  \det{}_1 = \det{}_2,
\end{equation}
where $det_i$ is the determinant of the sub-matrix of $R$ scattering only states from set $\Omega_i$.
Specifically, by direct evaluation using (\ref{eq:RLL}) one has
\begin{equation}
  \label{eq:det1-det2}
  \det{}_1 = R_{1111} R_{2222} = -1 = R_{1122} R_{2211} - R_{2112} (-R_{1221}) = \det{}_2
\end{equation}
The nice outcome of these identities lies in the following observation. By the general theory of Yangians, one has an expectation for conditions such as the equality of the two determinants in~\eqref{eq:det1-det2} and the centrality of~\eqref{eq:RE11-RE22} to hold. However, in the case centrally-extended superalgebras, checking these explicitly remains a highly non-trivial problem. In the case of $AdS_5$, for example, one needs to fully exploit the RTT formulation for the fundamental magnon S-matrix~\cite{Beisert:2014hya}. On the other hand, as we have shown above, in the case of small rank, we can quite easily demonstrate such relationships. This also gives a straightforward demonstration of how the $AdS_3$ massless sector can be accommodated inside the familiar algebraic framework.

From the algebraic point of view, there is still a challenge in proving the full equivalence of the RTT formulation of various Yangians appearing in AdS/CFT to the other formulations in which they are originally case-by-case discovered. In the $AdS_3$ case of small-rank algebras, the fact that we can prove these identities is a strong encouragement that one might be able to complete the circle and demonstrate the full equivalence with, in this instance, Drinfeld's second realisation. When combined with the recent results of~\cite{Stromwall:2016dyw}, this might shed light in particular on the complete algebraic structure governing the integrability of the massless sector, which seems to remain not only a profound physical occurrence, but also a fascinating representation-theoretical problem.

\section{Conclusions}

The $\AdS_3/\CFT_2$ dual pair with small $\superN=(4,4)$ superconformal 
symmetry is a pivotal example of the gauge/string correspondence. 
It features prominently in a range of key problems including 
black hole entropy counting~\cite{Strominger:1996sh},  instanton moduli space 
invariants~\cite{Dijkgraaf:1998gf},   
holography of two-dimensional super-QCD, and higher-spin theories. 
Its relatively small amount of supersymmetry and large moduli 
space have made it a particularly rich model in which 
holography can be investigated while retaining some control in 
the quantum regime. Recently, it has become apparent that a 
much stronger handle over this duality can be achieved through 
the use of integrability. This opens the possibility of computing 
generic unprotected quantities at all values of the coupling for 
the dual pair in the planar limit. In turn, this 
is likely to have important consequences on 
the set of problems closely linked with the $\AdS_3/\CFT_2$ duality.

An entirely novel feature of $\AdS_3/\CFT_2$  is the presence of 
massless modes. A symmetry-based approach to incorporate these into 
the integrable framework was developed in~\cite
{Borsato:2014exa,Borsato:2014hja,Lloyd:2014bsa,Borsato:2015mma}. 
However, a number of ingredients of the integrable construction 
remained undetermined. The way to address these issues was sketched briefly in~\cite{Borsato:2016kbm}. A more detailed explanation
and derivation of those findings was given in the present paper. 
In particular, we investigated the analytic structure of the 
massless modes and found minimal solutions of the 
crossing equations for the massless and mixed-mass dressing 
factors. We argued that the only natural candidate for a homogeneous 
solution to the crossing equations comes at the AFS-order of 
the massless dressing factor. Equipped with the complete
S matrix we showed that, just as in the relativistic case, no
massless bound states exist.

We derived all-loop Bethe equations for the complete spectrum of 
closed strings on $\AdS_3\times S^3\times \Torus^4$ and showed 
that it has the expected $\algPSU(1,1|2)^2$ global symmetry
as well as the  translation symmetries along the $\Torus^4$ directions. Our results
throughout this paper focus on the zero-winding, zero-momentum sector of the torus
and it would be interesting to extend them to include non-trivial winding and momentum. 

Subsequently, we investigated the weak-coupling 
limit of these equations where a spin-chain description of
the integrable system is expected to emerge~\cite{OhlssonSax:2011ms,Sax:2012jv}.
As will be shown elsewhere~\cite{forthcoming}, the weak-coupling limit
of our Bethe equations does indeed lead to a spin-chain of the type
argued for in~\cite{Sax:2012jv}, with the massless modes
being described by gapless modes and the resulting degeneracy of the
groundstates matching the protected supergravity spectrum
found in~\cite{deBoer:1998kjm}. A possible origin for such
a spin-chain on the $\CFT_2$ side was proposed in~\cite{Sax:2014mea} and
the relation of those results to the present work, particularly to
the massless modes should be explored.

As part of our Bethe equations analysis, we briefly discussed the 
generalisation of these to the mixed R-R and NS-NS flux background,
building on the massive-sector result~\cite{Babichenko:2014yaa} and 
related works~\cite{Hoare:2013pma,Hoare:2013ida,Hoare:2013lja,Lloyd:2014bsa}. 
Solving the mixed-flux crossing equations in this setting remains an 
important open problem.

Given the recent results on the massless L\"uscher terms~\cite{Abbott:2015pps}, 
it is important to understand better the role of finite-size
corrections to these Bethe equations. One may hope that
the construction of a thermodynamic Bethe 
ansatz~\cite{Ambjorn:2005wa,Arutyunov:2007tc,Arutyunov:2009zu,
Gromov:2009tv,Bombardelli:2009ns,Arutyunov:2009ur,Cavaglia:2010nm} and a
quantum spectral curve~\cite{Gromov:2013pga,Cavaglia:2014exa,Gromov:2014caa} 
for $\AdS_3/\CFT_2$ will provide a better understanding of these issues.

Recently, perturbative computations of the S matrix in the near-BMN limit 
were performed in~\cite{Sundin:2016gqe}. We compared our results 
with these calculations, and, for the most part, found agreement. 
In the two cases where our results appear to differ from those 
of~\cite{Sundin:2016gqe}, we proposed likely explanations for 
the origin of the discrepancies. As the interested processes involve IR divergences, it appears subtle to perform the comparison. At one-loop, this can be seen as a dependence on worldsheet IR regulator, or as an order-of-limits issue when expanding the all-loop results with vanishing mass and large coupling. At two loops, discrepancies are perhaps even expected, given that it is known that perturbation theory and symmetry arguments predict different dispersion relations. While some of these mismatches may be resolved by adding an \textit{ad-hoc} CDD factor, it is likely that they depend on subtleties of the massless near-BMN kinematics which are yet to be fully understood and should be further investigated. For example, since the discrepancies appear also in logarithmic terms, which in the massive case are cut-constructible, it would be interesting to generalise the methods of~\cite{Engelund:2013fja,Bianchi:2014rfa} to include massless modes in the asymptotic states.

We have also analysed the Yangian symmetry, 
which provides the algebraic manifestation of integrability 
in the massless sector. We were able to construct the Yangian 
algebra in the so-called Drinfeld second realisation, and 
to provide the associated evaluation representation consistent 
with crossing-symmetry. Furthermore, we have studied the hypercharge 
generator at all Yangian levels, and spelled out the rules for 
its coproduct and charge-conjugation. We finally went on to 
investigate certain determinantal identities and related Yangian 
central elements, taking advantage of the relative simplicity of 
the massless S-matrix entries. Such identities insert the algebraic 
problem into the framework of the RTT formulation of Yangians, 
enriching the arena of integrable techniques the massless sector 
of the $\AdS_3$ string is amenable to.

We expect that the results presented here will generalise to the 
$\AdS_3\times S^3\times S^3\times S^1$ background~\cite
{Gauntlett:1998kc,Elitzur:1998mm,deBoer:1999gea,Gukov:2004ym}, 
whose symmetry
is the \textit{large} $\superN=(4,4)$ supersymmetry algebra~\cite{Sevrin:1988ew}.
This algebra and its zero-mass limit have played
an important role in the higher-spin limit of the gauge/string
correspondence~\cite{Gaberdiel:2014cha,Datta:2016cmw} and relating our 
results to these works is an important
open problem. Perhaps, together with~\cite{Tong:2014yna}, these results might also
shed light on the apparently enigmatic $\CFT_2$ dual.

\section*{Acknowledgements}
We would like to thank Gleb Arutyunov,  Olalla Castro-Alvaredo, Benjamin Doyon, Paul Fendley, Kolya Gromov, Chris Hull, Fedor Levkovich-Maslyuk, Tom Lloyd, Arkady Tseytlin, Kostya Zarembo for interesting discussions. We are particularly grateful to Per Sundin and Linus Wulff for many insightful comments and conversations, and for sharing with us their results.
A.T.~would like to thank M.~Abbott, P.~Dorey, D.~Fioravanti, S.~Negro, F.~Ravanini and R.~Tateo for discussions. R.B.~was supported by the ERC
Advanced grant No.~290456. O.O.S.~was supported by
ERC Advanced grant No.~341222. A.S's research was
partially supported by the NCCR SwissMAP, funded
by the Swiss National Science Foundation. B.S.~acknowledges funding support from an STFC Consolidated
Grant ST/L000482/1. A.T.~thanks
the EPSRC for funding under the First Grant project EP/K014412/1 \textit{Exotic quantum groups,
Lie superalgebras and integrable systems}, and the STFC for support under the Consolidated
Grant project nr.~ST/L000490/1 \textit{Fundamental Implications of Fields, Strings and Gravity}. R.B., O.O.S., B.S.~and A.T.~would like
to thank ETH Zurich and all participants of the workshop \textit{All about AdS3} for providing a stimulating atmosphere where parts of this work were undertaken. R.B.,
A.S., B.S.~and A.T.~would like to thank Nordita for hosting us
during the final stages of this project.

No data beyond those presented and cited in this work are needed to validate this study.

\appendix

\section{Massless scattering in relativistic integrable systems}
\label{app:Zamolo}

In this section, we review the classic treatment of massless scattering in relativistic integrable system, as it was mainly developed in~\cite{Zamolodchikov:1992zr,Fendley:1993wq,Fendley:1993xa} (see also~\cite{Polyakov:1983tt,Polyakov:1984et,Fioravanti:1996rz}). This will include the massive-to-massless limit, considerations of analyticity, a discussion of bound states (or, rather, of the absence thereof) and the adaptation of the notions of crossing symmetry and unitarity. We also include an algebraic digression on the connections with Hopf algebras.

\subsection{Massless limit and (no) bound-states}
\begin{figure}[t]
  \centering
  \includegraphics{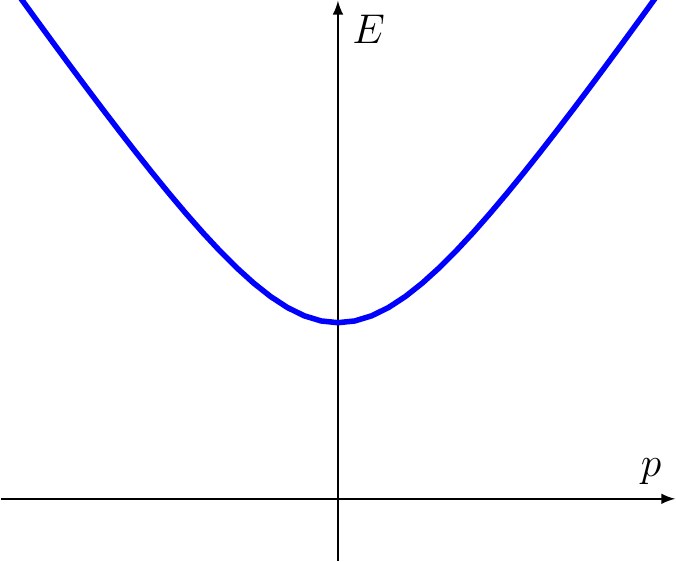}  

\caption{Relativistic dispersion relation (for $m=1$ in appropriate units).}
\label{fig:crossing-massive-rel}
\end{figure}

Let us start by recalling what happens in the relativistic case, when Zamolodchikov's massless scaling limit~\cite{Zamolodchikov:1992zr} is taken. One begins with a massive scattering, characterised by a dispersion relation
\begin{equation}
E = m \cosh \theta, \qquad p = m \sinh \theta, \qquad E^2 - p^2 = m^2,
\end{equation}
with $\theta \in (-\infty, \infty)$ a \emph{rapidity} variable. Lorentz boosts correspond to constant shifts in $\theta$. We can plot the dispersion relation as in Fig.~\ref{fig:crossing-massive-rel}

The transformation $\theta \to - \theta$ can be achieved by a finite Lorentz boost (equal to $- 2 \theta$), hence the two arms of~\ref{fig:crossing-massive-rel} are connected by admissible changes of relativistic frame.

\smallskip 

Let us then consider the massless limit $m\to 0$. As one may expect, the notion of scattering in the massless case is not well-defined. Furthermore, many theorems related to integrable scattering do not hold for massless particles. Nevertheless, the notion of factorised S-matrix is still sensible and one can use it as a tool to write down Bethe-ansatz equations, and derive a TBA system of equations for the finite-size spectrum (see for instance~\cite{Zamolodchikov:1994za,Mann:2004jr,Mann:2005ab}). In fact, massless integrable two-dimensional theories typically retain a scale (hence, they are not scale-invariant, see~\eqref{eq:scale} below) and describe the renormalisation group flow between two conformal field theories (UV and IR fixed points of the flow).      

\begin{figure}[t]
\centering
  \includegraphics{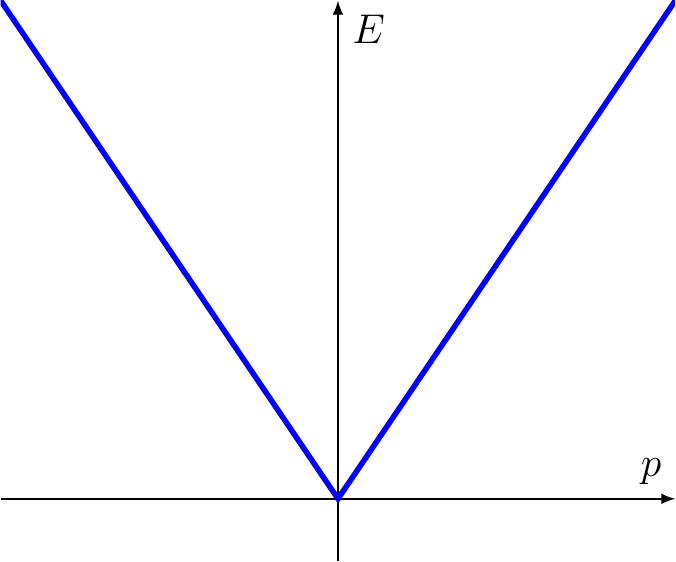}  

    \caption{Relativistic dispersion relation for $m=0$.}
\label{fig:reldisp22}
\end{figure}

In the massless limit we write $\theta = \theta_0 + \xi$, send $m \to 0$ and $\theta_0 \to \pm \infty$, while simultaneously keeping the combination 
\begin{equation}
\label{eq:scale}
m \, e^{|\theta_0|} = M 
\end{equation}
finite.
The limit naturally splits into two branches (corresponding to the two branches of the limiting dispersion relation $E = |p|$):
\begin{itemize}
\item \textit{$+$ handed} 
\begin{equation}
E = M e^{\xi_+}, \qquad p = M e^{\xi_+}, \qquad E = p, \qquad \xi_+ \in (-\infty, \infty)
\end{equation}
\item 
\textit{$-$ handed} 
\begin{equation}
E = M e^{- \xi_-}, \qquad p = - M e^{- \xi_-}, \qquad E = - p, \qquad \qquad \xi_- \in (-\infty, \infty) 
\end{equation}
\end{itemize}
At this point, $+$ handed particles move from left to right on the line ($p>0$) and $-$ handed particles move from right to left ($p>0$), both at the speed of light. There is no way to boost a frame to change the sign of either momenta. Equivalently, no transformation $\xi_- = \xi_- (\xi_+)$ connects the two branches. One has two types of particles in the massless limit, living on the two arms of Fig.~\ref{fig:reldisp22}.  
The rapidities $\xi_\pm$ can each be extended to the whole complex plane, and the particle-to-antiparticle map for rapidities is given by 
\begin{equation}
\xi_\pm \, \to \, \xi_\pm \, + \, i \pi.
\end{equation}

We can also understand why scattering loses its physical meaning, at least if we try and scatter two $+$ handed particles (or two $-$ handed ones). The particles have to move collinearly at the speed of light, hence a scattering is hard to attain (or resolve, if the two particles are at coincident points). Rather, the massless S-matrices of same handedness survive the conformal limits, and characterise the CFTs at the extrema of the flow~\cite{Bazhanov:1994ft,Bazhanov:1996dr,Bazhanov:1996aq,Bazhanov:1998dq}, while the  mixed scattering is affected by the scale $M$ and should ultimately be sensitive to the flow.

\smallskip

\begin{figure}[t]
  \centering
  \includegraphics{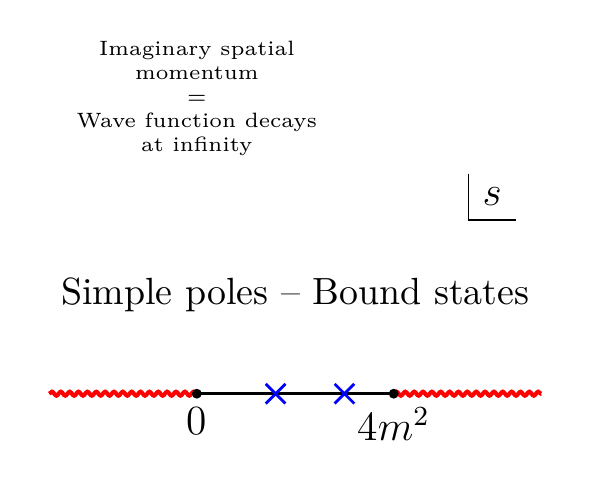} 










    \caption{Analytic structure of the massive relativistic S-matrix.}
\label{fig:Mandelstam}
\end{figure}

The analytic structure of the S-matrix is particularly subtle in the relativistic massless case. Before taking any massless limit, one generically has the picture described in Fig.~\ref{fig:Mandelstam}, where, for equal masses of the scattering particles, the Mandelstam variable $s$ satisfies
\begin{equation}
s = 2 m^2 [1 + \cosh (\theta_1 - \theta_2)].
\end{equation}
The two cuts are the $s$-channel (right) and $t$-channel (left), respectively. The visible part of the complex plane (upper Riemann sheet) is called the \emph{physical region}. 

When $m \to 0$ (and in the case of opposite handedness of the scattering particles), the allowed region where the bound-state simple poles can reside shrinks to zero. This is in accordance with the statement that no stable bound state of massless particles exist. In the context of the analytic S-matrix, this is rephrased by saying that the two branch-cuts ($s$- and $t$- channel) get into contact at the origin, eliminating the mass-gap in the spectrum. 

\begin{figure}[t]
  \centering
  \includegraphics{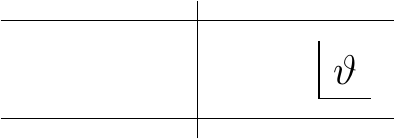}



\caption{The $\vartheta$-plane, with indicated the physical strip.}
\label{fig:rapido}
\end{figure}

In terms of the rapidity-difference 
\begin{equation}
\vartheta \equiv \theta_1 - \theta_2,
\end{equation}
the physical region is mapped onto the strip $\Im \vartheta \in [0,\pi]$, and the bound-state poles reside on the imaginary axis within the physical strip. When the massless limit is taken in the case of same handedness, the analyticity structure in the $\vartheta$-plane is only mildly affected\footnote{In this limit, we will then understand for example
\begin{equation}
\vartheta_{lim} = \xi_{+,1} - \xi_{+2}.
\end{equation}} since the parameter $m$ is already virtually factored out. However, the analysis of the bound-state poles (performed on the massless S-matrix) must be consistent with the existence of no stable bound states (see Fig.~\ref{fig:rapido}).
We are going to describe the implications of this for crossing symmetry and unitarity in the next subsection.

\subsection{Analyticity and the crossing-unitarity relation}

The shrinking, which we have just described, of the line where stable bound-state poles are allowed to reside, and the final merging of the $s$- and $t$-channel branch cuts, have an effect on the crossing and unitarity relations which the S-matrix has to satisfy.

\begin{figure}[t]
\centering
  \includegraphics{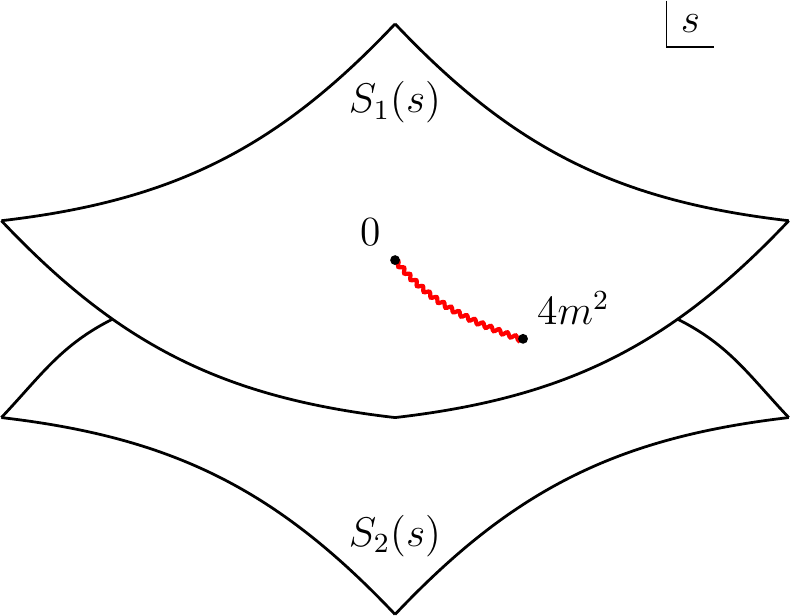}  

    \caption{The two-sheeted structure after the branch-cut switch}
\label{fig:reldispe}
\end{figure}

Closely following~\cite{Zamolodchikov:1994za}, let us first switch to a picture where the branch cut is in fact running on the real line \emph{inside} the region $\Re s \in [0,4m^2]$ (see Fig.~\ref{fig:reldispe}). Let the value of the S-matrix on the upper (lower) Riemann sheet be $S_1$ ($S_2$, respectively). We assume a square-root type of branch cut  for simplicity in this discussion.

This subdivision of sheets is not directly in one-to-one correspondence with the attribute of a \emph{physical} and \emph{not physical} sheet. Instead, after switching the branch cuts, the \emph{physical sheet} is composed of the upper half-plane of the upper sheet and of the lower half plane of the lower sheet. In other words, the physical values of the S-matrix are\footnote{For simplicity, we restrict to a scalar S-matrix for the scope of this subsection.}
\begin{equation}
S(s) = S_1(s) \qquad \mbox{if} \qquad \Im s >0, \qquad S(s) = S_2(s) \qquad \mbox{if} \qquad \Im s <0. \qquad
\end{equation}
The unitarity and crossing relations respectively read 
\begin{equation}
S_1(s) \, S_2(s) = 1 \qquad \mbox{and} \qquad S_1(s) = S_2(4 m^2 - s). 
\end{equation}

On the one hand, the first equation has the interpretation of an analytic continuation through the cut $[0,4 m^2]$. One is comparing the value of the function $S(s)$ on the upper sheet with the analytically-continued value at the same point $s$ on the lower sheet. If the starting point was in the physical region, the landing point is not. 

On the other hand, the second equation involves an analytic continuation through the cut, landing however on the point $4m^2 - s$ on the lower sheet (which means that one is never leaving the physical sheet).

The situation is rather different in the massless limit  (now thought of as a $+$ handed particle scattering against a $-$ handed one\footnote{ When formulated in the rapidity plane, the $++$ and $--$ S-matrices are virtually indistinguishable from their massive counterparts, and therefore can be taken to formally satisfy the usual massive-type relations in terms of the rapidities.}). The branch cut now shrinks to a single point, and one is left with two separate meromorphic functions $S_1(s)$ and $S_2(s)$, whose only way to communicate is \emph{via} a single algebraic relation
\begin{equation}
\label{eq:algebro}
S_1(s) = S_2(-s).
\end{equation}
This implies that, on each respective sheet, one has a combined \emph{crossing-unitarity} relation:
\begin{equation}
\label{eq:algebro2}
S_i(s) \, S_i(-s) = 1 \qquad \forall \, \, \, \, i=1,2. 
\end{equation}

\subsection{Connections with Hopf algebras}

In this section we present some Hopf-algebra considerations on the crossing and unitarity relations.  

Let us recall that the universal R-matrix\footnote{We will assume its existence for the purposes of the arguments in this section, although a mathematical proof of this statement is still lacking for $AdS/CFT$ (see however the recent progress obtained in~\cite{Beisert:2016qei} in the context of $q$-deformations of the $AdS_5$ integrable system).} $R$ of the Hopf algebra $\mathcal{A}$, abstractly defined to solve
\begin{equation}
\Delta^{op} (\alg{J}) \, R = R \, \Delta(\alg{J}) \qquad \forall \, \, \, \alg{J} \in \mathcal{A},
\end{equation}
and assumed to be invertible, satisfies
\begin{equation}
R_{12} \, R_{21} = \matId \otimes \matId \qquad \mbox{\textit{braiding unitarity}}, 
\end{equation}
where 
\begin{equation}
\label{eq:uni}
R_{21} = R^{op} = \Pi \, R
\end{equation}
($\Pi$ being the graded permutation operator), and
\begin{equation}
\label{eq:crossing}
(\Sigma \otimes \matId) R = R^{-1} \qquad \mbox{\textit{crossing symmetry}},
\end{equation}
with $\Sigma$ the Hopf-algebra antipode. Upon introducing a representation $\pi$, and assuming the existence of a representation $\bar{\pi}$ \emph{conjugate} to it, such that
\begin{equation}
\pi \big(\Sigma(\alg{J})\big) = C \, \bar{\pi}(\alg{J})^{str} C^{-1} \qquad \forall \, \, \, \alg{J} \in \mathcal{A},
\end{equation}
(the apex ${}^{str}$ denoting supertransposition and $C$ being the \emph{charge-conjugation} matrix), the crossing symmetry condition~\eqref{eq:crossing} reduces to the following matrix equation
\begin{equation}
(C \otimes \matId) \, [\bar{\pi}_1 \otimes \pi_2] R^{str_1} (C^{-1} \otimes \matId) = [\pi_1 \otimes \pi_2] R^{-1},
\end{equation}
where ${}^{str_1}$ denotes supertransposition in the first factor of the tensor product.

Let us now combine the crossing and braiding-unitarity conditions together. First, let us focus on a specific representation, namely, a massive relativistic bosonic particle. In this representation the R-matrix becomes the S-matrix of the scattering problem, and it depends only on the difference of the particle-rapidities,
\begin{equation}
[\pi_1 \otimes \pi_2] R = S(\vartheta) = S(\vartheta)_{bd}^{ac} E_a^b \otimes E_c^d, \qquad \vartheta = \theta_1 - \theta_2,
\end{equation}
where $E_a^b$ is the matrix which sends particle $a$ into particle $b$.
Straightforward manipulations of\eqref{eq:uni} and~\eqref{eq:crossing} imply the condition
\begin{equation}
\label{eq:repo}
S(i \pi - \vartheta)_{bd}^{ac} (\matId \otimes C) E_c^d \otimes E_b^a (\matId \otimes C^{-1}) = S(\vartheta)_{bd}^{ac}  E_a^b \otimes E_c^d.
\end{equation}
We have used the fact that under crossing
\begin{equation}
\theta_i \to \theta_i + i \pi, \qquad i=1,2.
\end{equation}
At the Hopf algebra level, this is echoed by the relation 
\begin{equation}
\label{eq:abs}
(\matId \otimes \Sigma) R_{21} = R,
\end{equation}
which can be obtained by applying the permutation operator to the crossing equation and performing simple algebraic rearrangings. In order to match~\eqref{eq:abs} to~\eqref{eq:repo}, it is useful to notice that\footnote{The operator $\Pi$ is taken to permute abstract elements of $\mathcal{A}$ as well as the corresponding matrices in specific representations.}
\begin{equation}
[\pi_1 \otimes \pi_2] R_{21} = \Pi \,  [\pi_2 \otimes \pi_1] R. 
\end{equation}

The requirement of \emph{braiding unitarity} is something different from the \emph{unitarity} discussed in the previous section, namely 
\begin{equation}
S_1(s) \, S_2(s) = \matId.
\end{equation} 
The latter is sometimes dubbed \emph{physical unitarity}, and it is connected to the property of \emph{real analyticity} of the S-matrix. In terms of the rapidity-difference, physical unitarity can be phrased as 
\begin{equation}
S(\vartheta) \, S(-\vartheta) = \matId,
\end{equation}
where we have removed the sheet-indices $1,2$ from $S(\vartheta)$ since the S-matrix is now a meromorphic function on the $\vartheta$-plane. Conversely, the crossing condition~\eqref{eq:crossing} becomes in this case in the rapidity-plane:
\begin{equation}
\label{eq:inde}
S(\vartheta) \, (C \otimes \matId) \, S(i \pi + \vartheta)(C^{-1} \otimes \matId) = \matId.
\end{equation}

From the Hopf-algebra viewpoint, the condition of physical unitarity can be obtained from the very same equation~\eqref{eq:uni}, by projecting it onto representations $\pi_1 \otimes \pi_2$ and applying the permutation operator $\Pi$. It may seem surprising that no new relations are needed at the abstract level to obtain the independent relation~\eqref{eq:inde}. The subtelty is that, when we apply the specific relativistic-particle representation $\pi$, we have to make a statement regarding the region in the complex $\vartheta$-plane we project our relation into, and different choices end up encoding different physical relations.

It is for this reason that the massless case will also be encoded in the same set of algebraic relations~\eqref{eq:uni} and~\eqref{eq:crossing}. Upon projection onto the massless representations, the interpretation in terms of the complex $\vartheta_{lim}$-plane will now be according to the previous subsection's arguments, and, in particular, one will recover~\eqref{eq:algebro} and the mixed crossing-unitarity condition~\eqref{eq:algebro2} directly from~\eqref{eq:crossing}.

\section{\texorpdfstring{Antisymmetry of $\theta^{\circ\circ}_{\mbox{\scriptsize min}}(x,y)$ and $\chi^{\mbox{\scriptsize HL}}_m$}{Antisymmetry of theta\^{}oo\string_min and chi\^{}HL\string_m}}
\label{app:chi-asym} 

To verify directly that $\theta_{\mbox{\scriptsize min}}^{\circ\circ}(x,y)$ given in equation~\eqref{eq:hl-massless} is antisymmetric note that
\begin{align}
  -\int\limits_{-1+i\epsilon}^{1+i\epsilon}\!\! \frac{dz}{4\pi}g(z,x)G_-(z,y)&=\!\!\!\int\limits_{-1+i\epsilon}^{1+i\epsilon}\!\!
\frac{dz}{4\pi}g(z,y)G_-(z,x) 
\\
-\int\limits_{-1-i\epsilon}^{1-i\epsilon}\!\!\frac{dz}{4\pi}g(z,\tfrac{1}{x})G_+(z,\tfrac{1}{y})&=\!\!\!\int\limits_{-1-i\epsilon}^{1-i\epsilon}\!\!\frac{dz}{4\pi}g(z,\tfrac{1}{y})G_+(z,\tfrac{1}{x}) 
\\
-\!\!\int\limits_{-1+i\epsilon}^{1+i\epsilon}\!\!\frac{dz}{4\pi}g(z,\tfrac{1}{x})G_-(z,y)-\frac{i}{2}G_-(x,y)&=\!\!\!\int\limits_{-1-i\epsilon}^{1-i\epsilon}\!\!\frac{dz}{4\pi}g(z,y)G_+(z,\tfrac{1}{x}) -\frac{i}{2}G_+(\tfrac{1}{y},\tfrac{1}{x})
\\
-\!\!\int\limits_{-1-i\epsilon}^{1-i\epsilon}\!\!\frac{dz}{4\pi}g(z,x)G_+(z,\tfrac{1}{y})+\frac{i}{2}G_+(x,y)&=
\!\!\!\int\limits_{-1+i\epsilon}^{1+i\epsilon}\!\!\frac{dz}{4\pi}g(z,\tfrac{1}{y})G_-(z,x) +\frac{i}{2}G_-(y,x).
\end{align}
The above identities hold for $\mbox{Im}(x)$, $\mbox{Im}(y)>0$ and can be used to show that
\begin{align}
\theta^{\circ\circ}_{\mbox{\scriptsize min}}(x,y)&=
\int\limits_{-1+i\epsilon}^{1+i\epsilon}\frac{dz}{8\pi}\left(\frac{1}{z-y}-\frac{1}{z-\tfrac{1}{y}}\right)G_-(z,x)
\nonumber \\ &
-\frac{1}{2\pi}\int\limits_{-1-i\epsilon}^{1-i\epsilon}\frac{dz}{8\pi}\left(\frac{1}{z-y}-\frac{1}{z-\tfrac{1}{y}}\right)G_+(z,\tfrac{1}{x})-\frac{i}{8}\left(G_-(y,x)-G_+(\tfrac{1}{y},\tfrac{1}{x})\right)
\nonumber \\&=-\theta_{\mbox{\scriptsize min}}^{\circ\circ}(y,x).
\end{align}
An almost identical proof of anti-symmetry applies to the `deformed contour' expression for $\theta_m^{\HL}$ given in equation~\eqref{eq:hl-generic}.

Let us next show that $\chi^{\mbox{\scriptsize HL}}_m$ in equation~\eqref{eq:def-chi-hl} is also anti-symmetric. By changing the integration variable $z\rightarrow\tfrac{1}{z}$ one can write $\chi^{\mbox{\scriptsize HL}}_m(x,y)$ as
\begin{equation}
\chi^{\mbox{\scriptsize HL}}_m(x,y)=
-\inturl \frac{dz}{4\pi}g(z,x)G_\pm(z,y)
=\intdlr \frac{dz}{4\pi}g(z,x)G_\pm(z,y)
.
\label{eq:hl-sing-sem-circ-int}
\end{equation}
Since $G_\pm(1,y)=G_\pm(0,y)=0$, one can integrate by parts to show that for $\mbox{Im}(x),\mbox{Im}(y)>0$ we have for example
\begin{equation}
\inturl \frac{dz}{4\pi}g(z,x)G_-(z,y)=-\inturl \frac{dz}{4\pi}g(z,y)G_-(z,x),
\end{equation}
while for $\mbox{Im}(x)>0$ and $\mbox{Im}(y)<0$ we have
\begin{equation}
\inturl \frac{dz}{4\pi}g(z,x)G_+(z,y)=-\inturl \frac{dz}{4\pi}g(z,y)G_-(z,x).
\end{equation}
As a result, the single-integral
expression~\eqref{eq:hl-sing-sem-circ-int} for $\chi^{\mbox{\scriptsize HL}}_m(x,y)$
and hence also~\eqref{eq:def-chi-hl} is anti-symmetric under $x\leftrightarrow y$
\begin{equation}
\chi^{\mbox{\scriptsize HL}}_m(x,y) = -\chi^{\mbox{\scriptsize HL}}_m(y,x).
\end{equation}

\section{Charge expansion of the phases}
\label{app:four-and-crs}
In this appendix we collect the formulas needed to expand the phases in terms of the charges $q_r$. We first review some standard Fourier expansions and then use these
to determine the $c_{r,s}$ expansion coefficients.

\subsection{Fourier expansions}
\label{app:fourier-exp}

\noindent We will use the following Fourier expansions, which are valid for
$y=e^{i\pi\theta}$ with $\theta\in[0,1]$
\begin{align}
\label{eq:f1}
\log(-i(y - z))&=
\log y-\frac{i\pi}{2}-\sum_{n=1}^{\infty}\frac{1}{n}\left(\frac{z}{y}\right)^n
&\qquad z\in\left[-1+i\epsilon,\,1+i\epsilon\right]
\\
\label{eq:f2}
\log(-i(y - \tfrac{1}{z}))
&=
-\log z+\frac{i\pi}{2}-\sum_{n=1}^{\infty}\frac{1}{n}\left(zy\right)^n
&\qquad z\in\left[-1+i\epsilon,\,1+i\epsilon\right]
\\
\label{eq:f3}
\log(i(\tfrac{1}{y} - z))
&=
-\log y+\frac{i \pi}{2}-\sum_{n=1}^{\infty}\frac{1}{n}\left(zy\right)^n
&\qquad z\in\left[-1-i\epsilon,\,1-i\epsilon\right]
\\
\label{eq:f4}
\log(i(\tfrac{1}{y} - \tfrac{1}{z}))
&=
-\log z-\frac{i\pi}{2}-\sum_{n=1}^{\infty}\frac{1}{n}\left(\frac{z}{y}\right)^n
&\qquad z\in\left[-1-i\epsilon,\,1-i\epsilon\right]
\end{align}
We also note that for $x=e^{i\pi\phi}$ with $\phi\in[0,1]$ and $z\in\left[-1,\,1\right]$
\begin{equation}
\frac{1}{z-x}-\frac{1}{z-1/x} =\sum_{n=0}^\infty z^{n}
\left(x^{n+1}-x^{-n-1}\right)
\end{equation}
It is also useful to recall certain Fourier series involving the Heaviside step function $\Theta(x)$ which is $0$ for $x<0$ and $1$ for $x>0$
\begin{equation}
\Theta(\theta-\phi)\Theta(1-\theta-\phi)=\sum_{m,n=1}^\infty 
a^{(1)}_{m,n}\sin(n\pi\phi)\sin(m\pi\theta),
\end{equation}
where
\begin{equation}
a^{(1)}_{m,n}=\frac{2}{\pi^2}\left\{\begin{array}{ll}(1-(-1)^m)
\frac{(-1)^{(m+n)/2}(1-(-1)^n)m+2n}{m(n^2-m^2)} & m\neq n \\
\frac{1-(-1)^n}{n^2} & m=n\end{array}\right.
\end{equation}
Similarly,
\begin{equation}
\Theta(\theta-\phi)\Theta(\theta+\phi-1)=\sum_{m,n=1}^\infty 
a^{(2)}_{m,n}\sin(n\pi\phi)\sin(m\pi\theta),
\end{equation}
where
\begin{equation}
a^{(2)}_{m,n}=\frac{2}{\pi^2}\left\{\begin{array}{ll}-(1-(-1)^n)
\frac{(-1)^{(m+n)/2}(1-(-1)^m)n+2(-1)^{m+n}m}{n(n^2-m^2)} & m\neq n \\
\frac{1-(-1)^n}{n^2} & m=n\end{array}\right.
\end{equation}
Next,
\begin{equation}
\phi=\sum_{m,n=1}^\infty 
a^{(3)}_{m,n}\sin(n\pi\phi)\sin(m\pi\theta),
\end{equation}
where
\begin{equation}
a^{(3)}_{m,n}=-\frac{4}{\pi^2}\frac{(-1)^n(1-(-1)^m)}{nm}.
\end{equation}
Similarly,
\begin{equation}
\theta=\sum_{m,n=1}^\infty 
a^{(4)}_{m,n}\sin(n\pi\phi)\sin(m\pi\theta),
\end{equation}
where
\begin{equation}
a^{(4)}_{m,n}=-\frac{4}{\pi^2}\frac{(-1)^m(1-(-1)^n)}{nm}.
\end{equation}
Finally,
\begin{equation}
1=\sum_{m,n=1}^\infty 
a^{(5)}_{m,n}\sin(n\pi\phi)\sin(m\pi\theta),
\end{equation}
where
\begin{equation}
a^{(5)}_{m,n}=\frac{4}{\pi^2}\frac{(1-(-1)^n)(1-(-1)^m)}{nm}.
\end{equation}
We also note
\begin{equation}
\Theta(\theta-\phi)=\sum_{m,n=1}^\infty 
a^{(6)}_{m,n}\sin(n\pi\phi)\sin(m\pi\theta),
\end{equation}
where
\begin{equation}
a^{(6)}_{m,n}=\frac{4}{\pi^2}\left\{\begin{array}{ll}
\frac{(-1)^m(1-(-1)^n)m^2+(1-(-1)^m)n^2}{mn(n^2-m^2)} & m\neq n \\
\frac{1-(-1)^n}{n^2} & m=n\end{array}\right.
\end{equation}

\subsection{\texorpdfstring{Finding $c^{\circ\circ}_{r,s}$}{Finding c\^{}oo\string_rs}}
\label{app:deriving-massless-crs}

In this appendix we present the details of the derivation that leads to equation~\eqref{eq:massless-crs}.
To obtain the coefficients $c^{\circ\circ}_{r,s}$ we insert the Fourier series~\eqref{eq:f1}--\eqref{eq:f4} into equation~\eqref{eq:hl-massless}
and perform the integrals. We will find that the resulting expressions naively are not anti-symmetric in $x\leftrightarrow y$. However, upon closer inspection, we show that the final expansion is afterall anti-symmetric as expected from the general results in Appendix~\ref{app:chi-asym}. 

When performing the 
integrals in equation~\eqref{eq:hl-massless}, for the majority of terms we can take the limit $\epsilon\rightarrow 0$  before performing the integral. Only in the case of the integrands proportional to $\log z$, coming from the expansions~\eqref{eq:f2} and~\eqref{eq:f4} is there a potential subtlety. Such terms contribute 
\begin{align}
&\int\limits_{-1-i\epsilon}^{1-i\epsilon}\!\!\frac{dz}{2\pi}
 \left(\frac{1}{z-x}-\frac{1}{z-\tfrac{1}{x}}\right)\log z -
\int\limits_{-1+i\epsilon}^{1+i\epsilon}\!\!\frac{dz}{2\pi}\left(\frac{1}{z-x}-\frac{1}{z-\tfrac{1}{x}}\right)\log z 
\nonumber \\
=&
\int\limits_{-1-i\epsilon}^{0-i\epsilon}\!\!\frac{dz}{2\pi}
\left(\frac{1}{z-x}-\frac{1}{z-\tfrac{1}{x}}\right)\log z 
-\int\limits_{-1+i\epsilon}^{0+i\epsilon}\!\!\frac{dz}{2\pi}
 \left(\frac{1}{z-x}-\frac{1}{z-\tfrac{1}{x}}\right)\log z 
\nonumber \\
\rightarrow&
-\int_{-1}^{0}\frac{dz}{2\pi} \left(\frac{1}{z-x}-\frac{1}{z-\tfrac{1}{x}}\right)2\pi i
\nonumber \\
=&
-i\log x
\end{align}
where we have used the fact that $\log z_+=\log z_-+2\pi i$. We also note that when $x$ is on the upper semi-circle
\begin{equation}
\int_{-1}^1 dz \left(\frac{1}{z-x}-\frac{1}{z-1/x}\right) = \pi i.
\end{equation}

Combining these observations, one finds that the 
part of $\theta^{\circ\circ}$ expressed as integrals in equation~\eqref{eq:hl-massless}
can be written as\footnote{The terms $-\pi-i\log y$ come from the $\log y$ and $i \pi$ terms in the expansions, while the $\log x$ term comes from the integrals involving $\log z$. The remaining terms come from the infinite sums.}
\begin{align}
&-\int\limits_{-1+i\epsilon}^{1+i\epsilon}\!\!\frac{dz}{2\pi}
\left(\frac{1}{z-x}-\frac{1}{z-\tfrac{1}{x}}\right)F_-(z,y)
+\int\limits_{-1-i\epsilon}^{1-i\epsilon}\!\!\frac{dz}{2\pi}\left(\frac{1}{z-x}-\frac{1}{z-\tfrac{1}{x}}\right)F_+(z,\tfrac{1}{y})
\nonumber \\ =&
-\pi-i\log y -2i\log x-\frac{1}{\pi}\sum_{m,n=1}^\infty
\frac{1-(-1)^{m+n}}{m(m+n)}(x^n-x^{-n})(y^m-y^{-m}).
\end{align}
As mentioned above, naively this expression is \emph{not} antisymmetric in $x\leftrightarrow y$. To investigate this apparent discrepancy let us denote the double-infinite sum above by $l(x,y)$.
Then we can write the above expression as
\begin{equation}
-\pi-i\log y -2i\log x +\frac{l(x,y)+l(y,x)}{2} +\frac{l(x,y)-l(y,x)}{2},
\label{eq:int-pt-of-phase-not-anti-symmetrised}
\end{equation}
One can show that
\begin{equation}
\frac{l(x,y)+l(y,x)}{2} =\frac{\pi + i\log x+i\log y}{2},
\label{eq:int-pt-of-phase-symmetrised}
\end{equation}
while the anti-symmetric part of $l(x,y)$ gives
\begin{align}
\frac{l(x,y)-l(y,x)}{2}&=-
\int\limits_{-1+i\epsilon}^{1+i\epsilon}\!\!\frac{dz}{4\pi}
\left(\frac{1}{z-x}-\frac{1}{z-\tfrac{1}{x}}\right)F(z,y)
\nonumber \\ &=
-\frac{1}{2\pi}\sum_{m,n=1}^\infty
\frac{1-(-1)^{m+n}}{m(m+n)}\left((x^n-x^{-n})(y^m-y^{-m})
-(x\leftrightarrow y)\right)\nonumber \\ &=
\frac{1}{2\pi}\sum_{m,n=1 }^\infty
(1-(-1)^{m+n})
\frac{m-n}{mn(m+n)}(x^n-x^{-n})(y^m-y^{-m}).
\label{eq:int-pt-of-phase-anti-symmetrised}
\end{align}
Soon we return to this expression, but next we turn to the non-integral part of the phase given in equation~\eqref{eq:hl-massless}
\begin{equation}
\frac{i}{2}\left(F_-(x,y)-F_+(\tfrac{1}{x},\tfrac{1}{y})\right).
\end{equation}
For $x=e^{i \pi \phi}$ and $y=e^{i\pi \theta}$, on the upper semi-circle this expression reduces to\footnote{This identity holds at almost all points $\theta,\,\phi\in\left[0,1\right]$ at
a finite number of points the two functions differ by a
finite amount. Since we will shortly be interested in the Fourier expansions of
these such differences will play no role.}
\begin{equation}
\pi+ i\log x-\pi \Theta(\theta-\phi),
\label{eq:non-int-pt-of-phase}
\end{equation}
where $\Theta(x)$ is the Heaviside step function which is $0$ for $x<0$ and $1$ for $x>0$.

If we combine equation~\eqref{eq:non-int-pt-of-phase} with equations~\eqref{eq:int-pt-of-phase-not-anti-symmetrised},~\eqref{eq:int-pt-of-phase-symmetrised} and~\eqref{eq:int-pt-of-phase-anti-symmetrised} we find for $x=e^{i \pi \phi}$ and $y=e^{i\pi \theta}$
\begin{align}
\theta_{\mbox{\scriptsize min}}^{\circ\circ}(x,y)=&
-\pi-i\log y -i\log x +\frac{\pi + i\log x+i\log y}{2}
+\pi+ i\log x-\pi \Theta(\theta-\phi)
\nonumber \\
&
+\frac{1}{2\pi}\sum_{m,n=1 }^\infty
(1-(-1)^{m+n})
\frac{m-n}{mn(m+n)}(x^n-x^{-n})(y^m-y^{-m})
\nonumber \\
=&
\frac{\pi+ i\log x-i\log y}{2}
-\pi \Theta(\theta-\phi)
\nonumber \\
&
+\frac{1}{2\pi}\sum_{m,n=1 }^\infty
(1-(-1)^{m+n})
\frac{m-n}{mn(m+n)}(x^n-x^{-n})(y^m-y^{-m})
\label{eq:theta-expansion}
\end{align}
Next, note the following identity that can be verified using the expressions in Appendix~\eqref{app:fourier-exp}
\begin{align}
\frac{1}{2\pi}&
\sum_{m,n=1 }^\infty(1-(-1)^{m+n})
\frac{m^2+n^2}{mn(m^2-n^2)}(x^n-x^{-n})(y^m-y^{-m})
\nonumber \\
  =&-\frac{\pi}{2}\left(\theta-\phi+1-2\Theta(\theta-\phi)\right).
\end{align}
Inserting this into equation~\eqref{eq:theta-expansion} we find\footnote{We have also used for $m\neq n$
\begin{equation}
\frac{m-n}{mn(m+n)}=\frac{m^2+n^2}{mn(m^2-n^2)}-\frac{2}{m^2-n^2}.
\end{equation}}
\begin{align}
\theta_{\mbox{\scriptsize min}}^{\circ\circ}(x,y)=&
\frac{\pi+ i\log x-i\log y}{2}
-\pi \Theta(\theta-\phi)
\nonumber \\ &
+\frac{1}{2\pi}\sum_{m,n=1 }^\infty(1-(-1)^{m+n})
\frac{m^2+n^2}{mn(m^2-n^2)}(x^n-x^{-n})(y^m-y^{-m})
\nonumber \\ &
-\frac{1}{\pi}\sum_{m,n=1 }^\infty(1-(-1)^{m+n})
\frac{1}{m^2-n^2}(x^n-x^{-n})(y^m-y^{-m})
\nonumber \\ =&
\frac{\pi}{2}(1-\phi+\theta)-\pi \Theta(\theta-\phi)
-\frac{\pi}{2}\left(\theta-\phi+1-2\Theta(\theta-\phi)\right)
\nonumber \\ &\!\!\!
-\frac{1}{\pi}\sum_{r,s=2 }^\infty
\frac{1-(-1)^{r+s}}{(r-s)(r+s-2)}\Bigl((x^n-x^{-n})(y^m-y^{-m})-(x\leftrightarrow y)\Bigr).
\nonumber \\
\end{align}
The terms on the penultimate line above cancel giving the expansion~\eqref{eq:massless-crs}. Note in particular that the above expansion is anti-symmetric in $x\leftrightarrow y$.

\section{Solutions of the crossing equations}

\subsection{Rewriting HL}
\label{app:deform-hl}
Let us define
\begin{equation}
  \begin{aligned}
    \varphi^{\HL}(x,y)&=\intdlr \frac{dz}{4\pi} \ g(z,x) \ G_\pm(z,y),\\
    &=\intulr \frac{dz}{4\pi} \ g(z,x) \ G_\pm(z,y),
  \end{aligned}
\end{equation}
which is obtained from~\eqref{eq:def-chi-hl} by redefining the integration variable $z \to 1/z$ in one of the two integrals of the second line. The freedom to integrate in the upper or in the lower semicircle is a consequence of the symmetry
\begin{equation}
  G_\pm(1/z,y) = - G_\pm(z,y).
\end{equation}
In the massive case the total result is obtained by summing the four contributions
\begin{equation}\label{eq:sum4}
  \Phi^{\HL}(x^\pm,y^\pm)=\varphi^{\HL}(x^+,y^+)-\varphi^{\HL}(x^-,y^+)-\varphi^{\HL}(x^+,y^-)+\varphi^{\HL}(x^-,y^-).
\end{equation}
Before doing this sum we will actually move the contour of integration to the real interval $[-1,+1]$. We will decide whether we want to start from the representation with the upper or the lower semicircle, depending on the position of the branch cuts of the integrand that we do not want to intersect when moving the contour. Let us look at the four possible cases.
\begin{itemize}
\item $(x^+,y^+)$
  We start from integrating on the upper semicircle, and when moving the contour we do not pick any pole.
  \begin{equation}
    \intulr \frac{dz}{4\pi} \ g(z,x^+) \ G_-(z,y^+)=\int\limits_{-1+i\epsilon}^{1+i\epsilon}\frac{dz}{4\pi} \ g(z,x^+) \ G_-(z,y^+).
  \end{equation}

\item $(x^-,y^+)$
  We start from integrating on the upper semicircle, and when moving the contour we pick a pole at $1/x^-$.
  \begin{equation}
    \intulr \frac{dz}{4\pi} \ g(z,x^-) \ G_-(z,y^+)=\int\limits_{-1+i\epsilon}^{1+i\epsilon}
    \frac{dz}{4\pi} \ g(z,x^-) \ G_-(z,y^+) +\frac{i}{2} G_-(\tfrac{1}{x^-},y^+).
  \end{equation}

\item $(x^+,y^-)$
  We start from integrating on the lower semicircle, and when moving the contour we pick a pole at $1/x^+$.
  \begin{equation}
    \intdlr \frac{dz}{4\pi} \ g(z,x^+) \ G_+(z,y^-)=\int\limits_{-1-i\epsilon}^{1-i\epsilon}
    \frac{dz}{4\pi} \ g(z,x^+) \ G_+(z,y^-) -\frac{i}{2} G_+(\tfrac{1}{x^+},y^-).
  \end{equation}

\item $(x^-,y^-)$
  We start from integrating on the lower semicircle, and when moving the contour we do not pick any pole.
  \begin{equation}
    \intdlr \frac{dz}{4\pi} \ g(z,x^-) \ G_+(z,y^-)=\int\limits_{-1-i\epsilon}^{1-i\epsilon}
    \frac{dz}{4\pi} \ g(z,x^-) \ G_+(z,y^-) .
  \end{equation}

\end{itemize}
Summing up the four contributions as in~\eqref{eq:sum4} we obtain the result in~\eqref{eq:hl-generic}.

\subsection{Proof for the solution to the HL crossing equation}
\label{app:proof-cr-hl}
In this appendix we will show that the phase~\eqref{eq:hl-generic} satisfies the crossing equation that follows from~\eqref{eq:massive-hl-cross} by studying its analytic continuation to the crossed region.
We will do that without assuming whether the variables involved correspond to massive or massless particles.
Consider
\begin{equation}\label{eq:hl-generic-Phi}
\begin{aligned}
\Phi^{\HL} (x^\pm,y^\pm) &=
\int\limits_{-1+i\epsilon}^{1+i\epsilon}
\frac{dz}{4\pi}G_-(z,y^+)\left( g(z,x^+)-g(z,x^-) \right)
 \\ &\qquad
- \int\limits_{-1-i\epsilon}^{1-i\epsilon}
 \frac{dz}{4\pi}G_+(z,y^-) \left( g(z,x^+)-g(z,x^-) \right)\\
& \qquad-\frac{i}{2} \left( G_-(\tfrac{1}{x^-},y^+)-G_+(\tfrac{1}{x^+},y^-) \right),
\end{aligned}
\end{equation}
which we want to identify with $\theta^{\HL}$ in the physical region $\text{Im }x^+>0,\text{Im }x^-<0$ (and similarly for $y^\pm$).
First we note that it satisfies the ``homogeneous crossing equation''
\begin{equation}
\Phi^{\HL} (\tfrac{1}{x^\pm},y^\pm)+\Phi^{\HL} (x^\pm,y^\pm)=0,
\end{equation}
which is simply a consequence of the identities $G_\pm(\tfrac{1}{x},y)+G_\pm(x,y)=0$, and $g(z,\tfrac{1}{x})+g(z,x)=\tfrac{2}{z}$.
The function jumps for $x^\pm$ or $y^\pm$ going through the real line. We therefore study the analytic continuation in order to get a continuous $\theta^{\HL}$. The path is chosen according to the prescription of~\cite{Borsato:2013hoa}, where massive parameters are continued by crossing the \emph{lower semicircle} from the outside to the inside. 
It is clear that this is equivalent to saying that we continue the variable $x^+$ across the long interval $]-\infty,-1]\cup [+1,+\infty[$, and the variable $x^-$ across the short one $[-1,+1]$.

The discontinuities come from the integrals appearing in the first line of~\eqref{eq:hl-generic-Phi}. For the purpose of this computation the two integrals $\int\limits_{-1\pm i\epsilon}^{1\pm i\epsilon}$ can be considered together. We study separately the cases of continuation of $x^+$ and $x^-$.
\begin{itemize}
\item
  $x^+$\\
  We are crossing the long interval from above. Take $x_0^+ \in \mathbb{R}$ with $|x_0^+|>1$.
  The discontinuity is given by the function evaluated\footnote{The $\varepsilon$ that we are using here is different from the $\epsilon$ used to regulate the contours of integration.} at $x^+_0+i\varepsilon$ minus that at $x^+_0-i\varepsilon$
  \begin{equation}
    \left( \int\limits_{-1+i\epsilon}^{1+i\epsilon}
      \frac{dz}{4\pi}G_-(z,y^+)- \int\limits_{-1-i\epsilon}^{1-i\epsilon} \frac{dz}{4\pi}G_+(z,y^-)\right) \left( g(z,x^+_0+i\varepsilon)-g(z,x^+_0-i\varepsilon) \right).
  \end{equation}
  We find a pole at $z=1/x_0^+$ with residue $-1$ and since we are integrating clockwise we find that the discontinuity is
  \begin{equation}
    -\frac{i}{2}\left( G_-(x_0^+,y^+) -G_+(x_0^+,y^-)\right).
  \end{equation}
\item
  $x^-$\\
  Now we are crossing the short interval from below. Take $x_0^- \in \mathbb{R}$ with $|x_0^-|<1$.
  The discontinuity is given by the function evaluated at $x^-_0-i\varepsilon$ minus that at $x^-_0+i\varepsilon$
  \begin{equation}
    -\left( \int\limits_{-1+i\epsilon}^{1+i\epsilon}
      \frac{dz}{4\pi}G_-(z,y^+)- \int\limits_{-1-i\epsilon}^{1-i\epsilon}
      \frac{dz}{4\pi}G_+(z,y^-)\right) \left( g(z,x^-_0-i\varepsilon)-g(z,x^-_0+i\varepsilon) \right).
  \end{equation}
  We find a pole at $z=x_0^-$ with residue $+1$ and since we are integrating clockwise we find that the discontinuity is
  \begin{equation}
    -\frac{i}{2}\left( G_+(x_0^-,y^-) -G_-(x_0^-,y^+)\right).
  \end{equation}
\end{itemize}
From the above calculation we conclude that in order to get a function continuous across the real line we need to define it in the crossed region $\text{Im }x^+<0,\text{Im }x^->0$ as
\begin{equation}
  \Phi^{\HL} (x^\pm,y^\pm)-\frac{i}{2}\left( G_-(x_0^+,y^+)+ G_+(x_0^-,y^-) -G_+(x_0^+,y^-) -G_-(x_0^-,y^+)\right).
\end{equation}
Checking the crossing equation amounts then to computing
\begin{equation}
  \begin{aligned}
    \theta^{\HL} (x^\pm,y^\pm)+\theta^{\HL} (\tfrac{1}{x^\pm},y^\pm) &=
    \Phi^{\HL} (x^\pm,y^\pm)+\Phi^{\HL} (\tfrac{1}{x^\pm},y^\pm)
    \\ & \quad
    -\frac{i}{2} \bigl( G_-(\tfrac{1}{x^+},y^+)+ G_+(\tfrac{1}{x^-},y^-) 
    \\ & \quad
    \phantom{{}-\frac{i}{2}\bigl(}
    -G_+(\tfrac{1}{x^+},y^-) -G_-(\tfrac{1}{x^-},y^+) \bigr)\\
    &= \frac{i}{2}\bigl( G_-(x^+,y^+)+ G_+(x^-,y^-) 
    \\ & \phantom{{} = \frac{i}{2}\bigl(}
    -G_+(x^+,y^-) -G_-(x^-,y^+) \bigr) .
  \end{aligned}
\end{equation}
which is consistent with the right-hand-side of equation~\eqref{eq:massive-hl-cross}.
Since we did not need to specify whether the starting point of $x^\pm$ is outside or on the unit circle (and similary for $y^\pm$), the proof is valid for the massive, the massless and the mixed mass cases.

\section{\texorpdfstring{The massless limit of $\sigma^-$}{The massless limit of sigma\^{}-}}
\label{app:sigma-minus}

In this appendix we show that deforming the contour and imposing massless kinematics on the dressing factor $\sigma^-$ in a manner analogous to Sections~\ref{sec:def-cont}
and~\ref{sec:mass-phase-crossing-soln} trivializes $\sigma^-$. To see this note the following identities
\begin{align}
\inturl \frac{dz}{4\pi}\frac{H(z,y)}{x-z}&=\intdlr \frac{dz}{4\pi}\frac{H(z,y)}{z(xz-1)}
\\
\intdlr \frac{dz}{4\pi}\frac{H(z,y)}{x-z}&=\inturl \frac{dz}{4\pi}\frac{H(z,y)}{z(xz-1)}.
\end{align}
When deforming the integration contour to the interval $[-1,1]$ we want to avoid
the branch-cut of the logarithm in $H(z,y)$ and to account for the poles located inside the unit circle. Defining
\begin{equation}
h(z,x)=\frac{d}{dz}H(z,x)=\frac{1}{z-\tfrac{1}{x}}+\frac{1}{z-x}-\frac{1}{z},
\end{equation}
one can show that deforming the contour and imposing massless kinematics gives
\begin{align}
\chi^-_m(x^+,y^+)=&\inturl \frac{dz}{4\pi}\left(H(z,x^+)h(z,y^+)-H(z,y^+)h(z,x^+)\right)\nonumber \\
=&\int\limits_{-1+i\epsilon}^{1+i\epsilon}\frac{dz}{4\pi}\left(H(z,x^+)h(z,y^+)-H(z,y^+)h(z,x^+)\right)
\nonumber \\
\rightarrow&
\int\limits_{-1+i\epsilon}^{1+i\epsilon}\frac{dz}{4\pi}\left(H(z,x)h(z,y)-H(z,y)h(z,x)\right),
\\
\chi^-_m(x^+,y^-)=&\inturl \frac{dz}{4\pi}H(z,x^+)h(z,y^-)
+\intdlr \frac{dz}{4\pi}H(z,y^-)h(z,x^+)
\nonumber \\
=&\int\limits_{-1+i\epsilon}^{1+i\epsilon}\frac{dz}{4\pi}H(z,x^+)h(z,y^-)+\tfrac{i}{2}H(\tfrac{1}{y^-},x^+)
\nonumber \\
&
+\int\limits_{-1-i\epsilon}^{1-i\epsilon}\frac{dz}{4\pi}H(z,y^-)h(z,x^+)+\tfrac{i}{2}H(\tfrac{1}{x^+},y^-)
\nonumber \\
\rightarrow&
\int\limits_{-1+i\epsilon}^{1+i\epsilon}\frac{dz}{4\pi}H(z,x)h(z,\tfrac{1}{y})+\tfrac{i}{2}H(y,x)
\nonumber \\
&
+\int\limits_{-1-i\epsilon}^{1-i\epsilon}\frac{dz}{4\pi}H(z,\tfrac{1}{y})h(z,x)+\tfrac{i}{2}H(\tfrac{1}{x},\tfrac{1}{y})
,
\end{align}
with similar results for $\chi^-_m(x^-,y^+)$ and $\chi^-_m(x^-,y^-)$. Combining the expressions as in equation~\eqref{eq:def-of-theta-as-chis} it is easy to show that all the terms cancel upon using the identity
\begin{equation}
H(\tfrac{1}{x},\tfrac{1}{y})=H(x,y),
\end{equation}
and so $\sigma^-\rightarrow 1$.

\section{Singularities of the massless dressing factors}
\label{app:singularities-dressing}

We have seen that we do not expect any bound-states to appear in the massless kinematics by analysing the matrix part of the $\algSU(1|1)_{\text{c.e.}}$ S~matrix. Here we will check that no singularities arise from the proposed dressing factors. Starting from the AFS order, we have
\begin{equation}
\theta_{\text{AFS}}(x,y)=\big(x+\frac{1}{x}-y-\frac{1}{y}\big)\log \frac{(x-y)^2}{(1-x y)^2}.
\end{equation}
Clearly all poles of the resulting dressing factor should come from logarithmic singularities of the phase. However, any singularities of the logarithmic yield $\theta_{\text{AFS}}\approx 0\log0$ which is regular.

As for the Hernandez-Lopez order, it is convenient to use the dilogarithm expression of eqs. (3.4) and (3.9) in~\cite{Beisert:2006ib}. As usual, the dressing factor is given by products and ratios of the form $\exp(i\chi_{\text{HL}}(x^\pm_p,x^\pm_q))$. We introduce the short-hand notation
\begin{equation}
\chi^{\pm\pm}_{pq}\equiv \chi_{\text{HL}}(x^\pm_p,x^\pm_q),
\end{equation}
and we have
\begin{equation}
\label{eq:hl-dilog-expression}
\begin{aligned}
&2 \pi\chi^{\alpha \beta}_{pq}=
+\mbox{Li}_2 \frac{e^{i \alpha \frac{p_1}{4}}+ e^{- i \beta \frac{p_2}{4}}}{e^{i \alpha \frac{p_1}{4}}- e^{i \beta \frac{p_2}{4}}}
+ \mbox{Li}_2 \frac{e^{i \alpha \frac{p_1}{4}}- e^{- i \beta \frac{p_2}{4}}}{e^{i \alpha \frac{p_1}{4}}+ e^{i \beta \frac{p_2}{4}}}\\
&\qquad\qquad\qquad
- \mbox{Li}_2 \frac{e^{i \alpha \frac{p_1}{4}}- e^{- i \beta \frac{p_2}{4}}}{e^{i \alpha \frac{p_1}{4}}- e^{i \beta \frac{p_2}{4}}}
- \mbox{Li}_2 \frac{e^{i \alpha \frac{p_1}{4}}+ e^{- i \beta \frac{p_2}{4}}}{e^{i \alpha \frac{p_1}{4}}+ e^{i \beta \frac{p_2}{4}}}
-\Big(p_1\leftrightarrow p_2\Big),
\end{aligned}
\end{equation}
where $\alpha,\beta =\pm 1$. 
Singularities may arise when the argument of $\mbox{Li}_2 (z)$ approaches infinity, at which point the function diverges as $- \frac{1}{2} \log^2 (-z)$. Therefore  potential singularities come from 
\begin{equation}
\label{eq:amo}
e^{\frac{i}{4} (\alpha p_1-\beta p_2)} = \pm 1,\qquad \text{with}\quad 0\leq \Re p_i <2\pi.
\end{equation}
Let us start with the case when the right-hand-side is $-1$. Then, looking at the real part of the equation, we are requiring that the sum or difference of two numbers between $0$ and $\pi/2$ equals~$\pi$. This can only happen when both momenta lie at the boundary of the physical region. When the right-hand-side is $+1$, instead, we find that it must be $p_1=\pm p_2$ depending on $\alpha,\beta$. The case $p_1=-p_2$ corresponds to the singlet singularity at $x=1/y$ described in Section~\ref{sec:boundstates}, and it is easy to check that for generic values of the momenta there is no singularity due to cancellations among the different terms in~\eqref{eq:hl-dilog-expression}. Finally, the case while $p_1=p_2$ is a new configuration, which makes sense for real momenta, and where the dressing phase must be regular and indeed vanish in an appropriate branch due to antisymmetry.

\section{On the nesting procedure}\label{app:nesting}
In this appendix we collect some details of the derivation of the Bethe equation and the nesting procedure. For brevity, we omit some of the basic steps, for which we refer to~\cite{Borsato:2016hud}. In Appendix~\ref{sec:nesting-massless} we focus on the nesting procedure when including massless excitations. The nesting procedure at weak coupling is discussed in Appendix~\ref{sec:nest-proc-weak}. In Appendix~\ref{sec:nest-su2}, for completeness, we write the Bethe equations with a non-trivial $\algSU(2)_\circ$ factor of the S matrix.

\subsection{Nesting and the massless excitations}\label{sec:nesting-massless}
The nesting procedure for the massive sector was carried out in the spin-chain frame in~\cite{Borsato:2013qpa}.
Since here we use the string-frame S matrix of~\cite{Borsato:2015mma}, we prefer to rederive the results in the massive sector, which will now have some different frame-dependent factors.
This will make it simpler to compare to the calculations with massless excitations.

We choose to work in the bosonic grading of Section~\ref{sec:bos-grad-Bethe}, so that a level-I vacuum with only Left massive excitations is made up of $Y^{\smallL}$'s.
In order to diagonalise the S-matrix involving also excitations created by the action of $\gen{Q}^{\smallL 1}$, we start with the following ansatz for the two-particle state
\begin{equation}
\ket{\mathcal{Y}_{22}} = f_2(y,p_1) \ket{\eta^{\smallL 1}_{p_1} Y^{\smallL}_{p_2}} + f_2(y,{p_2}) S_{22}^{\text{II,I}}(y,p_1) \ket{Y^{\smallL}_{p_1} \eta^{\smallL 1}_{p_2}}.
\end{equation}
Above, the subscripts ``2'' 
are used to indicate that these level-I excitations correspond to node ``2''.
An eigenstate $\ket{\Psi}$ of the S-matrix can be found by making the ansatz  $\ket{\Psi}=\ket{\mathcal{Y}_{22}}+\mathcal{A}_{p_1p_2}\ket{\mathcal{Y}_{22}}_{\mathbf{\pi}}$ and imposing the compatibility condition
\begin{equation}\label{eq:compat-cond-level-II-22}
\mathcal{S} \ket{\mathcal{Y}_{22}} = \mathcal{A}^{22}_{p_1p_2} \ket{\mathcal{Y}_{22}}_{\mathbf{\pi}}.
\end{equation}
Here $\mathcal{A}^{22}_{p_1p_2}=\bra{Y^{\smallL}_{p_2}Y^{\smallL}_{p_1}}\mathcal{S}\ket{Y^{\smallL}_{p_1}Y^{\smallL}_{p_2}}$ is the scattering element between two level-I excitations, and $\ket{\mathcal{Y}_{22}}_{\mathbf{\pi}}$ is obtained from $\ket{\mathcal{Y}_{22}}$ by permuting the momenta $p_1$ and $p_2$.
The solution of equation~\eqref{eq:compat-cond-level-II-22} is
\begin{equation}\label{eq:sol-nest-22}
f_2(y,p_j)=  \frac{g_2(y) \eta_{p_j}}{h_2(y) - x^+_{p_j}},
\qquad
S_{22}^{\text{II,I}}(y,p_j)= \left(\frac{x^+_{p_j}}{x^-_{p_j}}\right)^{1/2} \frac{h_2(y) - x^-_{p_j}}{h_2(y) - x^+_{p_j}}  ,
\end{equation}
for some undetermined functions $g_2,h_2$ of the auxiliary root $y$.

This procedure can also be applied to the case when the two-particle state is made up of massless excitations. Now the level-I excitation\footnote{We may choose $\chi^1$ or $\chi^2$, since the $\algSU(2)_\circ$ part of the massless S-matrix was chosen to be trivial. See~\cite{Borsato:2016hud} for the derivation without this assumption.} is $\chi^1$ and the ansatz for the two-particle state containing level-I and level-II excitations is
\begin{equation}
\ket{\mathcal{Y}_{00}} = f_0(y,p_1) \ket{T^{1 1}_{p_1} \chi^{1}_{p_2}} - f_0(y,{p_2}) S_{00}^{\text{II,I}}(y,p_1) \ket{\chi^{1}_{p_1} T^{1 1}_{p_2}},
\end{equation}
where the minus sign comes from moving the supercharge past the fermion $\chi^1$.
Now the compatibility condition is
\begin{equation}\label{eq:compat-cond-level-II-00}
\mathcal{S} \ket{\mathcal{Y}_{00}} = \mathcal{A}^{00}_{p_1p_2} \ket{\mathcal{Y}_{00}}_{\mathbf{\pi}},
\end{equation}
where $\mathcal{A}^{00}_{p_1p_2}=\bra{\chi^{1}_{p_2}\chi^{1}_{p_1}}\mathcal{S}\ket{\chi^{1}_{p_1}\chi^{1}_{p_2}}$.
The solution to this equation turns out to be essentially the same as the previous one\footnote{This is a consequence of the basis we are using for massless excitations. In this basis the massless module is isomorphic to (two copies of) the Left module in a different grading.}
\begin{equation}\label{eq:sol-nest-00}
f_0(y,p_j)=  \frac{g_0(y) \eta_{p_j}}{h_0(y) - z^+_{p_j}},
\qquad
S_{00}^{\text{II,I}}(y,p_j)= \left(\frac{z^+_{p_j}}{z^-_{p_j}}\right)^{1/2} \frac{h_0(y) - z^-_{p_j}}{h_0(y) - z^+_{p_j}}  ,
\end{equation}
where $z^\pm_p$ are spectral parameters for massless excitations. We have introduced functions $g_0,h_0$ of $y$ which are in principle different from $g_2$ and $h_2$.

We can relate the unconstrained functions $g_0,h_0,g_2,h_2$ by looking at a level-I vacuum which contains at the same time massless and massive excitations.
The ansatz for the two-particle state and its permutation are now
\begin{equation}
\begin{aligned}
\ket{\mathcal{Y}_{20}} &= f_2(y,p_1) \ket{\eta^{\smallL 1}_{p_1} \chi^{1}_{p_2}} + f_0(y,{p_2}) S_{20}^{\text{II,I}}(y,p_1) \ket{Y^{\smallL}_{p_1} T^{1 1}_{p_2}},
\\
\ket{\mathcal{Y}_{20}}_{\mathbf{\pi}} &= f_0(y,p_2) \ket{T^{1 1}_{p_2} Y^{\smallL}_{p_1}} - f_2(y,{p_1}) S_{02}^{\text{II,I}}(y,p_2) \ket{\chi^{1}_{p_2} \eta^{\smallL 1}_{p_1}}.
\end{aligned}
\end{equation}
The equation
\begin{equation}\label{eq:compat-cond-level-II-20}
\mathcal{S} \ket{\mathcal{Y}_{20}} = \mathcal{A}^{20}_{p_1p_2} \ket{\mathcal{Y}_{20}}_{\mathbf{\pi}},
\end{equation}
with $\mathcal{A}^{20}_{p_1p_2}=\bra{\chi^{1}_{p_2}Y^{\smallL}_{p_1}}\mathcal{S}\ket{Y^{\smallL}_{p_1}\chi^{1}_{p_2}}$ is solved by identifying
\begin{equation}
\begin{aligned}
& h_0(y)=h_2(y)\equiv y,\qquad
&& S_{20}^{\text{II,I}}(y,p_j)=S_{00}^{\text{II,I}}(y,p_j)\\
&g_0(y)=g_2(y)=g(y),
&& S_{02}^{\text{II,I}}(y,p_j)=S_{22}^{\text{II,I}}(y,p_j).
\end{aligned}
\end{equation}
Above, $g(y)$ is an arbitrary function of $y$ which does not appear in the Bethe equations, and we have used the freedom in parameterising $h_0=h_2$ to define the auxiliary root $y$.

\subsection{Nesting at weak coupling}\label{sec:nest-proc-weak}
It is interesting to note that at weak coupling the S-matrix in the mixed sector greatly simplifies, and this has consequences on the nesting procedure. 
Let us start the discussion from the level-I vacuum $\ket{Y^{\smallL}Y^{\smallL}}$ and repeat the diagonalisation procedure for the S matrix that we obtain when keeping only the leading order in the $h\sim 0$ expansion.
In this limit it is useful to introduce $u$ variables for massive excitations defined as
\begin{equation}
p=2\, \text{arccot}(2 u).
\end{equation}
It is then easy to check that the results for the diagonalisation at weak coupling are essentially the ones we obtain by taking the weak coupling limit of the all-loop results. Because $x^\pm_p \sim \order(1/h)$ for $h\to 0$, we then need to rescale $h_2(y)\to h_2(y)/h$, $g_2(y)\to g_2(y)/h$ in equation~\eqref{eq:sol-nest-22}. In particular, one obtains at weak coupling
\begin{equation}
S_{22}^{\text{II,I}}(y,u_j)= \left(\frac{u_j+\tfrac{i}{2}}{u_j-\tfrac{i}{2}}\right)^{1/2} \frac{h_2(y)-u_j+\tfrac{i}{2}}{h_2(y)-u_j-\tfrac{i}{2}},
\end{equation}
where the first factor is $e^{ip/2}$ expressed in terms of $u$.

On the other hand, the massless S matrix has \emph{no explicit $h$-dependence}. Therefore, the diagonalisation procedure works exactly as above, with no need to introduce $u$-variables for massless excitations. Because $\eta_p$ is proportional to $\sqrt{h}$ for massless excitations, we do need to rescale $g_0(y) \to g_0(y)/\sqrt{h}$ in equation~\eqref{eq:sol-nest-00}.
Nevertheless, unlike $h_2(y)$, $h_0(y)$ is not rescaled with powers of $h$.

Looking at the mixed level-I vacuum, one discovers that now the compatibility condition~\eqref{eq:compat-cond-level-II-20} imposes weaker constraints. In particular, it is not necessary anymore to identify  the functions $h_0$ and $h_2$. The immediate consequence of this is that at weak coupling one has distinct sets of auxiliary roots for massive and massless excitations, as given in equation~\eqref{eq:aux-roots-weak}.
Moreover, one finds that the scattering processes between level-II and level-I excitations in the mixed-mass case do not depend on the auxiliary variables\footnote{As before, these results may be obtained by doing a weak-coupling expansion of the all-loop results, after the rescalings of $h_j, g_j$ with powers of $h$ explained previously and without using the identifications of $h_0,h_2$ and $g_0,g_2$.}
\begin{equation}
S_{20}^{\text{II,I}}(y,p_j)= e^{ip_2/2},
\qquad\qquad
S_{02}^{\text{II,I}}(y,u_j)=\left(\frac{u_j+\tfrac{i}{2}}{u_j-\tfrac{i}{2}}\right)^{-1/2},
\end{equation}
and these factors of $e^{ip}$ can be reabsorbed in the redefinition of the length $L$ in the Bethe equations, as in~\eqref{eq:Bethe-weak-2}.

\vspace{12pt}

A simplification at weak coupling happens also in the LR massive sector. 
In that case, we may consider the level-I vacuum $\ket{Z^{\smallR}Z^{\smallR}}$ and study level-II excitations obtained by acting on it with the lowering supercharge $\gen{Q}^{\smallL 1}$, as in Section~\ref{sec:nesting-massless}. At all loops we find a scattering element between level-II and level-I which depends on the function $h_{\bar{2}}(y)$. Demanding compatibility with the mixed vacuum $\ket{Y^{\smallL}Z^{\smallR}}$ one finds that we should again identify $h_{\bar{2}}(y)=h_{2}(y)$.

At weak coupling, instead, the scattering elements between level-II and level-I which involve Right excitations are just factors of $e^{ip}$, and hence we do not even need to introduce the function $h_{\bar{2}}(y)$. Therefore, at weak coupling the auxiliary roots for $\gen{Q}^{\smallL 1}$ do not represent Right massive excitations.
The situation is obviously reversed when we look at the case of the lowering supercharge $\overline{\gen{Q}}_{\smallR}{}^{1}$, whose auxiliary roots only represents Right excitations an not Left ones.
This shows that at weak coupling we cannot combine anymore the auxiliary roots associated to $\gen{Q}^{\smallL a}$ and $\overline{\gen{Q}}_{\smallR}{}^{a}$ supercharges, as done at all loops in Section~\ref{sec:bos-grad-Bethe}.

A way to obtain the result at weak coupling from the all-loop result of Section~\ref{sec:bos-grad-Bethe} is to allow auxiliary roots $y_{J,j}$ to scale differently with powers of $h$: scalings as $1/h$ or $h$ would correspond to Left massive and Right massive excitations respectively, while no scaling with $h$ corresponds to massless excitations.

\subsection{\texorpdfstring{Nesting and the $\algSU(2)_{\circ}$ root}{Nesting and the su(2)\string_o root}}\label{sec:nest-su2}
In the main text we have assumed that the $\algSU(2)_\circ$ factor of the S matrix is trivial, as suggested by perturbation theory. To achieve this we have sent the function of the momentum $w_p \to \infty$. If we had not done that, scattering between the massless fermions $\chi^1$ and $\chi^2$ would have non-trivial transmission and reflection contributions depending on $w_p$. Therefore only one of these excitations should be assigned to level I, and we would need to use the nesting procedure also to diagonalise the action of the $\algSU(2)_\circ$ lowering operator.
If we decide to put $\chi^1$ in level I and regard $\chi^2$ as a level-II excitation, then the Bethe equations for massless excitations~\eqref{eq:BA-0} would contain an additional factor on the right-hand-side
\begin{equation}
    \label{eq:BA-0-withsu2}
      \left(\frac{z_k^+}{z_k^-}\right)^L =
      \cdots
	\prod_{j=1}^{N_{4}}  \frac{w_k - y_{4,j} - i/2}{w_k -y_{4,j} + i/2} ,
\end{equation}
where the dots stand for the other factors in~\eqref{eq:BA-0}, which we do not repeat.
Here $y_{4,j}$ are new auxiliary roots, associated to the action of the $\algSU(2)_\circ$ lowering operator.
They satisfy their own Bethe equations, 
\begin{equation}    \label{eq:BA-su(2)}
   1  
   = \prod_{\substack{j = 1\\j \neq k}}^{N_{4}} \frac{y_{4,k} - y_{4,j} + i}{y_{4,k} - y_{4,j} - i}
    \prod_{j=1}^{N_{0}} \frac{y_{4,k} - w_j - i/2}{y_{4,k} - w_j + i/2}  ,
\end{equation}
which contain the interaction between level I and level II, as well as a level II self-interaction. Notice further that the $y_{4,j}$ do not couple to the massive roots.

\bibliographystyle{nb}
\bibliography{refs}

\end{document}